\documentclass{article}
\pdfoutput=1
\usepackage{amsmath}
\usepackage{amssymb}
\usepackage{amscd}
\usepackage{pb-diagram}
\usepackage{graphicx}
\usepackage{epstopdf}

\def\be{\begin{eqnarray}}
\def\ee{\end{eqnarray}}
\def\nn{\nonumber}

\def\p{\partial}

\def\l[{\phantom.[}

\def\Rede{Redemeister}
\def\lD{{\cal D}}

\def\span{{\rm span}}
\def\bu{\bullet}
\def\w{\circ}
\def\dq{{\rm dim}_q}
\def\ker{{\rm Ker}}
\def\im{{\rm Im}}
\def\coim{{\rm CoIm}}

\def\l[{\phantom.\![}

\hoffset= - 1.0in         % switch off for draft style
\begin{document}

\hfill ITEP/TH-31/13

\bigskip

\centerline{\Large{Introduction to Khovanov Homologies.
}}
\centerline{\Large{III.
%Pedestrian derivation
A new and simple tensor-algebra construction
}}
\centerline{\Large{of Khovanov-Rozansky invariants
}}

\bigskip

\centerline{V.Dolotin and A.Morozov}

\bigskip

\centerline{\it ITEP, Moscow, Russia}

\bigskip

\centerline{ABSTRACT}

\bigskip

{\footnotesize
We continue to develop the tensor-algebra approach
to knot polynomials with the goal to present
the story in elementary and comprehensible form.
The previously reviewed description
of Khovanov cohomologies for the gauge group of rank $N-1=1$
was based on the cut-and-join calculus of the
planar cycles, which are involved rather artificially.
We substitute them
by alternative and natural set of cycles,
not obligatory planar.
Then the whole construction is straightforwardly lifted from
$SL(2)$ to $SL(N)$ and reproduces Khovanov-Rozansky (KR) polynomials,
simultaneously for all values of $N$.
No matrix factorization and related tedious calculations are
needed in such approach, which can therefore become not only
conceptually, but also practically useful.
}

\bigskip

\bigskip

\tableofcontents

\section{Introduction}

The theory of knot polynomials \cite{knp,Wit} is nowadays one of the
fast-developing branches of theoretical and mathematical physics,
unifying critical ideas from many different other subjects,
from topology to localization and AGT relations.
A special part of the story is relation to index theorems
and homological algebra.
In physical language this is the possibility to reinterpret
the averages of characters in Chern-Simons theory (HOMFLY polynomials)
as Euler characteristics of certain complexes, invariant
under infinitesimal deformations of curves,
and further promote them to Poincare polynomials of the same
complexes in a way, which preserves the invariance.
Poincare polynomial is a generating function of cohomologies,
therefore the task is to use topological invariance to
reduce the functional integral to the infrared --
to the zero-modes of differential operators, which can be
rewritten as acting on parameters ("times") of the
low-energy effective action.
Remarkably, these operators have a typical form of cut-and-join
operators -- or, what is nearly the same, of the Hamiltonians
of integrable systems.

We reviewed the main technical parts of this
Khovanov's categorification program \cite{Kho1}-{\cite{Khol} in
\cite{DM1,DM2} with the main example of $GL(2)$ gauge group
and the fundamental representation.
This theory of Jones superpolynomials is very transparent and
clear, especially after the pedagogical presentation of
D.Bar-Natan \cite{BN} and following advances in computerization.
The problem is, however, far more severe for higher $GL(N)$
groups, where the main results are obtained with the help
of Khovanov-Rozansky construction \cite{KhR},
based on the additional technique of matrix factorization,
which makes the story obscure both conceptually and technically.
We return to this construction -- quite beautiful by itself --
in one of the papers of our review series,
but before that we prefer to present a natural generalization
of the $N=2$ story, immediately implied by the tensor algebra
approach, which was our starting point in \cite{DM1},
and {\it without any direct reference or use of matrix factorization}.

Technically the peculiarity of $N=2$ was that representations
of $SL(2)$ are real, and this allowed to substitute the naturally
appearing cycle decomposition, true for all $N$, by
that into {\it planar} cycles -- and {\it such} construction
seemed and was un-generalizable to arbitrary $N$.
However, as we show in the present paper, the naive one \cite{DM1},
with non-planar cycles,
actually reproduces all the results for $N=2$ -- and works just the
same way for {\it all} values of $N$.
Moreover, it directly provides the answers as explicit
functions of $N$.
We restrict in this paper to the simplest examples,
and reproduce just the very first items of the currently available
list of Khovanov-Rozansky polynomials
(both reduced and unreduced)
in the fundamental representation, worked out by terribly
complicated calculations in \cite{CM}.\footnote{
While knowing ordinary knot polynomials for generic $N$
immediately provides HOMFLY polynomials, depending on $A=q^N$
instead of $N$, the story is more complicated in the case
of {\it superpolynomials}.
The basic difference is that Khovanov-Rozansky polynomials
depend on the quantum numbers
$[N-k] = \frac{Aq^{-k}-A^{-1}q^k}{q-q^{-1}}$ and therefore
themselves are {\it not} Laurent polynomials of $A$ with
{\it positive} coefficients, as superpolynomials are
requested to be (at least in the fundamental representation).
In fact, the $N$-dependent Khovanov-Rozhansky polynomials
live in a seemingly non-trivial factor-space
of the $A$-dependent superpolynomials -- and the lifting
to superpolynomials can be a little tricky \cite{DGR}-\cite{art}.
However, the study of this lifting is still severely restricted
by the lack of diverse examples -- this adds to the need
of developing technical means to effectively produce arbitrary
Khovanov-Rozansky polynomials.
}
Our feeling is that with the alternative technique,
suggested in the present paper, the list can be reproduced
and substantially enlarged -- to the same extent as it is
available for Jones superpolynomials in \cite{katlas}.

\section{From knots to knot diagrams and tensor algebras
\label{frokno}}

The very first step in the theory of knot polynomials
is to reformulate the problem in terms of knot diagrams --
the graphs with colored vertices.
In knot theory the graphs

(i) \ \ are planar,

(ii) \ have vertices of valence $(2,2)$,

(iii) just two colors are allowed.

\noindent
The tensor-algebra construction
in the style of \cite{nonlin}, which we are going to use
is in no way restricted
by these choices -- still in the present paper we discuss this
standard setting, and some technicalities will depend on it.

For the way to reduce to this form the more conventional formulations
of the problem -- either in terms of knot theory
or in those of Chern-Simons correlators in temporal gauge --
see \cite{MoSmi} and references therein.
In one word, knot diagrams (planar graphs) appear when
oriented lines in $3$ dimensions (oriented knots or links)
are projected on $2$-dimensional plane.
Not to loose information,
one should distinguish which of the two lines was above another,
when their projections cross -- this means that there are
two different types of vertices, which we call black and white
(in ${\cal R}$-matrix formalism they would be associated with
${\cal R}$ and ${\cal R}^{-1}$).
To keep topological invariance -- equivalence of different
projections of the same link/knot -- one should consider only
\Rede-invariant functions on the graphs.

\bigskip

Our starting point is just the theory of $(2,2)$-valent
planar graphs $\lD_c$ with vertices of two kinds (colors),
and our main claim is that the "physical input"
implies simply the need to consider {\it invariant tensors}
of the tensor algebra $T_N$ -- this condition alone will lead
us to \Rede-invariant knot polynomials.
Exactly like in the case of $N=2$, for all $N$
the HOMFLY invariants will just count the number of cycles
in the resolution of knot diagram
(which however, still need to be properly defined),
while Khovanov-Rozansky ones will count  cohomologies (Poincare polynomials)
of associated complexes, made in a nearly canonical way
from the vector spaces (actually,  for generic $N$
these are factor-spaces).
HOMFLY are their Euler characteristics, and depend only
on {\it dimensions} of the vector spaces, not on the morphisms between
them, and these dimensions are just made from the (graded) numbers of
above-mentioned cycles.
In other words,
{\bf direct application of tensor-algebra ideas {\it a la}
\cite{nonlin} provides a natural but previously unnoticed
construction of the commutative quiver on the hypercube,
involving vector spaces $V=C^N$ of arbitrary dimension $N$,
so that the properly normalized Poincare polynomial of associated
quiver complex reproduces Khovanov-Rozansky polynomials.
}
This brief description implies the familiarity with either
\cite{BN} or \cite{DM1,DM2} -- for the sake of completeness
we repeat that standard construction in the case of $N=2$
in section \ref{basics} below,
while now we return to tensor algebra.

If no other structures are introduced, the tensor algebra
$T_N$ itself has just two $SL(N)$-invariant tensors:
\be
\delta^i_j \ \ \ \ \ \ \ {\rm and} \ \ \ \ \ \ \
\epsilon^{i_1\ldots i_N}
\ \ \ \ \ \ \ \ \ \ \ \
i,j=1,\ldots,N
\ee
The covariant $\epsilon_{i_1\ldots i_N}$ is made out of those two.
The fact that there are exactly {\it two} invariant tensors
appears to match perfectly with the desire to have vertices
of exactly {\it two} kinds (colors).
But to make the contact, we need first to get the proper valences.
Valence $(2,2)$ means that we need tensors with two upper and two lower
indices. Clearly, there are exactly three options:
\be
\delta^i_k \delta^j_l, \ \ \ \ \
\epsilon^{ij m_1\ldots m_{N-2}}\epsilon_{kl m_1\ldots m_{N-2}}
\ \ \ \ \ \ {\rm and} \ \ \ \ \
\delta^i_l\delta^j_k
\ee
The first two of them are planar, the third is not --
this is the reason why for $N=2$ one uses the linear combinations
of $\delta^i_k \delta^j_l$ and $\epsilon^{ij}\epsilon_{kl}$.
Still this choice is not so obvious.
For integer $N$ (or in the case of no $q$-deformation, if one prefers
this language) the three structures are linearly dependent:
\be
\epsilon^{ij m_1\ldots m_{N-2}}\epsilon_{kl m_1\ldots m_{N-2}}
= (N-2)!\ \Big(\delta^i_k \delta^j_l - \delta^i_l\delta^j_k\Big),
\ \ \ \ \ \ \ q=1
\label{epdd}
\ee
For $q\neq 1$ things are not so simple,
(see sec.\ref{Kauf} below),
still the dilemma of which
two of the three vertices to choose remains.

\bigskip

The conventional approach is to take
$\delta^i_k \delta^j_l$ and $\epsilon^{ij}\epsilon_{kl}$ for $N=2$,
this decomposes {\it resolved} knot diagrams into units
of {\it planar} cycles
-- but is well known not to work for $N\neq 2$
(does not give anything besides Jones polynomials, at best).
Starting from sec.\ref{modif}, {\bf we switch to
alternative choice:}
\be
\delta^i_k \delta^j_l \ \ \ \ \  and
\ \ \ \ \ \ \delta^i_k \delta^j_l-\delta^i_l \delta^j_k
\label{dddd}
\ee
Now the cycles are not all planar,
moreover some of them enter with negative signs,
but instead the construction appears to work
not only for $N=2$, but for arbitrary $N$.
"Works" means that it provides \Rede-invariant answers,
which depend non-trivially on $N$, moreover,
these answers coincide with those from \cite{CM},
derived with the help of the standard
matrix-factorization-induced Khovanov-Rozansky construction.

\newpage

The plan of this paper follows as close as possible
the main logic of Khovanov's approach:

\bigskip

\bigskip

\begin{tabular}{cclcccccc}
&&&&&& knot/link \\  \\
&&&&&& $\downarrow$ \\    \\
&&&&&& knot/link diagram $\lD_c$ && {\it sec.\ref{frokno}}\\  \\
&&&&&& $\downarrow$ \\     \\
&&&&&& $c$-independent hypercube $H(\lD)$\\
&&&&&& of resolutions && {\it sec.\ref{basics}}\\
&&&&&& and cycle decomposition \\  \\
&&&&&$\swarrow$ & $\downarrow$ \\ \\
&{\it sec.\ref{Jones}} &{\footnotesize (Jones, $N=2$)}
&&HOMFLY polynomial && $q$-graded vector \\
& &&& ${\cal H}^{(\lD_c)}(N|q)$, &&  factor-spaces, && {\it sec.\ref{KRco}}  \\
&{\it sec.\ref{modif}}&{\footnotesize (HOMFLY)}
&&counting cycles&& associated with cyclvaes\\  \\
&&&& $\uparrow$ && $\downarrow$ \\    \\
&{\it sec.\ref{KRco}}&&& $q$-Euler characteristic&& linear maps (morphisms)
&&{\it sec.\ref{KRco}}\\
&&&& of $K(\lD_c)$ && between vector spaces, making \\
&&&& && from $H(\lD_c)$ a commutative quiver \\ \\
&&&& &$\nwarrow$& $\downarrow$ \\   \\
&&&& && associated complex $K(\lD_c)$&& {\it sec.\ref{KRco}}\\  \\
&&&& && $\downarrow$ \\  \\
&&&& && Khovanov-Rozansky polynomial\\
&&&& && ${\cal P}^{(\lD_c)}(N|q|T),$ && {\it sec.\ref{KRco}}\\
&&&& && counting its cohomologies \\ \\
\end{tabular}
%}

\bigskip

Section \ref{specu} describes the first steps towards
a similar systematization of the results of sec.\ref{KRco}
for KR superpolynomials,
but this story has more subtleties and interesting
deviations -- it will be continued in more detail elsewhere.

\bigskip

The "global" approach, suggested in the present paper is to
follow the chain:
\be
{\rm knot/link} \ \longrightarrow \ {\rm knot\ diagram}\ \lD_c
\ \longrightarrow\ {\rm hypercube}\ H(\lD) \ ({\rm cycles\ diagram})
\ \longrightarrow \nn \\
\longrightarrow \ {\rm quantization\ of\ dimensions\ of\ vector\ spaces\
at\ the\ vertices\ of} \ H(\lD)\ \longrightarrow \nn \\
\longrightarrow\ c-{\rm dependent\ morphisms\ between\ the\ spaces\ which\
decrease\ the\ gradation\ by\ one} \ \longrightarrow \nn \\
\longrightarrow \
{\rm cohomologies\ of\ associated\ complex} = {\rm KR\
superpolynomial\ for\ arbitrary}\ N
\ee
As presented in this paper, our construction is not fully algorithmic.
The two points, where some art is applied, are the quantization of
dimensions -- here one can control the choice by comparison
with the HOMFLY polynomials -- and adjustment of morphisms:
at this stage we use a very appealing "maxmal-subtraction" rule.
We also do not {\it fully} prove here the \Rede\ invariance.
It does not look too difficult to formalize all these details,
but our goal in this paper is rather to demonstrate the spirit
of our radical modification of Khovanov-Rozansky approach
and its impressive effectiveness and simplicity in concrete
examples.

\newpage

\section{Basic ideas in the case of $N=2$ and beyond \label{basics}}

We begin by describing the general ideas
of Khovanov approach, following \cite{BN} and \cite{DM1,DM2}.

\subsection{Hypercube $H(\lD)$ of colorings}

Consider not just a given link diagram $\lD_c$,
but the whole set with all possible colorings,
i.e. the given graph $\lD$ with all possible
colors at its vertices.\footnote{
There always exists
one particular coloring, which trivializes (unties)
the link, and knowing connections between
different colorings can provide information
about original link (with original coloring).
This idea can look similar to that of the skein
relations, but those are explicitly exploiting
the properties  of the
quantum ${\cal R}$-matrix (the knowledge of its eigenvalues),
which we do not use here. Of course, both approaches
are intimately related -- but not all the details
are yet understood about this relation.
}

If just two colors are allowed, we number
of colorings of the $n$-vertex graph $\lD$
is $2^n$, and what we get is an
$n$-dimensional hypercube $H(\lD)$, where
each vertex represents one particular coloring
$c$ of $\lD$:
\be
H(\lD) = \{c\}
\ee
Edges of the hypercube are naturally associated
with the elementary flips -- inversions of
color at one particular vertex of $\lD$.

Original link has particular coloring,
thus it is associated with one particular
"initial" vertex $c_0$ of the hypercube.
Once it is specified, edges become arrows,
pointing away from $c_0$.

\begin{picture}(100,110)(-290,-65)
\put(-180,0){\circle{40}}
\put(-160,0){\circle{40}}
\put(-170,16){\circle*{6}}
\put(-170,-16){\circle*{6}}
\put(-200,0){\vector(0,1){2}}
\put(-140,0){\vector(0,1){2}}
\put(0,0){\vector(2,1){40}}
\put(0,0){\vector(2,-1){40}}
\put(40,20){\vector(2,-1){40}}
\put(40,-20){\vector(2,1){40}}
\put(-10,-4){\circle*{5}}
\put(-10,4){\circle*{5}}
\put(35,27){\circle{5}}
\put(35,35){\circle*{5}}
\put(35,-27){\circle{5}}
\put(35,-35){\circle*{5}}
\put(90,-4){\circle{5}}
\put(90,4){\circle{5}}
\put(-172,-45){\mbox{$D$}}
\put(30,-55){\mbox{$H(D)$}}
\end{picture}

\Rede moves are associated with duplication of
the hypercube. For example, adding an elementary
loop in $R1$ introduces one extra vertex in $\lD$,
what implies that the new hypercube consists
of two copies of the original one.
Similarly, $R2$ adds two vertices of different
color to $\lD$ -- then the new hypercube consists
of four copies of the original,
while $R3$ relates the result of adding three
vertices to $D$ in two different ways
and thus involves a three-dimensional
sub-cube in $H(\lD)$.

\begin{picture}(100,150)(-250,-80)
\put(-180,0){\circle{40}}
\put(-160,0){\circle{40}}
\put(-170,16){\circle*{6}}
\put(-170,-16){\circle*{6}}
\put(-200,0){\vector(0,1){2}}
%\put(-140,0){\vector(0,1){2}}
\put(-125,0){\circle{30}}
\put(-140,0){\circle{6}}
\put(0,20){\vector(4,1){40}}
\put(0,20){\vector(4,-1){40}}
\put(40,30){\vector(4,-1){40}}
\put(40,10){\vector(4,1){40}}
\put(-10,20){\circle*{5}}
\put(-10,27){\circle*{5}}
\put(-10,36){\circle{5}}
\put(25,-16){\circle*{5}}
\put(25,-25){\circle*{5}}
\put(25,-32){\circle*{5}}
\put(30,52){\circle{5}}
\put(30,43){\circle*{5}}
\put(30,36){\circle{5}}
%\put(35,-27){\circle{5}}
%\put(35,-35){\circle*{5}}
\put(90,36){\circle{5}}
\put(90,27){\circle{5}}
\put(90,20){\circle{5}}
\put(125,-16){\circle*{5}}
\put(125,-25){\circle{5}}
\put(125,-32){\circle{5}}
\put(88,-33){\circle*{5}}
\put(88,-41){\circle{5}}
\put(88,-48){\circle*{5}}
\put(75,62){\circle{5}}
\put(75,53){\circle{5}}
\put(75,46){\circle*{5}}
\put(69,40){\vector(-1,-1){24}}
\put(140,-53){\circle*{5}}
\put(140,-61){\circle*{5}}
\put(140,-68){\circle{5}}
\put(129,-58){\vector(-1,1){44}}
\put(40,-20){\vector(4,1){40}}
\put(40,-20){\vector(4,-1){40}}
\put(80,-10){\vector(4,-1){40}}
\put(80,-30){\vector(4,1){40}}
\put(0,20){\vector(1,-1){40}}
\put(80,20){\vector(1,-1){40}}
\put(40,30){\vector(1,-1){40}}
\put(40,10){\vector(1,-1){40}}
\put(-172,-45){\mbox{$\lD$}}
\put(36,-60){\mbox{$H(\lD)$}}
\end{picture}

\subsection{Colorings as resolutions}

Now, the question is -- what are we going to
associate with this hypercube $H(\lD)$
in such a way that the \Rede\ duplications
do not affect this quantity?

Original Khovanov construction deals with the
planar graphs of valence $(2,2)$ and
is based on counting connected cycles
in the resolutions of $\lD$.
The black and white vertices $\bullet$ and $\circ$
are interpreted as two different resolutions.

Initial knot diagram $\lD_{c_0}$  corresponds to a particular
coloring/resolution, and the corresponding vertex ${c_0}$
of the hypercube becomes "initial".
After that any other hypercube vertex $c\in H(\lD)$ is characterized
by a number $h_0(c) = h(c)-h(c_0)$ of edges,
separating it from initial $c_0$.

\subsection{Morphisms between the resolutions}

The edges of the hypercube connect resolved diagrams of the same
topology, but with one vertex resolved differently.
In this sense an edge is naturally associated with the morphisms
between the two resolutions at one vertex of $\lD$.

\bigskip

The questions then are what is the relevant choice of resolutions
and morphisms.

Actually at the level of HOMFLY polynomials the morphisms do not
matter, only the choice of resolutions is important.
Thus we begin from them.

\section{Standard approach to Jones polynomials: planar cycles
\label{Jones}}

\subsection{Resolutions, leading to planar cycles
\label{plare}}

In this standard approach
the two different resolutions, associated with
the vertices $\bullet$ and $\circ$
are  the following:

\begin{picture}(100,100)(-100,-50)
\put(5,5){\vector(1,1){15}}
\put(-5,5){\vector(-1,1){15}}
\put(-20,-20){\vector(1,1){15}}
\put(20,-20){\vector(-1,1){15}}
\put(0,0){\circle*{10}}
\put(40,0){\mbox{$=$}}
\put(75,-20){\vector(0,1){40}}
\put(95,-20){\vector(0,1){40}}
\put(205,5){\vector(1,1){15}}
\put(195,5){\vector(-1,1){15}}
\put(180,-20){\vector(1,1){15}}
\put(220,-20){\vector(-1,1){15}}
\put(200,0){\circle{10}}
\put(235,0){\mbox{$=$}}
\qbezier(260,20)(270,0)(280,20)
\qbezier(260,-20)(270,0)(280,-20)
\put(263,-15){\vector(1,1){2}}
\put(277,-15){\vector(-1,1){2}}
\put(263,15){\vector(-1,1){2}}
\put(277,15){\vector(1,1){2}}
\end{picture}

\noindent
They are dictated in an obvious way \cite{DM1} by the existence of two
invariant tensors $\delta^i_j$ and $\epsilon_{ij}$
in the tensor algebra with $N=2$.
This choice is difficult to generalize {\it literally} to $N>2$,
though its minor modification is easily generalizable -- as we shall
see in the next section \ref{modif}.
But first we proceed with the standard approach.

Once resolutions are chosen,
the planar graph $\lD_c$ at the vertex $c$ of the hypercube
decomposes into $\nu_c$ disconnected cycles.
Thus with each vertex $c\in H(D)$
one associates two numbers: this $\nu_c$ and $h_c=h_0(c)$,
which is the distance between $c$ and initial $c_0$.

The crucial observation is that the \Rede\ moves
change $\nu_c$ and $h_c$ in a simple way.

\begin{picture}(100,150)(-250,-80)
%
%\put(-180,0){\circle{40}}
%\put(-160,0){\circle{40}}
%\put(-170,16){\circle*{6}}
%\put(-170,-16){\circle*{6}}
\qbezier(-160,30)(-120,0)(-160,-30)
%\put(-200,0){\vector(0,1){2}}
%\put(-140,0){\vector(0,1){2}}
\put(-125,0){\circle{30}}
\put(-140,0){\circle{6}}
\qbezier(0,55)(30,15)(40,40)
%\qbezier(25,40)(30,30)(40,40)
\qbezier(40,40)(60,60)(60,30)
\qbezier(40,20)(60,0)(60,30)
\qbezier(0,5)(30,35)(40,20)
\qbezier(0,-10)(40,-35)(0,-60)
\put(45,-35){\circle{30}}
\put(-90,15){\vector(3,1){70}}
\put(-90,-15){\vector(3,-1){70}}
\put(-25,50){\circle{5}}
\put(-25,-25){\circle*{5}}
\end{picture}

The first \Rede\ move $R1$ duplicates
the hypercube:
$H(\lD) \rightarrow H(\lD) \cup H'(\lD)$,
so that the corresponding vertices of $H'(D)$
have $\nu'_c = \nu_c +1$ and $h'_c = h_c + 1$.
This is because
when the white resolution is chosen,
the number of cycles does not change at all,
while for the black resolution exactly one
cycle is added.
Now we can easily write down an invariant of $R1$:
\be
I\{\nu_c,h_c\} \ \stackrel{R1}{\longrightarrow}\
I\{\nu_c \cup \nu_c+1,\ h_c \cup h_c+1\}: \nn \\
(-)^n \sum_{c=1}^{2^n} (-)^{h_c} 2^{\,\nu_c}
\ \ \stackrel{R1}{\longrightarrow}\ \
(-)^{n+1}\left(\sum_c (-)^{h_c} 2^{\,\nu_c} +
\sum_c (-)^{h_c+1} 2^{\,\nu_c+1}\right)\ =\
(-)^n \sum_{c}^{} (-)^{h_c} 2^{\,\nu_c}
\ee
\noindent
Similarly one can check that this is also invariant of
$R2$ and $R3$.

\subsection{Towards knot/link polynomials}

From here one can go in different directions.
One can generalize to other types of graphs
(non-planar, non-oriented, with other types of
vertices and colorings) -- for this one needs
to modify the idea to associate colors with
resolutions of $\lD$ and cycles.
Instead one can extend the invariant from just
a number to function of several variables --
i.e. to something closer to the knot polynomials.
In what follows we proceed in this second direction.

One step seems obvious.
Since invariance under $R1$ above was based
on the identity
\be
\underline{\underline{(-1)}}\Big(\underline{\underline{\underline{\underline{1}}}}
+\underline{(-1)}\cdot \underline{\underline{\underline{2}}}\Big) = 1
\label{relR1}
\ee
it is natural to deform any of the underlined
four parameters.
Actually, they are all of different nature
and can be deformed independently.
Since so far we have only one relation (\ref{relR1}),
one can expect at least {\it three} independent deformations.
As we shall see, this is the right expectation,
but actually it will not be quite so simple to
find all the three.

The problem is that so far we looked only at $R1$,
moreover, even this we did not do exhaustively.
When we wrote that
\be
R1_\circ:\ \ \ \ \{\nu_c,h_c\} \longrightarrow \{\nu_c,h_c\} \cup
\{\nu_c+1,h_c+1\}
\label{R1act}
\ee
it was true only for inserting an elementary loop with a
white vertex, so that its resolution does not add a new
disconnected cycle. However, the resolution of the black
vertex does the opposite: adds the cycle and increases
$\nu_c$ by one.
when we insert the white vertex, the white resolution goes
first, and the black one the second -- thus we obtain (\ref{R1act}).
However, if we insert an elementary loop with a black vertex,
then black resolution goes first and white resolution second,
so that (\ref{R1act}) will be substituted by
\be
R1_\bullet:\ \ \ \ \{\nu_c,h_c\} \longrightarrow \{\nu_c+1,h_c\} \cup
\{\nu_c,h_c+1\}
\label{R1act*}
\ee
Our invariant now changes sign:
\be
(-)^{n+1}\left(\sum_c (-)^{h_c} 2^{\,\nu_c+1} + \sum_c (-)^{h_c+1} 2^{\,\nu_c}\right)
= -(-)^n \sum_c (-)^{h_c}2^{\,\nu_c}
\ee
moreover this time the relevant identity is slightly different:
\be
(-1)\Big(2 + (-1)\cdot 1\Big) = -1
\ee
To get rid of the sign difference we can assume that the overall sign
factor in fact counts only white vertices of $\lD$, while black ones enter
instead with the factor unity, i.e. invariant of both $R1_\circ $ and
$R1_\bullet$ is
\be
J^{c_{_0}}(\lD) = (-)^{n_\bullet} \sum_{c=1}^{2^{n_\circ+n_\bullet}}
(-)^{h_c-h_{c_{_0}}}\cdot 2^{\,\nu_c}
\label{ininv}
\ee
We also made notation more adequate:
invariant depends on the link diagram $\lD$ and initial coloring $c_0$
and the "height" $h_c$ is counted as the distance from $c_0$.

It is this quantity that we are going to deform.
The most general expression that we can write down,
preserving the structure of (\ref{ininv}) is
\be
J^{c_{_0}}(\lD) = \alpha_\bullet^{n_\bullet}\alpha_\circ^{n_\circ}
\sum_{c=1}^{2^{n_\circ+n_\bullet}}
(-q)^{h_c-h_{c_{_0}}}\cdot D^{\,\nu_c}
\label{defoinv}
\ee
The two constraints that we already know state that
\be
\alpha_\circ (1 -q D) = 1, \nn \\
\alpha_\bullet(D -q ) =1
\label{r1cons}
\ee
what defines $\alpha_\circ$ and $\alpha_\bullet$ through the other two parameters.

\subsection{Implication of $R2$ invariance}

The \Rede\ move $R2$ substitutes each vertex of the hypercube $H(\lD)$ by
a square:

\begin{picture}(100,160)(-200,-80)
%
%\put(-180,0){\circle{40}}
%\put(-160,0){\circle{40}}
%\put(-170,16){\circle*{6}}
%\put(-170,-16){\circle*{6}}
\qbezier(-160,40)(-80,0)(-160,-40)
\qbezier(-100,40)(-180,0)(-100,-40)
%\put(-200,0){\vector(0,1){2}}
%\put(-140,0){\vector(0,1){2}}
%\put(-125,0){\circle{30}}
\put(-130,20){\circle{6}}
\put(-130,-20){\circle*{6}}
\qbezier(-30,40)(0,10)(30,40)
\qbezier(-10,-10)(-30,20)(0,20)
\qbezier(-10,-10)(10,-20)(-30,-40)
\qbezier(10,-10)(30,20)(0,20)
\qbezier(10,-10)(-10,-20)(30,-40)
\qbezier(130,-40)(160,-10)(190,-40)
\qbezier(150,10)(130,-20)(160,-20)
\qbezier(150,10)(170,20)(130,40)
\qbezier(170,10)(190,-20)(160,-20)
\qbezier(170,10)(150,20)(190,40)
\qbezier(60,70)(80,50)(100,70)
\put(80,50){\circle{15}}
\qbezier(60,30)(80,50)(100,30)
\qbezier(60,-30)(90,-40)(70,-45)
\qbezier(70,-45)(60,-50)(70,-55)
\qbezier(60,-70)(90,-60)(70,-55)
\qbezier(100,-30)(70,-40)(90,-45)
\qbezier(90,-45)(100,-50)(90,-55)
\qbezier(100,-70)(70,-60)(90,-55)
\put(30,0){\vector(3,1){50}}
\put(30,0){\vector(3,-1){50}}
\put(80,-17){\line(3,1){50}}
\put(80,17){\line(3,-1){50}}
\put(100,-10){\vector(3,1){2}}
\put(100,10){\vector(3,-1){2}}
\put(25,4){\circle{4}}
\put(25,-4){\circle*{4}}
\put(80,23){\circle{4}}
\put(80,30){\circle{4}}
\put(80,-23){\circle*{4}}
\put(80,-30){\circle*{4}}
\put(135,4){\circle*{4}}
\put(135,-4){\circle{4}}
\put(-105,0){\vector(1,0){70}}
\end{picture}

Only at the vertex $\ \bullet \, \bullet\ $  we have
the resolved $\lD$ with the same set of planar cycles
as it had before the \Rede\ move. The other three vertices
correspond to some other resolution of $\lD$ -- and they
should cancel among themselves.
This means that now we have two requirements:
\be
(-q)\alpha_\circ\alpha_\bullet = 1, \nn \\
\alpha_\circ\alpha_\bullet( 1 - q D + q^2) = 0
\label{r2cons}
\ee
From (\ref{r1cons}) and (\ref{r2cons}) it follows that
\be
D = q+\frac{1}{q} = \l[2]_q, \nn \\
\alpha_\bullet = q,  \ \ \ \ \ \alpha_{\w} = -\frac{1}{q^2}
\ee
and this is just a {\it one}-parametric family.
In order to get {\it more} we need to further modify the structure
of (\ref{defoinv}).

%\newpage

\section{From $2$ to $N$, HOMFLY polynomials
\label{modif}}

\subsection{Another system of cycles}

The first step of this modification introduces a new parameter $N$,
such that $D = [N]_q$ -- and the question is what should
be done with (\ref{defoinv}) in order to allow such a deformation
from $N=2$ to arbitrary $N$.

As we already mentioned, our suggestion is to abandon (\ref{epdd})
and use (\ref{dddd}) instead.
This means that instead of the two resolutions at the beginning
or s.\ref{plare} we use another pair:

\begin{picture}(300,100)(-50,-50)
\put(5,5){\vector(1,1){15}}
\put(-5,5){\vector(-1,1){15}}
\put(-20,-20){\vector(1,1){15}}
\put(20,-20){\vector(-1,1){15}}
\put(0,0){\circle*{10}}
\put(40,0){\mbox{$=$}}
\put(75,-20){\vector(0,1){40}}
\put(95,-20){\vector(0,1){40}}
\put(205,5){\vector(1,1){15}}
\put(195,5){\vector(-1,1){15}}
\put(180,-20){\vector(1,1){15}}
\put(220,-20){\vector(-1,1){15}}
\put(200,0){\circle{10}}
\put(240,0){\mbox{$=$}}
\put(280,-20){\vector(0,1){40}}
\put(300,-20){\vector(0,1){40}}
\put(320,-2){\mbox{--}}
\qbezier(340,20)(360,0)(380,-20)
\qbezier(340,-20)(360,0)(380,20)
\put(345,-15){\vector(1,1){2}}
\put(375,-15){\vector(-1,1){2}}
\put(345,15){\vector(-1,1){2}}
\put(375,15){\vector(1,1){2}}
\end{picture}

\noindent
In fact, this choice seems much more natural from the
point of view of the tensor algebra -- and it indeed
is much easier deformed.
The price to pay is that now the second resolution
gives rise not only to planar cycles, moreover,
it provides not a single cycle, but a linear combination,
moreover, with some coefficients negative.
When in the next sections \ref{KRco} we further substitute cycles
with vector spaces, this means that some of those
will actually be factor-spaces.

As to the present stage, this means that eq.(\ref{defoinv})
associates not just a single power $D^{\nu_c}$ with each
vertex $c$ of the hypercube $H(\lD)$:  when $c$ involves white
vertices, there is a linear combination instead.
As we shall see below, this actually implies that powers  $D^{\nu_c}$
are substituted by less trivial products ${D}^{(c)}$ of "differentials"
\be
D_{-k} = [N-k] = \frac{\{Aq^{-k}\}}{\{q\}}
\ee
which are known to play a big role in other branches of
knot theory \cite{DGR}-\cite{art}.
Here $A=q^N$ and $\{x\}=x-x^{-1}$ and $[k]_q = \frac{q^k-q^{-k}}{q-q^{-1}}
= \frac{\{q^k\}}{\{q\}}$.
Note that only "negative" differentials appear,
reflecting the negative sign in the definition of the white
resolution.
Also note that, despite there are negative contributions,
the total contribution of each vertex $c$ is positive:
negative contributions are always smaller than the positive ones.

Finally the deformation of (\ref{defoinv}) which we are going
to discuss in this section looks like
\be
H^{c_{_0}}(\lD) = \alpha_\bullet^{n_\bullet}\alpha_\circ^{n_\circ}
\sum_{c=1}^{2^{n_\circ+n_\bullet}}
(-q)^{h_c-h_{c_{_0}}}\cdot {D}^{(c)}
\label{defoinvH}
\ee
where ${D}^{(c)}$ now depends not only on $q$, but also on
additional parameter $N$ (or $A$).
As we shall see, the \Rede\ invariance requires that
\be
\alpha_\bu = q^{N-1}, \ \ \ \ \ \alpha_\w= -q^{-N}
\label{weightsEu}
\ee
We denote this invariant by $H$, because it actually is nothing
but a HOMFLY polynomial.
{\it A priori} HOMFLY of a knot is an average of a character (Wilson loop)
in Chern-Simons theory \cite{Wit,MoSmi}, and -- since we consider
only knot polynomials in the fundamental representation -- at $q=1$
it reduces to $N$. Likewise a link is an averaged product of characters,
so that in general
\be
H^{c_0}(\lD|N|q=1) = \sum_{c=1}^{2^{n_\circ+n_\bullet}}
(-)^{h_c-h_{c_{_0}}}\cdot {D}^{(c)}
= N^{\#({\rm link\ components})}
\ \ \ \ \ \ \  \forall\ \lD\ {\rm and} \ c_0
\label{normH}
\ee
(see also eq.(136) or ref.\cite{evo}, saying that {\it reduced}
HOMFLY for a knot is always $1+O(\log q)$, provided $A=q^N$ with $N$ fixed,
-- and unreduced polynomial is $N$ times larger in this limit).
This will be always true in our construction.

The rest of this section is just a collection of examples,
which tell much more about the story than any formal definitions.
Those will be provided elsewhere.

\subsection{$1$-dimensional hypercube and the $R1$-invariance
\label{eightgraphdim}}

We begin with the knot diagram $\lD$ with a single vertex, i.e.
of the shape of eight.
The corresponding hypercube is one dimensional, i.e. just a segment
with two vertices:

\begin{picture}(200,120)(-200,-100)
\put(-60,0){\circle{20}}\put(-40,0){\circle{20}}\put(-50,0){\circle*{5}}
\put(40,0){\circle{20}}\put(60,0){\circle{20}}\put(50,0){\circle{5}}
\put(-62,-40){\circle{20}}\put(-38,-40){\circle{20}}
\put(8,-40){\circle{20}}\put(32,-40){\circle{20}}
\put(48,-42){\mbox{$-$}}
\put(70,-40){\circle{20}}\put(90,-40){\circle{20}}
\put(-50,-65){\line(1,0){100}}\put(-50,-65){\circle*{3}}\put(50,-65){\circle*{3}}
\put(-53,-22){\mbox{$||$}}\put(47,-22){\mbox{$||$}}
\put(-85,-85){\mbox{$N^2 \longrightarrow [N]^2$}}
\put(0,-85){\mbox{$N^2-N=N(N-1) \longrightarrow [N][N-1]$}}
\put(-55,-65){\mbox{$\boxed{\phantom.}$}}
\end{picture}

%\noindent
It is very useful to represent this hypercube as arising in three steps.
At the first step we just insert a cross $X$ instead of the true
resolution $\ || - X\ $ at all white vertices of $\lD_c$
and draw what we call the cycle diagram (boxed in (\ref{cydeight}) below).
The result of $\ ||\ $ insertion is naturally obtained from   $\ X\ $
by {\it cutting}, and we use the arrow in cycle diagram to show
the direction of this cut procedure: in the present case it maps
the white vertex into the black one.
The vertices where all arrows are only terminating are called {\it drain},
and the vertex with all resolutions black, is called the
{\it main} one, it is always among the drain vertices,
and we often put it into a box.
Alternatively it could be called {\it Seifert} vertex, because the
corresponding decomposition is in planar Seifert cycles.
At the level of HOMFLY polynomials {\it drain} vertices do not play
any interesting role, but the {\it Seifert} vertex does.

At the second step we construct the "classical"  $D^{(c)}$, which are
just the linear combinations
of powers $D^{\nu_c}$: $D^\bu = N^2$, $D^\w= N^2-N$.
Then we note that they can be naturally rewritten
as products, and then, at the third step apply
the "obvious" quantization rules for these $D^{(c)}$:
\be
\begin{array}{cc|crcc|cccc}
\boxed{\ \boxed{2}\ \longleftarrow\ 1\ }  &&&  N^2 \ \longleftarrow \ N^2-N &=N(N-1) & &&
\l[N]^2 \ \longleftarrow \ [N][N-1] \\ &&&&&&&&\\
{\rm step}\ 1 &&& {\rm step}\ 2 &&&&  {\rm step}\ 3
\end{array}
\label{cydeight}
\ee

\noindent
Thus  from  (\ref{defoinvH}) and (\ref{weightsEu})
we obtain the answers for the single-vertex ${\cal D}$
with the black and white vertices respectively:
\be
{\cal H}^{\bullet}_{_\Box} = q^{N-1}\Big([N]^2 - q[N][N-1]\Big)
= [N] = \frac{\{A\}}{\{q\}}, \nn \\
{\cal H}^{\circ}_{_\Box} = -q^{-N}\Big([N][N-1]-q[N]^2\Big)
= [N] = \frac{\{A\}}{\{q\}}
\label{HOeight}
\ee
i.e. reproduce the HOMFLY polynomial for the unreduced unknot -- as they should.
Since reduced HOMFLY differ just by division over $[N]$ we do not consider
them separately and from now on denote unreduced HOMFLY by $H$.

\bigskip

Just the same calculation explains invariance of so constructed HOMFLY polynomials
under the first \Rede\ move:
$R1$ doubles the hypercube and multiplies the answer by
\be
\bullet\longrightarrow\circ \ \ \ &
N - (N-1) = 1 & \ \ \ \ \ \
q^{N-1}\Big(\l[N] - q[N-1]\Big) = 1 \nn \\ \nn \\
\circ\longrightarrow\bullet \ \ \ &
-\Big((N-1)-N\Big)= 1 & \ \ \
-q^{-N}\Big([N-1]-q[N] \Big) = 1
\ee
This actually follows from analysis of the following example.

\subsection{Double eight
\label{dei}}

Adding one more vertex converts our single-vertex "eight" into the
the two-vertex knot diagram $\lD$
with the shape of a "double eight":

\begin{picture}(100,35)(-200,-15)
\put(-20,0){\circle{20}}\put(0,0){\circle{20}}\put(20,0){\circle{20}}
\end{picture}

\noindent
Whatever the coloring, this is just an unknot (a result of two applications of
$R1^{\pm 1}$ to a circle) -- and this is immediately seen from
the answers for knot polynomials. Reading from the picture for the hypercube
$H(O\!\!\!\bullet\!\!\!O\!\!\!\bullet\!\!\!O)$,

\begin{picture}(250,140)(-50,-70)
% rombus
\put(130,0){\line(-3,1){50}}
\put(130,0){\line(-3,-1){50}}
\put(80,-17){\line(-3,1){50}}
\put(80,17){\line(-3,-1){50}}
\put(30,0){\circle*{3}}\put(130,0){\circle*{3}}
\put(80,17){\circle*{3}}\put(80,-17){\circle*{3}}
\put(25,0){\mbox{$\boxed{\phantom.}$}}
\put(70,0){\circle{10}}\put(80,0){\circle{10}}\put(90,0){\circle{10}}
%
% bb
\put(27,10){\circle*{3}}
\put(33,10){\circle*{3}}
\put(-20,0){\circle{10}}\put(-7,0){\circle{10}}\put(6,0){\circle{10}}
\put(-10,-20){\mbox{{ $N^3$}}}
%
% wb
\put(65,22){\circle{3}}
\put(71,22){\circle*{3}}
\put(39,35){\circle{10}}\put(52,35){\circle{10}}\put(65,35){\circle{10}}
\put(80,32){\mbox{$-$}}
\put(100,35){\circle{10}}\put(110,35){\circle{10}}\put(124,35){\circle{10}}
\put(30,52){\mbox{{ $  N^3-N^2=N^2(N-1)$}}}
%
% bw
\put(65,-22){\circle*{3}}
\put(71,-22){\circle{3}}
\put(39,-35){\circle{10}}\put(52,-35){\circle{10}}\put(65,-35){\circle{10}}
\put(80,-38){\mbox{$-$}}
\put(100,-35){\circle{10}}\put(114,-35){\circle{10}}\put(124,-35){\circle{10}}
\put(30,-55){\mbox{{ $N^3-N^2=N^2(N-1)$}}}
%
% ww
\put(127,10){\circle{3}}
\put(133,10){\circle{3}}
\put(154,0){\circle{10}}\put(167,0){\circle{10}}\put(180,0){\circle{10}}
\put(190,-3){\mbox{$-$}}
\put(210,0){\circle{10}}\put(220,0){\circle{10}}\put(234,0){\circle{10}}
\put(244,-3){\mbox{$-$}}
\put(264,0){\circle{10}}\put(278,0){\circle{10}}\put(288,0){\circle{10}}
\put(298,-3){\mbox{$+$}}
\put(318,0){\circle{10}}\put(328,0){\circle{10}}\put(338,0){\circle{10}}
\put(160,-20){\mbox{ $N^3-2N^2+N=N(N-1)^2$}}
\end{picture}

\noindent
we get the cycle diagram
\be
\begin{array}{cccccc|cccccc}
&& 2 && && && N^3-N^2 &&\\
&\swarrow  && \nwarrow &&&& \swarrow  && \nwarrow\\
\boxed{3} &&&& 1 \ \ \ \ \ \ \ &&\ \ \ \ \ \ \ \  N^3\!\!\!\! &&&& \!\!\!\!\!\!\! N^3-2N^2+N\\
&\nwarrow  && \swarrow &&&&\nwarrow  && \swarrow\\
&& 2 &&  && && N^3-N^2 &&
\end{array}
\ee
Note that $N^3-N^2$ is obtained by subtracting the
$N^{\nu}$ at the tail of the arrow from $N^{\nu}$
at its nose.
Likewise, $N^3-2N^2+N$ is the similar
alternative-summation along the two paths leading from
the given vertex to {\it main} one (boxed), where all vertices
are black and all resolutions are trivial
(it corresponds to decomposition of ${\cal D}$ intp
Seifert cycles).

Now we apply the obvious quantization rules
\be
bb = & N^3\ & \longrightarrow\ [N]^3, \nn \\
bw=wb = & N^3-N^2 = N^2(N-1)\ & \longrightarrow \ [N]^2[N-1], \nn \\
ww = & N^3-2N^2+N = N(N-1)^2 \ & \longrightarrow\ [N][N-1]^2
\label{dimsde}
\ee
and get the HOMFLY polynomials ($A=q^N$):
\be
H^{\bullet\bullet}_{_\Box}(A|q) =
q^{2(N-1)}\Big(bb-2q\cdot bw+q^2ww\Big) = q^{2N-2}[N]\Big([N]^2-2q[N][N-1]+q^2[N-1]^2\Big)=\nn\\
= q^{2N-2}[N]\Big([N]-q[N-1]\Big)^2=\ \ [N]
\ \ =\ \frac{\{A\}}{\{q\}} = \ H^{{\rm unknot}}_{_{\Box}}(A|q),
\ \ \ \ \ \ \ \ \ \ \  \ \ \ \ \ \ \ \ \ \ \ \ \ \ \ \ \ \ \ \ \ \  \ \ \ \ \ \  \nn \\ \nn \\
H^{\bullet\circ}_{_\Box}(A|q) = q^{N-1}\cdot(-q^{-N})\Big(bw-q(bb+ww)+q^2wb\Big)
= -q^{-1}[N]\Big((1+q^2)[N][N-1]-q([N]^2+[N-1]^2)\Big) = \nn
\ee
\vspace{-0.6cm}
\be
= [N]\Big([N]^2+[N-1]^2-[2][N][N-1])\Big)=\ \ [N]
\ \ = \ \frac{\{A\}}{\{q\}} = \ H^{{\rm unknot}}_{_{\Box}}(A|q)
\ee

\subsection{Hopf link and the $R2$-invariance:}

The simplest next example is the Hopf link.
The knot diagram has two vertices, the hypercube has $2^2=4$ vertices,
like in the case of the double eight, i.e.
it is again the 2-dimensional square (or rhombus):

\begin{picture}(200,200)(-35,-100)
\put(30,0){\line(3,1){50}}
\put(30,0){\line(3,-1){50}}
\put(80,-17){\line(3,1){50}}
\put(80,17){\line(3,-1){50}}
\put(30,0){\circle*{3}}\put(130,0){\circle*{3}}
\put(80,17){\circle*{3}}\put(80,-17){\circle*{3}}
\put(75,-17){\mbox{$\boxed{\phantom.}$}}
%\put(100,-10){\vector(3,1){2}}
%\put(100,10){\vector(3,-1){2}}
%
%wb
\put(-5,15){\circle{15}}\put(5,15){\circle{15}}
\put(0,20){\circle{3}}\put(0,10){\circle*{3}}
\put(-2,0){\mbox{$_{||}$}}
\put(-32,-15){\circle{15}}\put(-15,-15){\circle{15}}
\put(-3,-17){\mbox{$-$}}
\qbezier(10,-15)(10,-25)(20,-15)\qbezier(25,-15)(35,-25)(35,-15)
\qbezier(10,-15)(15,0)(25,-15)\qbezier(20,-15)(30,0)(35,-15)
%
%bw
\put(155,15){\circle{15}}\put(165,15){\circle{15}}
\put(160,20){\circle*{3}}\put(160,10){\circle{3}}
\put(158,0){\mbox{$_{||}$}}
\put(128,-15){\circle{15}}\put(145,-15){\circle{15}}
\put(157,-17){\mbox{$-$}}
\qbezier(195,-15)(195,-5)(185,-15)\qbezier(180,-15)(170,-5)(170,-15)
\qbezier(195,-15)(190,-30)(180,-15)\qbezier(185,-15)(175,-30)(170,-15)
%
%bb
\put(75,-35){\circle{15}}\put(85,-35){\circle{15}}
\put(80,-30){\circle*{3}}\put(80,-40){\circle*{3}}
\put(78,-50){\mbox{$_{||}$}}
\put(70,-60){\circle{15}}\put(90,-60){\circle{15}}
%
%ww
\put(75,35){\circle{15}}\put(85,35){\circle{15}}
\put(80,30){\circle{3}}\put(80,40){\circle{3}}
\put(78,50){\mbox{$_{||}$}}
\put(0,60){\circle{15}}\put(17,60){\circle{15}}
\put(30,58){\mbox{$-$}}
\qbezier(45,60)(45,50)(55,60)\qbezier(60,60)(70,50)(70,60)
\qbezier(45,60)(50,75)(60,60)\qbezier(55,60)(65,75)(70,60)
\put(77,58){\mbox{$-$}}
\qbezier(115,60)(115,70)(105,60)\qbezier(100,60)(90,70)(90,60)
\qbezier(115,60)(110,45)(100,60)\qbezier(105,60)(95,45)(90,60)
\put(120,58){\mbox{$+$}}
\put(140,60){\circle{15}}\put(150,60){\circle{15}}
%
%\put(70,60){\circle{15}}\put(90,60){\circle{15}}
%\put(-3,58){\mbox{$-$}}
%\qbezier(10,60)(10,50)(20,60)\qbezier(25,60)(35,50)(35,60)
%\qbezier(10,60)(15,75)(25,60)\qbezier(20,60)(30,75)(35,60)
%\put(157,80){\mbox{$-$}}
%\qbezier(195,60)(195,70)(185,60)\qbezier(180,60)(170,70)(170,60)
%\qbezier(195,60)(190,45)(180,60)\qbezier(185,60)(175,45)(170,60)
%\put(157,58){\mbox{$+$}}
%\put(75,60){\circle{15}}\put(85,60){\circle{15}}
%
%
\put(328,60){\mbox{$2$}}
\put(287,43){\mbox{$1$}}
\put(365,43){\mbox{$1$}}
\put(324,30){\mbox{$\boxed{2}$}}
\put(300,50){\vector(3,1){20}}
\put(360,50){\vector(-3,1){20}}
\put(300,40){\vector(3,-1){20}}
\put(360,40){\vector(-3,-1){20}}
\put(324,-65){\mbox{$N^2$}}
\put(257,-48){\mbox{$N^2-N$}}
\put(370,-48){\mbox{$N^2-N$}}
\put(307,-27){\mbox{$2(N^2-N)$}}
\put(300,-50){\vector(3,-1){20}}
\put(360,-50){\vector(-3,-1){20}}
\put(300,-40){\vector(3,1){20}}
\put(360,-40){\vector(-3,1){20}}
\end{picture}

\noindent
In this example there are two drain vertices in the hypercube.
Hopf link {\it per se} corresponds to choosing as initial one hypercube
vertex with two identical colors, e.g. the {\it main}
vertex $\bu\bu$ at the bottom.
Then
\be
H^{[2,2]}_{_{\Box\times\Box}} = q^{2(N-1)}
\Big(\underbrace{[N]^2}_{\bu\bu} - 2q\underbrace{[N][N-1]}_{\bu\w\,=\,\w\bu} + q^2
\underbrace{[2][N][N-1]}_{\w\w}\Big)
= \frac{q^{2N}[N]}{[2]}\Big(q^{-2}[N+1] + q^2[N-1]\Big)
%= q^{2(N-1)}[N]\Big([N] - 2q[N-1] + q^2[2][N-1]\Big)
\label{HOMFLYhopf}
\ee
In terms of $A=q^N$ this is
\be
H^{[2,2]}_{_{\Box\times\Box}}(N|q) =
q^{-2}A^2\frac{\{A\}\Big(
\{A\}- 2q\{A/q\}+q^2[2]\{A/q\}\Big)}{\{q\}^2} =
A^2\frac{\{A\}\Big(q^{-2}\{Aq\}+q^2\{A/q\}\Big)}{\{q\}\{q^2\}}
\ee
This is the right answer for the HOMFLY polynomial.
Note that it is reproduced if we accept the quantization rule
$2N(N-1) \longrightarrow [2][N][N-1]$ at the vertex $\w\w$.
Of course, another $2$ in (\ref{HOMFLYhopf}),
which arises just from adding the two
identical contributions at vertices $\bu\w$ and $\w\bu$
is {\it not} quantized.

Taking as initial the white-white vertex we obtain the mirror-symmetric answer:
\be
 q^{2(-N)}\Big(\w\w-q(\bu\w\,+\w\bu)+q^2\bu\bu\Big)
= q^{-2N}\Big([2][N][N-1] - 2q[N][N-1]+q^2[N]^2\Big)
= H^{[2,2]}_{_{\Box\times\Box}}\Big(N\Big|-\!q^{-1}\Big)
\ee

If instead we start from the black-white or white-black vertices the
answer will be different -- as it should be, because in this case we
get the two unlinked unknots:
\be
q^{N-1}\cdot(-q^{-N})\Big(\bu\w - q(\bu\bu+\w\w) + q^2\w\bu\Big)
= -q^{-1} \Big((1+q^2)[N][N-1] - q([N]^2+[2][N-1][N])\Big) = \nn \\
= -[N]\Big([2][N-1]-[N]-[2][N-1]\Big) = [N]^2 =
\left(H^{{\rm unknot}}_{_{\Box}}\right)^2
\ee
This decomposition into a product of two unknots is the simplest
illustration of $R2$ invariance of our construction.

\subsection{Other 2-strand knots and links}

\subsubsection{Trefoil in the $2$-strand realization}

The first non-trivial knot is the trefoil.
It has two standard braid representations: $2$-strand and $3$-strand.
In the $2$-strand case the knot diagram $\lD$ has three vertices,
the hypercube $H(\lD)$ is three-dimensional, with $2^3=8$ vertices,
the cycles diagram is

\bigskip

\be
\begin{array}{cccccccc|ccccccccc}
&& 1 &\longrightarrow & 2 &&  &&  && N(N-1) &\longrightarrow & 2N(N-1) && \\
&\swarrow && \nearrow \!\!\!\!\!\! \searrow && \nwarrow &&&
&\swarrow && \nearrow \!\!\!\!\!\! \searrow && \nwarrow\\
\boxed{2} & \longleftarrow & 1 & &2 &\longleftarrow & 1 &&
\ \ \ \ \ N^2 & \longleftarrow & N(N-1) & & 2N(N-1) &\longleftarrow & 4N(N-1)\\
&\nwarrow && \searrow \!\!\!\!\!\! \nearrow && \swarrow &&&
&\nwarrow && \searrow \!\!\!\!\!\! \nearrow && \swarrow \\
&& 1 &\longrightarrow & 2 &&   &&
&& N(N-1) &\longrightarrow & 2N(N-1) &&
\end{array}
\ee

\bigskip

\noindent
We remind that classical dimensions in the right-hand-side diagram are
obtained from the cycles diagram by the simple rule -- taking alternated sum
along all paths connecting the given vertex with the boxed {\it main} one:
$N^2-N=N(N-1)$, $\ N^2-2N+N^2 = 2N(N-1)$, $\ N^2-3N + 3N^2 - N = 4N(N-1)$.

\bigskip

The knot polynomial, obtained by our rules is:
\be
H^{[2,3]}_{_\Box}(A|q) = q^{3(N-1)}
\Big([N]^2 - 3q[N][N-1] + 3q^2[2][N][N-1]-q^3[2]^2[N][N-1]\Big)
= \nn \\
= q^{2N}[N]\Big(q^2+q^{-2} - q^{2N}\Big) = \frac{q^{3N}[N]}{[2]}\Big(q^{-3}[N+1]-q^3[N-1]\Big)
= \nn \\
= \frac{A^3\{A\}}{\{q\}\{q^2\}}\Big(q^{-3}\{Aq\} - q^3\{A/q\}\Big)
= \frac{\{A\}}{\{q\}}\Big(1-A^2\{Aq\}\{A/q\}\Big)
\label{trefH}
\ee
what is the standard answer.
Note that it is obtained, if in the $www$ vertex we use the following quantization rule
$4N(N-1)\longrightarrow [2]^2[N][N-1]$ (instead of, say, $4\longrightarrow [4]$).

\bigskip

For another coloring $bbw$ we have instead:
\be
q^{2(N-1)}\cdot(-q^{-N})\cdot\left(bbw - q\Big(bbb+(bww+wbw)\Big)
+ q^2\Big((wbb+bwb)+www\Big)-q^3wwb\right) = \nn \\
= -q^{N-2}[N]\left([N-1] - q\Big([N]+2[2][N-1]\Big) + q^2\Big(2[N-1]+[2]^2[N-1]\Big)
-q^3[2][N-1]\right) = \nn \\
= q^{N-2}[N]\left(q[N] - [N-1]\Big(1-2q[2]+q^2(2+[2]^2)-q^3[2]\Big)\right)
= \nn \\
= q^{N-1}[N]\Big([N]-q[N-1]\Big) = [N]= H^{{\rm unknot}}_{_{\Box}}
\ee
The same unknot will be obtained, if initial vertex is $bww$.
For $www$ the answer is mirror-symmetric trefoil:
\be
-q^{-3N}\Big(www-q(bww+wbw+wwb)+q^2(bbw+bwb+wbb)-q^3bbb\Big)
= H^{[2,3]}_{_\Box}\Big(A^{-1}\,\Big|\,q^{-1}\Big)
\ee
Coincidence with the unknot is guaranteed by the right quantization
rule $4\longrightarrow [2]^2$, thus one can say, that this rule
is {\it derived} from the \Rede \ invariance.

\bigskip

\subsubsection{Generic knot/link $[2,k]$}

Unknot Hopf link and the trefoil are the members of entire series
of $k$-folds -- the closures of a 2-strand braid.
It is  instructive to perform our calculation for entire series
at once.

The cycle diagram is actually a sequence
$$
\boxed{2} \Longrightarrow k\otimes 1 \Longrightarrow \frac{k(k-1)}{2} \otimes 1
\Longrightarrow \frac{k(k-1)(k-2)}{6} \otimes 2 \Longrightarrow \ldots
\ \ \ \ \ \ \ \ \ \ \ \ \ \ \ \ \ \ \ \ \ \ \ \ \ \ \ \ \ \ \ \ \
$$
i.e. consists of alternated two- and -single cycle vertices
taken with the
multiplicities $C^j_k$ and connected
by arrows, which form the  $k$-dimensional hypercube.
All vertices with $2$ cycles are drain.
The corresponding classical hypercube is
$$
N^2 \Longrightarrow k\otimes N(N-1) \Longrightarrow \frac{k(k-1)}{2} \otimes
2N(N-1) \Longrightarrow
\ \ \ \ \ \ \ \ \ \ \ \ \ \ \ \ \ \ \ \ \ \ \ \ \ \ \ \ \ \ \ \ \
$$
$$
\Longrightarrow \frac{k(k-1)(k-2)}{6} \otimes 4N(N-1) \Longrightarrow \ldots
%$$
%$$
%\ldots
\Longrightarrow C^j_k \otimes 2^{j-1}N(N-1) \Longrightarrow \ldots
\ \ \ \ \ \ \ \ \ \ \ \ \ \ \ \ \ \ \ \ \ \ \ \ \ \ \ \ \ \ \ \ \
\ \ \ \ \ \ \ \ \ \ \ \ \ \ \ \ \ \
$$
and the quantization prescription, validated by
the known answer for \\ the HOMFLY polynomial and/or
the \Rede\ invariance is
$$
2^{j-1}N(N-1) \longrightarrow [2]^{j-1}[N][N-1]
\ \ \ \ \ \ \ \ \ \ \ \ \ \ \ \ \ \
%\label{quan2strand}
$$
\begin{picture}(100,160)(-435,-220)
\put(-6,20){\mbox{$\ldots$}}
\put(0,0){\circle{20}}
\put(0,-20){\circle{20}}
\put(0,-40){\circle{20}}
\put(0,-60){\circle{20}}
%\put(-6,-80){\mbox{$\ldots$}}
\put(-10,0){\vector(0,1){2}}
\put(-10,-20){\vector(0,1){2}}
\put(-10,-40){\vector(0,1){2}}
\put(-10,-60){\vector(0,1){2}}
\put(10,0){\vector(0,1){2}}
\put(10,-20){\vector(0,1){2}}
\put(10,-40){\vector(0,1){2}}
\put(10,-60){\vector(0,1){2}}
\put(0,10){\circle*{3}}
\put(0,-10){\circle*{3}}
\put(0,-30){\circle*{3}}
\put(0,-50){\circle*{3}}
\put(0,-70){\circle*{3}}
\put(0,40){\circle{20}}\put(0,50){\circle*{3}}\put(0,30){\circle*{3}}
\put(-10,40){\vector(0,1){2}}\put(10,40){\vector(0,1){2}}
\qbezier(0,50)(-10,65)(-20,50)\qbezier(0,-70)(-10,-95)(-20,-70)
\put(-20,50){\vector(0,-1){120}}
\qbezier(0,50)(10,65)(20,50)\qbezier(0,-70)(10,-95)(20,-70)
\put(20,50){\vector(0,-1){120}}
\end{picture}

\vspace{-5.5cm}

In result the answer for the HOMFLY polynomial is $q^{k(N-1)}[N]$ times
\be
\l[N] - kq[N-1] + \frac{k(k-1)}{2}\cdot q^2[2][N-1] - \frac{k(k-1)(k-2)}{6}\cdot q^3
[2]^2[N-1] + \ldots = \nn \\
\l[N] - [N-1]\frac{1-(1-q[2])^k}{[2]}
= A \left(1-\frac{1-(-q^2)^k}{q[2]}\right)
- \frac{1}{A}\left(1-q\,\frac{1-(-q^2)^k}{[2]}\right)
\ee
This is indeed the same as the well known \cite{RJ}-\cite{MMSS}
\be
\frac{1}{A}\Big(q^{k+1} - q^{-k-1}\Big) - A\Big(q^{k-1}-q^{-k+1}\Big)
& {\rm for\ odd}\ k \ ({\rm knots}) \\
\frac{1}{A}\Big(q^{k+1} + q^{-k-1}\Big) - A\Big(q^{k-1}+q^{-k+1}\Big)
& {\rm for\ even}\ k \ ({\rm links})
\ee

\bigskip

\subsubsection{Towards Kauffman-like formalism \label{Kauf}}

The ${\cal R}$-matrix approach \cite{MoSmi},\cite{TR}-\cite{AnoMMM21}
is to simply write down
explicit matrices at place of vertices of the link diagram $\lD_{c_0}$:

\begin{picture}(300,70)(-100,-40)
\put(5,5){\vector(1,1){15}}
\put(-5,5){\vector(-1,1){15}}
\put(-20,-20){\vector(1,1){15}}
\put(20,-20){\vector(-1,1){15}}
\put(0,0){\circle*{10}}
\put(40,0){\mbox{$=$}}
%\put(75,-20){\vector(0,1){40}}
%\put(95,-20){\vector(0,1){40}}
%
\put(205,5){\vector(1,1){15}}
\put(195,5){\vector(-1,1){15}}
\put(180,-20){\vector(1,1){15}}
\put(220,-20){\vector(-1,1){15}}
\put(200,0){\circle{10}}
\put(240,0){\mbox{$=$}}
\put(-27,-15){\mbox{$k$}}
\put(-27,15){\mbox{$i$}}
\put(23,-15){\mbox{$l$}}
\put(23,15){\mbox{$j$}}
\put(173,-15){\mbox{$k$}}
\put(173,15){\mbox{$i$}}
\put(223,-15){\mbox{$l$}}
\put(223,15){\mbox{$j$}}
\put(70,-2){\mbox{${\cal R}_{kl}^{ij}$}}
\put(275,-2){\mbox{$\overline{{\cal R}}_{kl}^{ij}$}}
\end{picture}

\noindent
\Rede\ invariance is guaranteed by the properties
\be
{\rm R1:}  & \sum_{i=1}^N {\cal R}^{ij}_{kj} = \delta^i_k
\nn\\ \nn \\
{\rm R2:}  & \sum_{k,l=1}^N {\cal R}^{ij}_{kl}\overline{{\cal R}}^{kl}_{mn} = \delta^i_m\delta^j_n
 \nn\\ \nn \\
{\rm R3:} & \sum_{b,c,e=1}^N\overline{\cal R}^{jk}_{bc}{\cal R}^{ib}_{le}{\cal R}^{ec}_{mn}
= \sum_{a,b,e=1}^N
{\cal R}^{ij}_{ab} {\cal R}^{bk}_{en} \overline{{\cal R}}^{ae}_{lm} \nn \\\nn \\
{\rm skein:}  & q^{-N}{\cal R}^{ij}_{kl} - q^N\overline{{\cal R}}^{ij}_{kl}
= -(q-q^{-1})\delta^i_k \delta^j_l
\ee
plus various permutations and inversions.
%Here
%\be
%I^i_k = q^{N+1-2i}\delta^i_k, &
%\sum_{i=1}^N delta^i_i = [N] = \frac{q^N-q^{-N}}{q-q^{-1}}
%\ee
%is a graded unit matrix, since it is not quite unit, we avoid calling
%$\overline{{\cal R}}$ inverse of ${\cal R}$.

As a generalization of (\ref{epdd}) and (\ref{dddd}),
\be
{\cal R}^{ij}_{kl} = q^{N-1}\Big(\delta^i_k\delta^j_l -qX^{ij}_{kl}\Big),\nn \\
\overline{{\cal R}}^{ij}_{kl} = -q^{-N}\Big( X^{ij}_{kl} - q\delta^i_k \delta^j_l\Big)
\ee
where $X$ is a graded version of contraction of two $\epsilon$-tensors.

The $N=2$ version of this construction (Kauffman's ${\cal R}$-matrix \cite{KaufR})
is presented in detail in sec.1 of \cite{DM1}, where afterwards
numerous examples are considered (see also sec.4 of \cite{evo} for more
advanced applications).
Specifics of $N=2$ was that one could actually deal with ordinary
$\delta$ and $\epsilon$ tensors, and $q$ can be introduced only in traces,
by "analytic continuation" from $D=2$ to $D=[2]=q+q^{-1}$.
For general $N$ such simple approach does not seem to work.

Still, if one allows to $q$-deform $\epsilon$-tensors,
the situation is not so pessimistic.
Here we just report a few simple observations,
relevant for the case of the 2-strand knots,
which can imply that some kind of generalization to
arbitrary $N$ can still be possible.

When $N=2$, we can
consider a rank $(2,0)$ tensor $\tilde\epsilon$ with components
\be
\tilde\epsilon_{12}=1,\ \tilde\epsilon_{21}=-q
\ee
and as its dual tensor $\tilde\epsilon^*$ of rank $(0,2)$ with:
\be
\tilde\epsilon^{12}=1/q,\ \tilde\epsilon^{21}=1
\ee
Then, the "vertex" tensor of rank $(2,2)$
\be
X_{ij}^{kl}=\tilde\epsilon^{ij}\tilde\epsilon^{kl}
\ee
satisfies
\be
X_{i_1j_1}^{i_2j_2}X_{i_2j_2}^{i_3j_3}\dots X_{i_nj_n}^{i_1j_1}=[2]^n
\ee

Passing to $N=3$ we can
take a rank $(3,0)$ tensor $\tilde\epsilon$ with non-vanishing components

\be
\tilde\epsilon_{123}=1,\ \tilde\epsilon_{213}=-q,\ \tilde\epsilon_{312}=q^2,
\ \tilde\epsilon_{132}=-q,\ \tilde\epsilon_{231}=q^2,\
\tilde\epsilon_{321}=-q^3
\ee
and its dual of rank $(0,3)$ tensor $\tilde\epsilon^*$ with:
\be
\tilde\epsilon^{123}=1/q^3,\ \tilde\epsilon^{213}=-1/q^2,\
\tilde\epsilon^{312}=1/q,\ \tilde\epsilon^{132}=-1/q^2,\  \tilde\epsilon^{231}=1/q,\
\tilde\epsilon^{321}=-1
\ee
Then for the 6 ways of getting a scalar from the pair $\tilde\epsilon,\tilde\epsilon^*$ we have:
\be
\tilde\epsilon_{ijk}\tilde\epsilon^{ijk}=[3][2]\nn \\
\tilde\epsilon_{ijk}\tilde\epsilon^{ikj}=\tilde\epsilon_{ijk}\tilde\epsilon^{jik}=2[3]\nn \\
\tilde\epsilon_{ijk}\tilde\epsilon^{jki}=\tilde\epsilon_{ijk}\tilde\epsilon^{kij}=3[2]\nn \\
\tilde\epsilon_{ijk}\tilde\epsilon^{kji}=6
\ee
Clearly, the contraction corresponding to the "eight" graph is:
$\ \epsilon_{ijk}\epsilon^{ijk}=[3][2]$.

For the $(2,2)$ tensor $X$ we also have a number of choices:
\be
X(1)=\tilde\epsilon_{rij}\tilde\epsilon^{klr},\
X(2)=\tilde\epsilon_{irj}\tilde\epsilon^{rkl},\
X(3)=\tilde\epsilon_{ijr}\tilde\epsilon^{krl}
\ee
For a pair of rank $(2,2)$ tensors $X,Y$ denote by
$X*Y=X_{ij}^{rs}Y_{rs}^{kl}$
their straightforward multiplication of rank $(2,2)$.
Then
\be
{\rm tr}(X(2)*X(3))=[2]^2[3]\nn \\
{\rm tr}(X(2)*X(3)*X(1)^n)=[2]^{2+n}[3]\nn \\
X(2)_{ij}^{kl}X(3)_{kl}^{ij}=[2]^2[3]\nn \\
X(2)_{ij}^{kl}X(3)_{kl}^{rs}X(1)_{rs}^{ij}=-[2]^3[3]\nn \\
X(2)_{ij}^{kl}X(3)_{kl}^{rs}X(1)_{rs}^{pq}X(2)_{pq}^{ij}=[2]^4[3]
\ee
Generalization to higher $N$ is straightforward.
Clearly, contractions of the $q$-deformed $\epsilon$ tensors
are capable to reproduce the peculiar
structures $[2]^n[N][N-1]$, providing our quantities
$D^{(c)}$, at least for the 2-strand  knots.

A question is, however, if one can make these observations
into a working formalism, which would not
just coincide with the standard quantum-$R$-matrix technique
\cite{TR,MMMkn12},
using explicitly the additional Lie-algebra-induced structure
on the tensor algebra.

\subsection{Trefoil and the figure-eight knot in the $3$-strand realization
\label{fe3str}}

The 3-strand braid with four vertices, depending on the coloring,
describes both the trefoil $3_1$ (if all the four vertices are black or all white)
and the figure-eight knot $4_1$ (if colors are alternating).

$$
\begin{array}{ccccccccc}
&&&&3 \\
&&2&&1&&2\\
&&2&&1&&2 \\
\boxed{3}&&&&&&&&1 \\
&&2&&1&&2\\
&&2&&1&&2 \\
&&&&3
\\
\end{array}
\ \ \ \ \ \ \ \ \ \ \ \ \ \ \ \ \ \ \ \ \ \
$$
\\
We do not show the arrows, which form the $4d$ hypercube.\\
Clearly there are three drain vertices, each with $3$ cycles.
$$
\begin{array}{ccccccccc}
\\
&&&&2N^2(N-1) \\
&&N^2(N-1)&&N(N-1)^2&&2N(N-1)^2\\
&&N^2(N-1)&&N(N-1)^2&&2N(N-1)^2 \\
N^3&&&&&&&&N(N-1)(3N-5) \\
&&N^2(N-1)&&N(N-1)^2&&2N(N-1)^2\\
&&N^2(N-1)&&N(N-1)^2&&2N(N-1)^2 \\
&&&&2N^2(N-1)
\end{array}
\ \ \ \ \ \ \ \ \ \ \ \ \ \ \ \ \ \ \ \ \ \
\ \ \ \ \ \ \ \ \ \ \ \ \ \ \
$$
\begin{picture}(100,190)(-430,-240)
\put(-20,0){\vector(0,1){160}}
\put(20,0){\vector(0,1){160}}
\qbezier(0,0)(0,10)(20,20)
\put(20,20){\vector(-1,1){40}}
\put(-20,60){\vector(1,1){40}}
\put(20,100){\vector(-1,1){40}}
\qbezier(-20,140)(0,150)(0,160)
\end{picture}

\vspace{-7cm}

\noindent
The classical dimensions are given by the general rules
(one should only imagine the right configuration of arrows,
suppressed in our diagrams):
\be
N^3-N^2 = N(N-1), \nn \\
N^3-2N^2+N^3 = 2N^2(N-1), \ \ \ \ \  N^3-2N^2+N = N(N-1)^2, \nn \\
N^3-3N^2 + (N^3+2N) - N^2 = 2N(N-1)^2, \nn \\
N^3 - 4N^2 + (2N^3+4N) - 4N^2 + N = N(N-1)(3N-5)
\ee
It follows, that
\be
H^{[3,2]}_{_\Box} =
q^{4(N-1)}\Big(\underbrace{[N]^3}_{bbbb} - 4q\underbrace{[N]^2[N-1]}_{bbbw}
+ 4q^2\underbrace{[N][N-1]^2}_{bbww} + 2q^2\underbrace{[2][N]^2[N-1]}_{bwbw}
-4q^3\underbrace{[2][N][N-1]^2}_{bwww}
+ \nn \\
+ q^4\underbrace{[N][N-1]\cdot\boxed{(3N-5)}}_{wwww}\Big)
= \nn \\
=
q^{4(N-1)}\Big(bbbb - q(wbbb+bwbb+bbwb+bbbw)
+q^2\Big((wwbb+wbbw+bbww+bwwb)+(wbwb+bwbw)\Big)
-\nn
\ee
\vspace{-0.5cm}
\be
-q^3(wwwb+wwbw+wbww+bwww) -q^4wwww\Big)
= \frac{\{A\}}{\{q\}}\Big(1-A^2\{Aq\}\{A/q\}\Big)
\ee
what is indeed the right answer for the trefoil, coinciding with
(\ref{trefH}).
All quantizations are obvious, except for the factor in the box.
If we quantize
\be
\boxed{3N-5} \longrightarrow 2[N-1]+[N-3] = [N-1]+[2][N-2],
\label{3N-5}
\ee
then the 3-strand and 2-strand
expressions for the trefoil $3_1$ are related by
\be
H^{[3,2]}_{_\Box} = H^{[2,3]}_{_\Box} = H^{3_1}_{_\Box}
\ee
thus above quantization rule can be justified by the \Rede\ invariance.

Also obvious are the degeneracies $bbbw=bbwb=bwbb=wbbb$, $bwbw=wbwb$ etc.

For alternating colors -- {\it and the  same expression for
quantum dimensions at the  hypercube vertices} --
we get the right expression for HOMFLY of the figure-eight knot $4_1$:
\be
H^{4_1}_{_{\Box}} = q^{2N-2}\cdot q^{-2N}\Big(
wbwb - q\Big((wwwb+wbww)+(wbbb+bbwb)\Big) + \nn \\
+q^2\Big(wwww+(wwbb+wbbw+bbww+bwwb)+bbbb\Big)
- q^3\Big((bbbw+bwbb)+(bwww+wwbw)\Big) + q^4bwbw\Big) = \nn \\
=q^{-2}[N]\Big([2][N][N-1]-2q\Big([2][N-1]^2+[N][N-1]\Big) + \nn \\
+q^2\Big(([N-1]\big(2[N-1]+[N-3]\big) + 4[N-1]^2+[N]^2\Big)
- 2q^3\Big([N][N-1]+[2][N-1]^2\Big)+ q^4[2][N][N-1]
\Big) = \nn
\ee
\vspace{-0.5cm}
\be
= \frac{\{A\}}{\{q\}}\Big(1+\{Aq\}\{A/q\}\Big)
\ee

Finally, two other types of colorings provide unknots:
\be
H^{{\rm Unknot}}_{_\Box} =
-q^{2N-3}\Big(  bbbw - q\Big(bbbb+(bbww+wbbw)+bwbw\Big)
+ q^2\Big((bbwb+bwbb+wbbb)+\nn \\ +(bwww+wbww+wwbw)\Big)
 - q^3\Big(wwww+(bwwb+wwbb)+wbwb) +q^4wwwb\Big) = [N]
\nn \\ \nn \\
H^{{\rm Unknot}}_{_\Box} = q^{-2}\Big(  bbww - q\Big((bbbw+bbwb)+(wbww+bwww)\Big)
+ q^2\Big((bbbb+(bwbw+wbwb)+\nn \\ +(wbbw+bwwb)+wwww\Big) - q^3\Big((bwbb+wbbb)+(wwwb+wwbw)\Big)
+q^4wwbb\Big) = [N]
\ee

\subsection{Twist knots
\label{twist}}

Twist knots is in a sense the simplest 1-parametric family
(see, for example, sec.5.2 of \cite{evo}),
which includes unknot, trefoil and the figure-eight knot $4_1$.
They are made out of the $2$-strand braid, only
-- in variance with the torus knots -- anti-parallel:

\begin{picture}(200,100)(-80,-50)
\put(0,0){\circle{30}}
\put(-50,-10){\line(1,0){37.5}}
\put(-50,10){\line(1,0){37.5}}
\put(80,-10){\line(-1,0){67.5}}
\put(60,10){\line(-1,0){47.5}}
\put(-50,-30){\line(1,0){140}}
\qbezier(-50,-10)(-60,-10)(-60,-20)
\qbezier(-50,-30)(-60,-30)(-60,-20)
\qbezier(91,-8)(100,-10)(100,-20)
\qbezier(90,-30)(100,-30)(100,-20)
\qbezier(-50,10)(-60,10)(-60,20)
\qbezier(-50,30)(-60,30)(-60,20)
\qbezier(90,10)(100,10)(100,20)
\qbezier(90,30)(100,30)(100,20)
\put(-50,30){\line(1,0){140}}
\qbezier(90,10)(80,10)(80,0)
\qbezier(80,0)(80,-5)(84,-6)
\qbezier(80,-10)(90,-10)(90,0)
\qbezier(90,0)(90,5)(86,7)
\qbezier(80,9)(70,10)(60,10)
\put(-40,30){\vector(-1,0){2}}
\put(-40,10){\vector(1,0){2}}
\put(-40,-10){\vector(-1,0){2}}
\put(-40,-30){\vector(1,0){2}}
\put(70,30){\vector(-1,0){2}}
\put(55,10){\vector(1,0){2}}
\put(55,-10){\vector(-1,0){2}}
\put(70,-30){\vector(1,0){2}}
\put(-7,-3){\mbox{$\bar {\cal R}^{2k}$}}
%%%\put(105,0){\mbox{${\cal T}$}}
%
%%%\put(60,20){\line(1,0){60}}
%%%\put(60,-20){\line(1,0){60}}
%%%\put(60,-20){\line(0,1){40}}
%%%\put(120,-20){\line(0,1){40}}
%
%
%
\put(200,0){\circle{30}}
\put(170,-10){\line(1,0){17.5}}
\put(170,10){\line(1,0){17.5}}
\put(230,-10){\line(-1,0){17.5}}
\put(230,10){\line(-1,0){17.5}}
\put(195,-3){\mbox{$\bar {\cal R}^2$}}
\put(240,-3){\mbox{$=$}}
\put(253,10){\line(1,0){7}}
\put(253,-10){\line(1,0){7}}
\put(340,10){\line(1,0){10}}
\put(340,-10){\line(1,0){10}}
\qbezier(260,10)(270,10)(280,0)
\qbezier(280,0)(290,-10)(300,-10)
\qbezier(300,10)(310,10)(320,0)
\qbezier(320,0)(330,-10)(340,-10)
%\put(90,-30){\line(1,0){40}}
\qbezier(260,-10)(270,-10)(278,-2)
\qbezier(281.5,2)(290,10)(300,10)
\qbezier(300,-10)(310,-10)(318,-2)
\qbezier(321.5,2)(330,10)(340,10)
\put(180,10){\vector(1,0){2}}
\put(180,-10){\vector(-1,0){2}}
\put(225,10){\vector(1,0){2}}
\put(225,-10){\vector(-1,0){2}}
\put(256,10){\vector(1,0){2}}
\put(256,-10){\vector(-1,0){2}}
\put(345,10){\vector(1,0){2}}
\put(345,-10){\vector(-1,0){2}}
\end{picture}

\noindent
Here  $k$ can be both positive and negative.
If the number of crossing in the antiparallel braid is
odd, this changes orientation at the two-vertex "locking block".
The corresponding knot diagrams
(after rotation by $90^\circ$) are:

\begin{picture}(100,170)(-150,-90)
\put(-6,20){\mbox{$\ldots$}}
\put(0,0){\circle{20}}
\put(0,-20){\circle{20}}
\put(0,-40){\circle{20}}
\put(0,-60){\circle{20}}
%\put(-6,-80){\mbox{$\ldots$}}
\put(-10,0){\vector(0,1){2}}
\put(-10,-20){\vector(0,-1){2}}
\put(-10,-40){\vector(0,1){2}}
\put(-10,-60){\vector(0,-1){2}}
\put(10,0){\vector(0,-1){2}}
\put(10,-20){\vector(0,1){2}}
\put(10,-40){\vector(0,-1){2}}
\put(10,-60){\vector(0,1){2}}
\put(0,10){\circle*{3}}
\put(0,-10){\circle*{3}}
\put(0,-30){\circle*{3}}
\put(0,-50){\circle*{3}}
\put(0,-70){\circle*{3}}
\put(-60,30){\mbox{even $p$}}
%\put(28,-20){\mbox{$=$}}
\put(90,30){\mbox{odd $p$}}
\put(54,20){\mbox{$\ldots$}}
\put(60,0){\circle{20}}
\put(60,-20){\circle{20}}
\put(60,-40){\circle{20}}
%%\put(60,-60){\circle{20}}
%\put(54,-80){\mbox{$\ldots$}}
\put(50,0){\vector(0,1){2}}
\put(50,-20){\vector(0,-1){2}}
\put(50,-40){\vector(0,1){2}}
%%\put(50,-60){\vector(0,1){2}}
\put(70,0){\vector(0,-1){2}}
\put(70,-20){\vector(0,1){2}}
\put(70,-40){\vector(0,-1){2}}
%%\put(70,-60){\vector(0,-1){2}}
\put(60,10){\circle*{3}}
\put(60,-10){\circle*{3}}
\put(60,-30){\circle*{3}}
\put(60,-50){\circle*{3}}
%%\put(60,-70){\circle*{3}}
%
%
\put(0,60){\circle{20}}
\put(-10,61){\circle*{3}}\put(10,61){\circle*{3}}
\put(0,70){\vector(1,0){2}}\put(0,65){\vector(-1,0){2}}
%\put(-10,60){\vector(0,1){2}}\put(10,60){\vector(0,-1){2}}
\qbezier(-20,50)(0,80)(20,50)
\put(0,40){\circle{20}}\put(0,50){\circle*{3}}\put(0,30){\circle*{3}}
\put(-10,40){\vector(0,-1){2}}\put(10,40){\vector(0,1){2}}
%\qbezier(0,50)(-10,65)(-20,50)
\qbezier(0,-70)(-10,-95)(-20,-70)
\put(-20,50){\vector(0,-1){120}}
%\qbezier(0,50)(10,65)(20,50)
\qbezier(0,-70)(10,-95)(20,-70)
\put(20,-70){\vector(0,1){120}}
\put(60,60){\circle{20}}
\put(50,61){\circle*{3}}\put(70,61){\circle*{3}}
\put(60,70){\vector(1,0){2}}\put(60,65){\vector(1,0){2}}
%\put(-10,60){\vector(0,1){2}}\put(10,60){\vector(0,-1){2}}
\qbezier(40,50)(60,80)(80,50)
\put(60,40){\circle{20}}\put(60,50){\circle*{3}}\put(60,30){\circle*{3}}
\put(50,40){\vector(0,-1){2}}\put(70,40){\vector(0,1){2}}
%\qbezier(60,50)(50,65)(40,50)
\qbezier(60,-50)(50,-75)(40,-50)
\put(80,50){\vector(0,-1){100}}
%\qbezier(60,50)(70,65)(80,50)
\qbezier(60,-50)(70,-75)(80,-50)
\put(40,-50){\vector(0,1){100}}
\put(150,-20){\vector(-1,1){20}}\put(150,0){\vector(-1,-1){20}}
%\put(130,-20){\vector(1,1){10}}\put(140,-10){\vector(1,-1){10}}
\put(140,-10){\circle*{4}}
\put(155,-12){\mbox{$=$}}
\qbezier(170,0)(180,-16)(190,0) \put(180,-8){\vector(-1,0){2}}
\qbezier(170,-20)(180,-4)(190,-20) \put(180,-12){\vector(1,0){2}}
%\put(190,-20){\vector(-1,1){20}}\put(190,0){\vector(-1,-1){20}}
%
\put(150,-60){\vector(-1,1){20}}\put(150,-40){\vector(-1,-1){20}}
%\put(130,-60){\vector(1,1){10}}\put(140,-50){\vector(1,-1){10}}
\put(140,-50){\circle{4}}
\put(155,-52){\mbox{$=$}}
\qbezier(170,-40)(180,-56)(190,-40) \put(180,-48){\vector(-1,0){2}}
\qbezier(170,-60)(180,-44)(190,-60) \put(180,-52){\vector(1,0){2}}
%\put(190,-60){\vector(-1,1){20}}\put(190,-40){\vector(-1,-1){20}}
\put(198,-52){\mbox{$-$}}
\put(230,-60){\vector(-1,1){20}}\put(230,-40){\vector(-1,-1){20}}
%\qbezier(210,-40)(220,-56)(230,-40) \put(220,-48){\vector(-1,0){2}}
%\qbezier(210,-60)(220,-44)(230,-60) \put(220,-52){\vector(1,0){2}}
%
\end{picture}

\noindent
The two pictures correspond respectively to the cases when the number $p$
of circles -- and thus the number $p+2$ of vertices -- is even and odd.
The two cases are essentially different,
already when all the vertices are black, i.e. at the
{\it main} Seifert vertex of the hypercube,
configurations are not the same:

\begin{picture}(100,160)(-200,-80)

\put(60,50){\circle{15}}
\put(60,30){\circle{15}}
\put(54,15){\mbox{$\ldots$}}
\put(60,0){\circle{15}}
\put(60,-20){\circle{15}}
\put(60,-40){\circle{15}}
%\put(60,-60){\circle{15}}
\put(40,-40){\line(0,1){90}}
\put(80,-40){\line(0,1){90}}
\qbezier(40,-40)(60,-70)(80,-40)
\qbezier(40,50)(60,90)(80,50)
\put(60,40){\circle*{2}}
\put(60,20){\circle*{2}}
\put(60,10){\circle*{2}}
\put(60,-10){\circle*{2}}
\put(60,-30){\circle*{2}}
\put(60,-50){\circle*{2}}
%\put(60,-70){\circle*{2}}
%
\put(50,57){\circle*{2}}
\put(70,57){\circle*{2}}
\put(-71,35){\mbox{even $p$}}
\put(100,35){\mbox{odd $p$}}
\put(-84,20){\mbox{knot $(p+1)_2$}}
\put(90,20){\mbox{knot $(p+2)_2$}}
\put(91,7){\mbox{($3_2=$ trefoil)}}
\put(0,64){\circle{15}}
\put(0,30){\circle{15}}
\put(-6,15){\mbox{$\ldots$}}
\put(0,0){\circle{15}}
\put(0,-20){\circle{15}}
\put(0,-40){\circle{15}}
\put(0,-60){\circle{15}}
\put(-20,-60){\line(0,1){100}}
\put(20,-60){\line(0,1){100}}
\qbezier(-20,-60)(0,-90)(20,-60)
\qbezier(-20,40)(-20,65)(-10,50)\qbezier(-10,50)(0,36)(10,50)\qbezier(20,40)(20,65)(10,50)
\put(0,40){\circle*{2}}
\put(0,20){\circle*{2}}
\put(0,10){\circle*{2}}
\put(0,-10){\circle*{2}}
\put(0,-30){\circle*{2}}
\put(0,-50){\circle*{2}}
\put(0,-70){\circle*{2}}
\put(-10,57){\circle*{2}}
\put(10,57){\circle*{2}}
\end{picture}

\noindent
Still, the number of cycles in both  cases is $p+1$,
so that in both cases the  hypercube vertex $b^{p+2}$ contributes
$N^{p+1}\longrightarrow [N]^{p+1}$.
When some vertex is changed from black to white, one subtracts a contribution
with a crossing at this vertex, what changes the number of cycles:
for example, when there is just one white vertex, subtraction contains $p$ cycles,
and the contribution of $b^{p+1}w$ vertex in the hypercube is $N^{p+1}-N^p
\longrightarrow [N]^p[N-1]$.

When all vertices are of the same color then the knot is
$(p+1)_2$ for even $p$ and $(p+2)_2$ for odd $p$.
If the two vertices at the top (two "horizontal" vertices)
have the opposite color to the $p$ vertical ones,
then the knot is $(p+2)_1$ for even $p$ and $(p+1)_1$ for odd $p$.
When the two horizontal vertices are of different colors,
we get an unknot. If some vertical vertices have different colors,
what matters is their algebraic sum, $p \cong \#_\bu - \#_\w$.

The answer for HOMFLY polynomials of the twisted knots
is well known, see, for example, sec.5.2 of \cite{evo}:
\be
H_k = 1 +
F_k(A^2)\{Aq\}\{A/q\} =
1 + F_k\left(q^{2N} \right)\left(q^{2N}- q^2-q^{-2}+q^{-2N}\right)
\label{twisho}
\ee
with $F_k(A^2) = -A^{k+1}\{A^{k}\}/\{A\}$.
For $k=0$ and $F_0=0$ we get unknot,
for $k=1$ and $F_{1}=-A^2$ -- the trefoil $3_1$
and for $k=-1$ and $F_{-1}=1$ -- the figure eight knot $4_1$.
More generally, for positive $k$ we get the knots $(2k+1)_2$,
while for negative $k$ -- $(2-2k)_1$ in the Rolfsen notation,
see \cite{katlas}.
Note that trefoil $3_1$ gets its right place in the series
of twisted knots, if treated as $3_2$.

\bigskip

\be
\begin{array}{c|ccccccccccccccccccccccc}
p && 1 &\ \ & 2 &\ \ & 3 &\ \ & 4 &\ \ & 5 &\ \ & 6 &\ \ & 7 &\ \ & 8 &\ \ & 9 & \ \
& \ldots   \\   & \\ & \\
{\tiny \begin{array}{ccc} \bu\!\!\!\!\!\! &&\!\!\!\!\!\!\bu\\ &\bu &\\ &\bu &\\ &
\ldots & \\ &\bu & \end{array}}
&& 3_2 && 3_2 && 5_2 && 5_2 && 7_2 && 7_2 && 9_2 && 9_2 && 11_2 &&  \ldots  \\  & \\ & \\
{\tiny \begin{array}{ccc} \bu\!\!\!\!\!\! &&\!\!\!\!\!\!\bu\\ &\w &\\ &\w &\\ &
\ldots & \\ &\w & \end{array}}
&&{\rm unknot} && 4_1 && 4_1 && 6_1 && 6_1 && 8_1 && 8_1 && 10_1 && 10_1 && \ldots
\end{array}
\nn
\label{twiknots}
\ee

\bigskip

\bigskip

\bigskip

Now we proceed to cycle diagrams.
In the case of twist knots they have a very special structure.
The point is that there are two different types of vertices in $\lD$:
the two at the "locking block" and $p$ others, located at a vertical
axis in above knot diagrams. This implies the obvious block form
for the diagrams of cycles and of classical dimensions.

\subsubsection{$p=1$ \ (includes trefoil $3_1$)}

\be
{\rm number\ of\ cycles} &
\begin{array}{c|cccccc}
p=1&& \bu\bu && \bu\w+\w\bu && \w\w   \\
\hline
&  \\
\bu && \boxed{2}  && 2\times 1 && 2 \\
& \\
\w && 1&& 2\times 2 && 1\\
&
\end{array}
\nn \\ \nn \\
{\rm classical} \ (q=1)\ {\rm values\ of} \ D^{(c)}\ \  &
\begin{array}{c|cccccc}
p=1&& \bu\bu && \bu\w+\w\bu && \w\w   \\
\hline
&  \\
\bu &&  N^2  && 2\times (N^2-N) && N^2-2N+N^2 \\
&\\
\w && N^2-N && 2\times (N^2-2N+N^2) && N^2 - 3N + 3N^2- N\\
&
\end{array}
\nn \\ \nn \\
{\rm quantum} \ D^{(c)} &
\begin{array}{c|cccccc}
p=1&& \bu\bu && \bu\w+\w\bu && \w\w   \\
\hline
&  \\
\bu &&  [N]^2  && 2\times [N][N-1] && [2][N][N-1] \\
&\\
\w && [N][N-1] && 2\times [2][N][N-1] && [2]^2[N][N-1] \\
&
\end{array}
\nn
\ee

\bigskip

\noindent
The first table lists the numbers of cycles
in the resolution of $\lD$ at a given vertex of the hypercube $H(\lD)$.
The first numbers ($2$) in the second column are
multiplicities of the vertices of a given type:
in the example of $p=1$ the multiplicities are non-trivial only
because we do not make a difference between $\bu\w$ and $\w\bu$
colorings at the locking block.

The middle table of classical dimensions $D^{(c)}$ is obtained
from the data in the first table by our standard rule.
To understand what stands at the  given vertex $c$ of $H(\lD)$
consider the sub-hypercube in between $c$ and the Seifert vertex
(which is boxed in the picture).Then what we put at $c$
is the alternated sum of the cycles numbers at this sub-hypercube
vertices. This prescription should look clear from this and below
examples.

Finally, the last table is that of the quantum ($q$-graded)
dimensions $D^{(c)}$ --
in this case they are obtained by the rules, more-or-less familiar
from our previous examples.

\bigskip

Now we can take different vertices as initial:
\be
H^{\bu\bu}_\bu = q^{3N-3}\Big([N]^2-3q[N[N-1] + 3q^2[2][N][N-1] - [2]^2[N][N-1]\Big)
= H(3_1),
\nn \\
H^{\bu\bu}_\w = \frac{q^{2N-2}}{-q^N}\left([N][N-1] - q\Big(]N]^2+2[2][N][N-1]\Big)
+q^2\Big(2[N][N-1]+[2]^2[N][N-1]\Big) -q^3[2][N][N-1]\right) = \nn \\
= -[N] = -H(O),
\nn
\ee
\vspace{-1.1cm}
\be
H^{\bu\w}_\bu = [N] = H(O), \nn\\
\ldots
\ee

\subsubsection{$p=2$ \ (includes trefoil $3_1=\,3_2$ and figure eight $4_1$)}

\bigskip

$$
\begin{array}{c|cccccc}
p=2&& \bu\bu && \bu\w+\w\bu && \w\w   \\
\hline
&  \\
\begin{array}{c}\bu\\ \bu\end{array} && \boxed{3}  && 2\times 2 && 3 \\
&\\
\begin{array}{c}\bu\\ \w\end{array}+ \begin{array}{c}\w\\ \bu\end{array}
&& 2\times 2 && 2^2 \times 1 && 2\times 2\\
&\\
\begin{array}{c}\w\\ \w\end{array} && 3 &&2\times 2 && 1
\end{array}
%\ \ \ \ \ \ \ \ \ \
$$

\bigskip

$$
\begin{array}{c|cccccc}
p=2&& \bu\bu && \bu\w+\w\bu && \w\w   \\
\hline
&  \\
\begin{array}{c}\bu\\ \bu\end{array} && \boxed{N^3}  && 2\times N^2(N-1) && 2N^2(N-1) \\
&\\
\begin{array}{c}\bu\\ \w\end{array}+ \begin{array}{c}\w\\ \bu\end{array}
&& 2\times N^2(N-1) && 2^2 \times N(N-1)^2 && 2\times 2N(N-1)^2\\
&\\
\begin{array}{c}\w\\ \w\end{array} && 2N^2(N-1) &&2\times 2N(N-1)^2 &&
N(N-1)\cdot\boxed{(3N-5)}
\end{array}
%\ \ \ \ \ \ \ \ \ \
$$

\bigskip

$$
\begin{array}{c|cccccc}
p=2&& \bu\bu && \bu\w+\w\bu && \w\w   \\
\hline
&  \\
\begin{array}{c}\bu\\ \bu\end{array} && \boxed{[N]^3}  && 2\times [N]^2[N-1] && [2][N]^2[N-1] \\
&\\
\begin{array}{c}\bu\\ \w\end{array}+ \begin{array}{c}\w\\ \bu\end{array}
&& 2\times [N]^2[N-1] && 2^2 \times [N][N-1]^2 && 2\times [2][N][N-1]^2\\
&\\
\begin{array}{c}\w\\ \w\end{array} && [2][N]^2[N-1] &&2\times [2][N][N-1]^2 &&
[N][N-1]\cdot\boxed{\big([N-1]+[2][N-2]\big)}
\end{array}
%\ \ \ \ \ \ \ \ \ \
$$

\bigskip

\noindent
In the right lower corner we applied
the quantization rule (\ref{3N-5}) -- and this provides the right answers:
\be
H^{\bu\bu}_{_{\bu\atop \bu}}=
q^{4(N-1)}\Big([N]^3-4q[N]^2[N-1]+q^2(2[2][N]^2[N-1]+4[N][N-1]^2)
-4q^3[2][N][N-1]^2 +
\nn \\
+ q^4(2[N][N-1]^2+[N][N-1][N-3])\Big)
= [N]\Big((q^2+q^{-2})q^{2N} - q^{4N}\Big) = H({\rm trefoil}), \nn \\
H^{\bu\bu}_{_{\w\atop \w}}=\frac{q^{2(N-1)}}{q^{2N}}\Big((1+q^4)[2][N]^2[N-1]
-(q+q^3)\big(2[N]^2[N-1]+2[2][N][N-1]^2\big) +
\nn \\
+q^2\big([N]^3 + 4[N][N-1]^2
+ 2[N][N-1]^2+[N][N-1][N-3]\big)\Big)
= [N]\Big(q^{2N}-q^2+1+q^{-2}+q^{-2N}\Big) = H(4_1),
\nn
\ee
\vspace{-0.4cm}
\be
H^{\bu\bu}_{_{\bu\atop \w}}=\frac{q^{3(N-1)}}{-q^{N}}\Big([N]^2[N-1] -\ldots\Big)
= H(O)
\ee
In fact, this is literally the same calculation that we already
performed in the previous sec.\ref{fe3str} .

\subsubsection{$p=3$ \ (includes $5_2$ and $4_1$)}

\bigskip

$$
\begin{array}{c|cccccc}
p=3&& \bu\bu && \bu\w+\w\bu && \w\w   \\
\hline
&  \\
\begin{array}{c}\bu\\ \bu\\ \bu \end{array} && \boxed{4}  && 2\times 3 && 4 \\
&\\
\begin{array}{c}\bu\\ \bu\\ \w\end{array}+ \begin{array}{c}\bu \\ \w\\ \bu\end{array}
+ \begin{array}{c}\w \\ \bu\\ \bu\end{array}
&& 3\times 3 && 2\cdot 3 \times 2 && 3\times 3\\
&\\
\begin{array}{c}\bu\\ \w\\ \w\end{array}+ \begin{array}{c}\w \\ \bu\\ \w\end{array}
+ \begin{array}{c}\w \\ \w\\ \bu\end{array}
&& 3\times 2 && 2\cdot 3 \times 1 && 3\times 2\\
&\\
\begin{array}{c}\w\\ \w\\ \w\end{array} && 1 &&2\times 2 && 1
\end{array}
%\ \ \ \ \ \ \ \ \ \
$$

\bigskip

$$
\begin{array}{c|cccccc}
p=3&& \bu\bu && \bu\w+\w\bu && \w\w   \\
\hline
&  \\
\begin{array}{c}\bu\\ \bu\\ \bu \end{array} && \boxed{N^4}  && 2\times N^3(N-1) && 2N^3(N-1) \\
&\\
\begin{array}{c}\bu\\ \bu\\ \w\end{array}+ \begin{array}{c}\bu \\ \w\\ \bu\end{array}
+ \begin{array}{c}\w \\ \bu\\ \bu\end{array}
&& 3\times N^3(N-1) && 2\cdot 3 \times N^2(N-1)^2&& 3\times 2N^2(N-1)^2\\
&\\
\begin{array}{c}\bu\\ \w\\ \w\end{array}+ \begin{array}{c}\w \\ \bu\\ \w\end{array}
+ \begin{array}{c}\w \\ \w\\ \bu\end{array}
&& 3\times N^2(N-1)^2 && 2\cdot 3 \times N(N-1)^3 && 3\times 2N(N-1)^3\\
&\\
\begin{array}{c}\w\\ \w\\ \w\end{array} && N(N-1)^3 &&
2\times N(N-1)\cdot \boxed{(N^2-3N+4)} &&
2N(N-1)\cdot\boxed{(N^2-3N+4)}
\end{array}
%\ \ \ \ \ \ \ \ \ \
$$

\bigskip

$$
\begin{array}{c|cccccc}
p=3&& \bu\bu && \bu\w+\w\bu && \w\w   \\
\hline
&  \\
\begin{array}{c}\bu\\ \bu\\ \bu \end{array} && \boxed{[N]^4}  && 2\times [N]^3[N-1]
&& [2][N]^3[N-1] \\
&\\
\begin{array}{c}\bu\\ \bu\\ \w\end{array}+ \begin{array}{c}\bu \\ \w\\ \bu\end{array}
+ \begin{array}{c}\w \\ \bu\\ \bu\end{array}
&& 3\times [N]^3[N-1] && 2\cdot 3 \times [N]^2[N-1]^2&& 3\times [2][N]^2[N-1]^2\\
&\\
\begin{array}{c}\bu\\ \w\\ \w\end{array}+ \begin{array}{c}\w \\ \bu\\ \w\end{array}
+ \begin{array}{c}\w \\ \w\\ \bu\end{array}
&& 3\times [N]^2[N-1]^2 && 2\cdot 3 \times [N][N-1]^3 && 3\times [2][N][N-1]^3\\
&\\
\begin{array}{c}\w\\ \w\\ \w\end{array} &&[N][N-1]^3 &&
2\times [N][N-1]\cdot Y_3  &&
[2][N][N-1]\cdot Y_3
\end{array}
%\ \ \ \ \ \ \ \ \ \
$$

\bigskip

\noindent
$Y_3$ in the last line is a deformation (quantization) of $(N^2-3N+4)$,
and it turns out to be
\be
Y_3 = [N-1][N-2]+[2]
\ee
%Instead, to provide
Indeed. this provides the necessary relations
\be
H^{\bu\bu}_{\bu\atop{\bu\atop \bu}} = H(5_2) =
[N]\left(1-q^{3N}\{q\}^2\frac{[2N][N+1][N-1]}{[N]}\right), \nn \\
H^{\bu\bu}_{\w\atop{\w\atop \w}} = H(4_1) = -[N]\Big(1+\{q\}^2[N+1][N-1]\Big), \nn \\
H^{\bu\w}_{\bu\atop{\bu\atop \bu}} = H(O) = [N], \nn \\
\ldots
\label{H5case}
\ee
Note that the answer for $5_2$ depends on $q$ not only through
quantum numbers -- and thus is not invariant under the change
$q\longrightarrow q^{-1}$.
In fact the complementary Khovanov-Rozansky polynomial
$H^{\w\w}_{\w\atop{\w\atop\w}} = H(5_2|q^{-1})$
differs from $H^{\bu\bu}_{\bu\atop{\bu\atop \bu}} = (5_2|q)$
exactly by this change.

\subsubsection{Generic $p$}

In general we numerate the column in the cycle diagram by $i$ -- the
number of white vertices among the $p$ vertical ones. Then

\bigskip

$$
\begin{array}{c|cccccc}
p&& \bu\bu && \bu\w+\w\bu && \w\w   \\
\hline
&  \\
0 && p+1 && 2\times p && p+1 \\
& \\
p\times 1 && p && 2\cdot p \times (p-1)&& p\times p \\
& \\
\ldots & \\
& \\
C^i_p\times i && C^i_p\times(p+1-i)&&2\cdot C^i_p \times (p-i) && C^i_p\times (p+1-i) \\
& \\
\ldots & \\
& \\
p\times (p-1) && p\times 2 && 2\cdot p\times 1&& p\times 2 \\
&\\
p &&
\begin{array}{ccc} 1 & {\rm for \ odd}\ p \\  3 & {\rm for\ even} \ p \end{array}
&& 2\times 2 && 1
\end{array}
$$

\bigskip

\noindent
Note that there is a small difference between odd and even $p$:
it is in the left lower corner of the table.
The difference will be more pronounced in the table of dimensions,
where it touches all the three entries in the last line.

Actually The table of dimensions can also be immediately written for generic $p$ --
moreover, except for the very last line, they are straightforwardly quantized:

\bigskip

$$
\begin{array}{c|cccccc}
p&& \bu\bu && \bu\w+\w\bu && \w\w   \\
\hline
&  \\
0 && [N]^{p+1} && 2\times [N]^p[N-1] && [2][N]^p[N-1] \\
& \\
p\times 1 && p\times [N]^p[N-1] && 2\cdot p \times [N]^{p-1}[N-1]^2
&& p\times [2][N]^{p-1}[N-1]^2 \\
& \\
\ldots & \\
& \\
C^i_p\times i && C^i_p\times[N]^{p+1-i}[N-1]^i &&
2\cdot C^i_p \times [N]^{p-i}[N-1]^{i+1} && C^i_p\times [2][N]^{p-i}[N-1]^{i+1} \\
& \\
\ldots & \\
& \\
p\times (p-1) && p\times [N]^2[N-1]^{p-1} &&
2\cdot p\times [N][N-1]^p&& p\times [2][N][N-1]^p \\
&\\  & \\
p &  \left[\ \begin{array}{c} {\rm odd}\ p \\ \\ {\rm even}\ p \end{array}\!\!\!\!\!\!\!\right. &
\begin{array}{c} \l[N][N-1]^p \\ \\ \l[N]^2[N-1]\cdot Y_{p-1} \end{array} &&
\begin{array}{c} 2\times [N][N-1]\cdot Y_{p} \\ \\
2\times [N][N-1]^2\cdot Y_{p-1} \end{array}  &&
\!\!\!\!\! \begin{array}{c} [2][N][N-1]\cdot Y_p\\ \\
\l[N][N-1]\Big([N-1]^{p-1}+[N-2]\cdot Y_{p-1}\Big) \end{array}
\end{array}
$$

\bigskip

\noindent
Slightly non-trivial are only the quantities $Y_k$, defined for odd values $k$,
of which we already know $Y_1 = [2]$ and $Y_3=[N-1][N-2]+[2]$.
The classical values of  $Y_k$
are polynomials of degree $k-1$ in $N$,
equal to
\be
Y_k^{cl} = \frac{(N-1)^k+(N+1)}{N}
\label{Ykcl}
\ee
Since $k$ is odd, these are indeed a polynomials, and they satisfy recursion relation
$Y_{k+2}^{cl}-Y_k^{cl} = (N-1)^k(N-2)$,
which is straightforwardly quantized:
\be
Y_{k+2} = [N-1]^k[N-2]+Y_k= [2] + \sum_{i=0}^{\frac{k-1}{2}} [N-1]^{2i+1}[N-2]
\label{Yk}
\ee
This quantization rule leads to the standard answers \cite{evo} for HOMFLY polynomials
of the twisted knots:
\be
[N]^{p+1} - q\cdot (p+2)[N]^p[N-1] + \ \ \ \ \ \ \ \ \ \ \ \ \ \ \ \ \ \ \ \ \ \
\ \ \ \ \ \ \ \ \ \ \ \ \ \ \ \ \ \ \ \ \ \ \ \ \ \ \ \ \ \ \ \ \ \ \ \ \ \ \ \ \ \ \ \
\ \ \ \ \ \ \ \ \ \ \ \ \ \ \ \ \ \ \ \ \ \ \ \ \ \ \ \ \ \ \ \ \nn \\
+ \sum_{i=2}^p (-q)^i
\left\{\left(\frac{p!}{i!(p-i)!} + \frac{2\cdot p!}{(i-1)!(p+1-i)!}\right)[N]^{p+1-i}[N-1]^i
+ \frac{[2]\cdot p!}{(i-2)!(p+2-i)!}[N]^{p+2-i}[N-1]^i\right\} + \nn \\
+(-q)^{p+1}\Big\{2[N][N-1]\cdot Y_p + p\cdot[2][N][N-1]^p\Big\}
\ +\  (-q)^{p+2} [2][N][N-1]\cdot Y_p = \nn \\
= q^{-(p+2)(N-1)}[N]\left(1 -\frac{q^{\frac{p+1}{2}N}\cdot \left[\frac{p-1}{2}N\right]}{[N]}
\cdot\{q\}^2 [N+1][N-1]\right) = q^{-(p+2)(N-1)}\cdot H^{(p+2)_2}
\nn
\ee
and
\be
\!\!\!\!\!\!\!\!\!\!\!\!\!\!
[N][N-1]^p - q\Big\{p\cdot[N]^2[N-1]^{p-1}+2[N][N-1]Y_p\Big\}
+ q^2\left\{\frac{p(p-1)}{2}[N]^3[N-1]^{p-2} +2p\cdot [N][N-1]^p+[2][N][N-1]\cdot Y_p\right\}
+ \nn \\   \!\!\!\!\!\!\!\!\!\!\!\!\!\!
+\sum_{i=0}^{p-3} (-q)^{p-i}\left\{\frac{p!}{i!(p-i)!}[N]^{p+1-i}[N-1]^i
+ \frac{2\cdot p!}{(i+1)!(p-i-1)!}[N]^{p-i-1}[N-1]^{i+2}
+ \frac{[2]\cdot p!}{(i+2)!(p-i-2)!}[N]^{p-i-2}[N-1]^{i+3}\right\} + \nn \\
+ (-q)^{p+1}\Big\{2[N]^p[N-1]+p\cdot [2][N]^{p-1}[N-1]^2\Big\}\ + \ (-q)^{p+2}[2][N]^p[N-1]=\nn \\
\!\!\!\!\!\!\!\!\!\!\!\!\!\! = \frac{(-q)^{pN}}{q^{2(N-1)}}\cdot(-)^N[N]
\left(1+\frac{q^{\frac{3-p}{2}N}\cdot\left[\frac{p-1}{2}N\right]}{[N]}\cdot\{q\}^2[N+1][N-1]\right)
= q^{(p-2)N+2}\cdot H^{(p+1)_1}
\nn
\ee
for odd $p$
and
\be
[N]^{p+1} - q\cdot (p+2)[N]^p[N-1] + \ \ \ \ \ \ \ \ \ \ \ \ \ \ \ \ \ \ \ \ \ \
\ \ \ \ \ \ \ \ \ \ \ \ \ \ \ \ \ \ \ \ \ \ \ \ \ \ \ \ \ \ \ \ \ \ \ \ \ \ \ \ \ \ \ \
\ \ \ \ \ \ \ \ \ \ \ \ \ \ \ \ \ \ \ \ \ \ \ \ \ \ \ \ \ \ \ \ \nn \\
+ \sum_{i=2}^p (-q)^i
\left\{\left(\frac{p!}{i!(p-i)!} + \frac{2\cdot p!}{(i-1)!(p+1-i)!}\right)[N]^{p+1-i}[N-1]^i
+ \frac{[2]\cdot p!}{(i-2)!(p+2-i)!}[N]^{p+2-i}[N-1]^i\right\} + \nn \\
+\ (-q)^p\Big\{[N][N-1]^p+[N]^2[N-1\cdot Y_{p-1}\Big\} + \nn \\
\ +\ (-q)^{p+1}\Big\{2[N][N-1]^2\cdot Y_{p-1} + p\cdot[2][N][N-1]^p\Big\}
\ +\  (-q)^{p+2} \Big\{[N][N-1]^p + [N][N-1][N-2]\cdot Y_{p-1}\Big\} = \nn \\
= q^{-(p+2)(N-1)}[N]\left(1 -\frac{q^{\frac{p}{2}N}\cdot \left[\frac{p}{2}N\right]}{[N]}
\cdot\{q\}^2 [N+1][N-1]\right) = q^{-(p+2)(N-1)}\cdot H^{(p+1)_2}
\nn
\ee
\be
\!\!\!\!\!\!\!\!\!\!\!\!\!\!
[N]^2[N-1]\cdot Y_{p-1} - q\Big\{p\cdot[N]^2[N-1]^{p-1}+2[N][N-1]^2\cdot Y_{p-1}\Big\}+\nn \\
+ q^2\left\{\frac{p(p-1)}{2}[N]^3[N-1]^{p-2} +2p\cdot [N][N-1]^p+
[N][N-1]^p + [N][N-1][N-2]\cdot Y_{p-1}\right\}
+ \nn \\   \!\!\!\!\!\!\!\!\!\!\!\!\!\!
+\sum_{i=0}^{p-3} (-q)^{p-i}\left\{\frac{p!}{i!(p-i)!}[N]^{p+1-i}[N-1]^i
+ \frac{2\cdot p!}{(i+1)!(p-i-1)!}[N]^{p-i-1}[N-1]^{i+2}
+ \frac{[2]\cdot p!}{(i+2)!(p-i-2)!}[N]^{p-i-2}[N-1]^{i+3}\right\} + \nn \\
+ (-q)^{p+1}\Big\{2[N]^p[N-1]+p\cdot [2][N]^{p-1}[N-1]^2\Big\}\ + \ (-q)^{p+2}[2][N]^p[N-1]=\nn \\
\!\!\!\!\!\!\!\!\!\!\!\!\!\!\!\!\!\!\!\!\!\!\!\!\!\!\!\! = \frac{(-q)^{pN}}{q^{2(N-1)}}\cdot[N]
\left(1+\frac{q^{\frac{2-p}{2}N}\cdot\left[\frac{p}{2}N\right]}{[N]}\cdot\{q\}^2[N+1][N-1]\right)
= q^{(p-2)N+2}\cdot H^{(p+2)_1}
\nn
\ee
for even $p$.

\bigskip

Quantization rule (\ref{Ykcl}) $\longrightarrow$ (\ref{Yk}) can look somewhat artificial.
However, as we shall see in the next section, this is not quite true.
The gradation-diminishing morphisms are naturally defined for the chains
of vector spaces with quantum dimensions $[N] \longrightarrow [N-1] \longrightarrow [N-2]
\longrightarrow \ldots$  -- and this is exactly a structure, implicit in (\ref{Yk}).

\newpage

To finish the entire section \ref{modif}, devoted to our new HOMFLY calculus,
we note that the HOMFLY polynomials are obtained in it by a rather strange
two-step procedure: dimensions $D^{(c)}$
are some $q$-deformed alternated summations over subsets of rhe cycles diagram,
and then HOMFLY are alternated sums of these dimension.
These two repeated sums can probably be converted into a simpler
determinant-like structure, which can also help with the quantization
($q$-deformation) -- like it happens in the studies of closely related
\cite{3dAGT,Sdua} subject of spin-chain dualities in \cite{XXXdual}.
This, however, is a subject for a separate investigation.

\bigskip

As to now, we proceed to another deformation -- to Khovanov-Rozansky polynomials.

\section{Substitute of KR cohomologies
\label{KRco}}

\subsection{The idea}

The main idea of Khovanov's approach is to interpret
$D^{\nu_c}$ in (\ref{defoinv})
as dimensions of $q$-graded vector spaces $V^{\otimes\nu_c}$,
associated with the vertices of the hypercube $H({\lD})$,
promote the coloring flips at the edges to commuting
morphisms between the vector spaces, what converts the
hypercube into Abelian quiver.
Then with this quiver one associates the complex
$K(\lD_{c_0})$, where vector spaces
\be
C_i=\oplus_c V^{\otimes\nu_c}\,\delta_{ h_c-h_{c_0},\,i}
\ee
are direct sums of those at vertices of a given height
$h_c-h_{c_0}=i$,
and differentials $d_i:\ C_{i-1}\longrightarrow C_{i}$
are combinations of commuting morphisms,
taken with appropriate signs to ensure the nilpotency
$d_{i+1}d_i=0$.

Then the entire alternated sum (\ref{defoinv})
can be interpreted as the Euler characteristic
of the complex $K(\lD_{c_0})$, while  its  Poincare polynomial
provides a new \Rede\ invariant -- Khovanov's superpolynomial.

In the language of formulas this means that we first rewrite
(\ref{defoinv})
\be
J^{\,c_{_0}}(\lD) = \alpha_\bullet^{n_\bullet}\alpha_\circ^{n_\circ}
\sum_{c=1}^{2^{n_\circ+n_\bullet}}
(-q)^{h_c-h_{c_{_0}}}\cdot D^{\,\nu_c} =
\alpha_\bullet^{n_\bullet}\alpha_\circ^{n_\circ}
\sum_{i=0}^{n_\bu+n_\w} (-q)^i\ {\rm dim}_q\, C_i
\label{defoinv1}
\ee
as
\be
J^{\,c_{_0}}(\lD) = \alpha_\bullet^{n_\bullet}\alpha_\circ^{n_\circ}
\sum_{i=0}^{n_\bu+n_\w} (-q)^i H_i
\label{defoinvhom}
\ee
where $H_{i}={\rm dim}_q\Big({\rm Ker}\ d_{i+1}\Big/{\rm Im}\ d_{i}\Big)$
are dimensions of cohomologies (quantum Betti numbers)
of the complex $K(\lD_{c_0})$ --
and afterwards we promote it to Poincare polynomial
\be
{\cal J}^{\,c_{_0}}(\lD) = \alpha_\bullet^{n_\bullet}\alpha_\circ^{n_\circ}
\sum_{i=0}^{n_\bu+n_\w} (qT)^i H_i
\ee
depending on additional parameter $T$, not obligatory equal to $-1$.
Normalization $\alpha$-parameters can also depend on $T$, in fact
\be
\alpha_\bu = q^{N-1}, \ \ \ \ \ \ \ \
\alpha_\w =   \frac{1}{q^NT}
\ee

Equivalence between (\ref{defoinv1}) and (\ref{defoinvhom}) -- the two different
representations of the Euler characteristics of a complex -- is a simple
theorem of linear algebra, which lies in the basement  of cohomology theory.
It remains true after $q$-deformation.

\bigskip

Our goal in this section is to explain what happens with this Khovanov's
construction when we substitute (\ref{defoinv}) by the its $N$-dependent
version (\ref{defoinvH}):
\be
H^{c_{_0}}(\lD) = \alpha_\bullet^{n_\bullet}\alpha_\circ^{n_\circ}
\sum_{c=1}^{2^{n_\circ+n_\bullet}}
(-q)^{h_c-h_{c_{_0}}}\cdot {D}^{(c)}
\label{defoinvH1}
\ee

$\bu$ First of all, we interpret $D^{(c)}$ as dimensions of some new
graded vector spaces, associated with the vertices of the hypercube $H(\lD)$.
In fact, this is the only thing that changes: now the basic vector space $V$
is not two-, but $N$-dimensional, and $D^{(c)}$ are dimensions of some
more sophisticated factor spaces, made from various copies of $V$.
Actually, in the present section we manage without specifying the
origin of these spaces explicitly -- but for a better grounded approach
this should be done, see s.\ref{specu} below.

$\bu$ Second, with the edges of $H(\lD)$ we associate commuting morphisms between
these vector spaces. Like in original Khovanov construction, we require
that morphisms decrease grading by one.
With each edge we associate two morphisms, acting in two directions,
both are decreasing.
Which morphism actually works, depends on the choice of initial vertex $c_0$
-- all morphism are chosen to point away from $c_0$.

$\bu$ Third, since morphisms are commuting, $H(\lD)_{c_0}$ has a structure
of Abelian quiver -- therefore there is an associated complex ${\cal K}(\lD_{c_0})$.
Therefore all the other steps remain the same:
\be
H^{c_{_0}}(\lD) = \alpha_\bullet^{n_\bullet}\alpha_\circ^{n_\circ}
\sum_{c=1}^{2^{n_\circ+n_\bullet}}
(-q)^{h_c-h_{c_{_0}}}\cdot {D}^{(c)} =
\alpha_\bullet^{n_\bullet}\alpha_\circ^{n_\circ}
\sum_{i=0}^{n_\bu+n_\w} (-q)^i\ {\rm dim}_q\, C_i
= \alpha_\bullet^{n_\bullet}\alpha_\circ^{n_\circ}
\sum_{i=0}^{n_\bu+n_\w} (-q)^i H_i \nn \\
\Longrightarrow \
{\cal P}^{\,c_{_0}}(\lD) = \alpha_\bullet^{n_\bullet}\alpha_\circ^{n_\circ}
\sum_{i=0}^{n_\bu+n_\w} (qT)^i H_i
\ \ \ \ \ \ \ \ \ \ \ \ \ \ \ \ \ \ \ \ \ \ \ \
\ \ \ \ \ \ \ \ \ \ \ \ \ \ \ \ \ \ \ \ \ \ \ \
\label{superdef}
\ee
Moreover the Poincare polynomial  ${\cal P}^{\,c_{_0}}(\lD)$,
introduced in this straightforward way,
turns to coincide with the KR superpolynomial,
obtained via matrix factorization.

\subsection{Unknot.  The choice of the main vector space: $V=C^N$
\label{unkn} }

We begin our consideration from the simplest case,
when there are $0$ vertices in the knot diagram $\lD$,
i.e. just $1$ vertex in the hypercube $H(\lD)$.
This is the ordinary unknot.

\subsubsection{Unreduced superpolynomial}

With a single vertex of the hypercube we naturally associate a vector
space $V= C^N$ with distinguished basis
$\{e_1,\ldots,e_N\}$, \ $V=\span(e_1,\ldots,e_N)$, graded as
\be
q^{g(e_i)} = q^{N+1-2i},
\ \ \ \ \ \ \ \
{\rm i.e.}
\ \ \ \
g(e_1) = N-1, \ \ \ g(e_2) = N-3,\ \ldots, \ g(e_N)=1-N
\ee
Thus quantum dimension, which by definition is the Khovanov-Rozansky
superpolynomial for the unknot in the fundamental representation $\Box$ is
\be
{\cal P}_{_\Box}(O) = {\cal P}^{\rm unknot}_{_\Box}(N|q|T)
= {\rm din}_q V = \sum_{i=1}^N q^{g(e_i)} = \frac{q^N-q^{-N}}{q-q^{-1}}=[N]
= \frac{\{A\}}{\{q\}}
\label{uPunknot}
\ee
It does not depend on the new parameter $T$.

\subsubsection{Reduced superpolynomial
\label{redunknot}}

In the theory of HOMFLY polynomials it is often convenient to
divide the answer by HOMFLY for the unknot -- what arises
is called {\it reduced} knot polynomial (and original, undivided,
is {\it un}reduced).
For superpolynomials the procedure is not so innocent:
sometime reduced superpolynomial is very different from
the unreduced one -- and in \cite{CM} they are evaluated and
listed in separate practically unrelated tables.
As reviewed in detail in  \cite{DM2}, reduced superpolynomial is obtained
in Khovanov approach by the following "reduction" procedure.
In the knot diagram $D$ we pick up (mark) one particular edge
(in principle, the answer could depend on the choice of this
edge, but it does not).
Then when at a given vertex $v$ of the hypercube $H(D)$ we decompose
$D$ into a set of cycles, we mark the cycles (one per each vertex
$v\in H(D)$), and substitute the corresponding vector space $V$
($N$-dimensional $V=C^N$ in our approach) by a one-dimensional $E=C$.
In the case of unknot this simply means that reduced superpolynomial is unity:
\be
P_{_\Box}(O) = {\rm dim}_q E = 1
\ee

Since in this paper we deal only with the fundamental representations
we often omit the subscript $\Box$ in what follows.

\subsection{Betti numbers from Euler characteristic: naive approach}

A very naive, still rather powerful approach to evaluation of
superpolynomials is to try
to saturate the given Euler characteristic by Betti numbers $H^{(c)}$,
which have lower degree in $N$ than original dimensions $D^{(c)}$.

\bigskip

\centerline{
{\footnotesize
$
\begin{array}{c|c|c|c}
{\rm knot/link} & {\rm Euler\ char} & {\rm Euler\ char}   &  {\rm Poincare\ pol} =\\
& {\rm (HOMFLY\ pol)\ via}\ D^{(c)} & {\rm via\ Betti} \ \#'s &{\rm KR\ superpolynomial} \\
\hline
1-{\rm foil} & q^{N-1}\Big([N]^2-q[N][N-1]\Big) & =[N]&  [N] \\
\hline
{\rm double\ eight} & q^{2N-2}\Big([N]^3-2q[N]^2[N-1]+ [N][N-1]^2\Big) &
= [N] & [N]\\
\hline
2-{\rm foil} & q^{2N-2}\Big([N]^2-2q[N][N-1]+q^2[2][N][N-1]\Big)
&=[N]\Big(q^{N-1}+q^{2N+1}[N-1]\Big)
& [N]\Big(q^{N-1}+q^{2N+1}T^2[N-1]\Big)  \\
\hline
3-{\rm foil} & q^{3N-3}\Big([N]^2-3q[N][N-1] + & =q^{2N-2}\Big([N]+q^3(1-q^{2N})[N-1]\Big)&
q^{2N-2}\Big([N]+q^3T^2(1+q^{2N}T)[N-1] \\
& +3q^2[2][N][N-1]-q^3[2]^2[N][N-1] \Big)    \\
\hline
\ldots
\end{array}
$}}

\bigskip

\noindent
Transition between the second and the third columns is just an identity:
we rewrite the polynomial in $A=q^N$ in the second column as a combination
of differentials $D_{-k} = \{A/q^k\}/\{q\}$ of the minimal possible degree
-- or, if degree can no longer be diminished, with minimal possible
coefficients.
Transition from this minimal polynomial to its $T$-deformed version in the
third column is often straightforward -- but, strictly speaking, not unique.
Fixing this procedure requires explicit definition of morphisms.
However, before we pass to them, it is instructive to present the above
potentially-ambiguous procedure in one more form.

\subsection{Spaces and morphisms. Unknot as an eight}

\subsubsection{Listing}

Within Khovanov approach we should interpret the quantities $D^{(c)}$
from section \ref{modif} as dimensions of some vector spaces:
$D^{(c)} = {\rm dim}_q\, V^{(c)}$.
Whatever is the deep origin of these spaces, see s.\ref{specu} below,
knowing $D^{(c)}$ we can list their basis vectors with definite grading degrees.
For example:

\be
\begin{array}{c|cc}
{\rm grading} & {\rm unknot} \\
\hline
& \\
q^{N-1} &  1 \\
q^{N-3} &  1 \\
q^{N-5} &  1 \\
\ldots & \\
q^{3-N} & 1 \\
q^{1-N} & 1 \\
&\\
\hline
&\\
{\rm dim}_q & [N]
\end{array}
\nn\\
\ee
for the unknot {\it per se} and similarly for the unknot,
represented as an eight knot from sec.\ref{eightgraphdim}:
\be
\nn\\
\begin{array}{c|ccccc}
{\rm grading} &&& {\rm 1-foil} \\
\hline
& \\
q^{2N-2} &\boxed{\boxed{1}}&  1 \\
q^{2N-3} &&&& 1  \\
q^{2N-4} &\boxed{\boxed{1}}&  2 \\
q^{2N-5} &&&& 2  \\
q^{2N-6} &\boxed{\boxed{1}}&  3 \\
&\\
\ldots && \ldots \\
&\\
q^2 &\boxed{\boxed{1}}& N-1 \\
q^1&&&&N-1 \\
q^0 &{\boxed{\boxed{1}}}\over{\boxed{1}}& N \\
q^{-1} &&&&N-1  \\
q^{-2} &\boxed{1}& N-1\\
&\\
\ldots && \ldots \\
&\\
q^{6-2N} &\boxed{1}& 3 \\
q^{5-2N} &&&&2\\
q^{4-2N} &\boxed{1}& 2 \\
q^{3-2N} &&&& 1 \\
q^{2-2N} &\boxed{1}& 1 \\
&\\
\hline
&\\
{\rm dim}_q\,C_i && [N]^2 &
\begin{array}{c}d^\bu\\ \longrightarrow \\ \longleftarrow \\ d^\w\end{array}
& [N][N-1]
\end{array}
\label{unredeighttable}
\ee
The space $V^{\otimes 2}$ at the vertex $\bu$ for the 1-fold has dimension $[N]^2$,
and there is degeneracy in
gradation already within this space: there is just one element of degree $2N-2$,
two elements of degree $2N-4$ and so on.
Likewise the factor space $V^{\otimes 2}/V$, which actually stands over the
vertex $\w$, has dimension $[N][N-1]$,
the gradings are odd, there is a single vector of the highest degree is $2N-3$,
two of degree $2N-5$ and so on.

Listed in the tables are multiplicities of basis
vectors of the given gradation degree.
At the bottom we write the sums over entire columns,
these are quantum (graded) dimensions of the spaces,
but below we often use them also to denote the spaces
themselves -- in cases where this should not make
any confusion.

\subsubsection{Morphisms and differentials}

The  table (\ref{unredeighttable})
shows very clearly what the decreasing morphisms are:
they act along decreasing diagonals -- one from left to the right,
another from right to the left.
The first one has a kernel -- its elements of a given grading are obtained
by subtracting the multiplicities along the diagonal and the remnants
are listed as a column of boxes: clearly the kernel has dimension
\be
H^\bu_0={\rm dim}_q {\rm Ker}(d^\bu) =
1+q^{-2}+\ldots + q^{4-2N} + q^{2-2N} = q^{1-N}[N]
\ee
Similarly the second one has a coimage -- again controlled by the
algebraic sums along the opposite diagonals:
the corresponding deficits are put in double boxes and
dimension of coimage is
\be
H^\w_1={\rm dim}_q {\rm CoIm}(d^\w) =
q^{2N-2} + q^{2N-4} + \ldots + q^2 +1 = q^{N-1}[N]
\ee
In this particular of the eight knot differentials in the complex ${\cal K}$
are just the morphisms, therefor from (\ref{superdef})
we get:
\be
{\cal P}^\bu = q^{N-1}\cdot H^\bu_0 = [N] \ \stackrel{(\ref{uPunknot})}{=} \ {\cal P} (O),
\ \ \ \ \ \ \ \ \ \ \ \ \ \
{\cal P}^\w = \frac{1}{q^NT}\cdot (qT)H^\w_1 = [N]  \ \stackrel{(\ref{uPunknot})}{=} \ {\cal P} (O)
\ee
and this demonstrates the \Rede\ invariance of the superpolynomial (\ref{superdef}).

\subsubsection{Reduced case}

A similar table and calculation for {\it reduced} case are even simpler:
\be
\nn\\
\begin{array}{c|ccccc}
q^{N-1} && \boxed{\boxed{ 1}} \\
q^{N-2} &&&&1 \\
q^{N-3} && 1 \\
q^{N-4} &&&& 1 \\
\ldots && \ldots\\
&\\
q^{3-N} && 1\\
q^{2-N} &&&& 1 \\
q^{1-N} && \boxed{1} \\
&\\
\hline
&\\
&&[N] &\begin{array}{c}d^\bu\\ \longrightarrow \\ \longleftarrow \\ d^\w\end{array} & [N-1]
\end{array}
\nn\\
\label{redeighttable}
\ee
where again the boxed and double-boxed entries represent the
non-vanishing cohomologies of ${\cal K}({\rm eight}_{c_0})$
with $c_0 = \bu$ and $c_0=\w$ respectively. Therefore reduced
superpolynomials are:
\be
P^\bu = q^{N-1}\cdot q^{1-N} = 1 = P(O),\ \ \ \ \ \ \ \ \ \
P^\w = \frac{1}{q^NT}\cdot (qT)\cdot q^{N-1} = 1 = P(O)
\ee

\subsubsection{Drawing: reduced case}

Now we can switch from tables to pictures and {\it draw} our two basic
decreasing morphisms (in these pictures $N=4$, but they can be used to
write formulas for arbitrary $N$):

\begin{picture}(200,130)(-100,-50)
\put(0,0){\circle*{3}}
\put(0,16){\circle*{3}}
\put(0,32){\circle*{3}}
 \put(0,48){\circle*{3}}
\put(-5,-5){\line(1,0){10}}
\put(-5,-5){\line(0,1){60}}
\put(5,55){\line(-1,0){10}}
\put(5,55){\line(0,-1){60}}
\put(-5,5){\line(1,0){10}}
%
%\put(75,-8){\line(1,0){50}}
\put(-15,0){\line(1,0){230}}
\put(50,8){\line(1,0){75}}
\put(-15,16){\line(1,0){230}}
\put(-25,24){\line(1,0){250}}
\put(-15,32){\line(1,0){230}}
\put(50,40){\line(1,0){75}}
\put(-15,48){\line(1,0){230}}
\put(50,56){\line(1,0){75}}
\put(100,8){\circle*{3}}
\put(100,24){\circle*{3}}
\put(100,40){\circle*{3}}
 %\put(100,56){\circle*{10}}
\put(95,51){\line(1,0){10}}
\put(95,3){\line(1,0){10}}
\put(95,3){\line(0,1){60}}
\put(105,63){\line(-1,0){10}}
\put(105,63){\line(0,-1){60}}
\put(200,0){\circle*{3}}
\put(200,16){\circle*{3}}
\put(200,32){\circle*{3}}
\put(200,48){\circle*{3}}
\put(195,43){\line(1,0){10}}
\put(195,-5){\line(1,0){10}}
\put(195,-5){\line(0,1){60}}
\put(205,55){\line(-1,0){10}}
\put(205,55){\line(0,-1){60}}
\put(-6,-35){\mbox{{\footnotesize $[N]$}}}
\put(65,-35){\mbox{{\footnotesize $q\Big([N]-q^{N-1}\Big)=[N-1]$}}}
\put(194,-35){\mbox{{\footnotesize $[N]$}}}
\qbezier(10,47)(50,44)(90,41)
\put(89,41){\vector(3,-1){2}}
\qbezier(10,31)(50,28)(90,25)
\put(89,25){\vector(3,-1){2}}
\qbezier(10,15)(50,12)(90,9)
\put(89,9){\vector(3,-1){2}}
\qbezier(10,-1)(50,-4)(90,-7)
\put(89,-7){\vector(3,-1){2}}
\put(95,-10){\mbox{$0$}}
\qbezier(110,39)(150,36)(190,33)
\put(189,33){\vector(3,-1){2}}
\qbezier(110,23)(150,20)(190,17)
\put(189,17){\vector(3,-1){2}}
\qbezier(110,7)(150,4)(190,1)
\put(189,1){\vector(3,-1){2}}
\put(-30,-10){\vector(3,1){15}}
\put(-50,-22){\mbox{Ker($\pi$)}}
\put(225,57){\vector(-3,-1){15}}
\put(227,60){\mbox{CoIm($\sigma$)}}
\put(48,65){\mbox{$\pi$}}
\put(148,62){\mbox{$\sigma$}}
\put(-40,21){\mbox{$0$}}
\put(-27,31){\mbox{$1$}}
\put(-27,47){\mbox{$3$}}
\put(38,36){\mbox{$2$}}
\put(30,2){\mbox{$-2$}}
\put(-35,10){\mbox{$-1$}}
\end{picture}

\noindent
In other words, $\pi$ acts as a shift down,
accompanied by multiplication by $q$, so that the grading changes
by $-1$
\be
\pi: \ \ \ \ \  [N] \ \longrightarrow \ q\Big([N]-q^{N-1}\Big) = [N-1]
\ee
while $\sigma$ is just multiplication by $q^{-1}$:
\be
\sigma: \ \ \ \ \ \left\{
\begin{array}{cccc}[N-1] & \longrightarrow &q^{-1}[N-1] = [N]-q^{N-1} & \in [N] \\ \\
\phantom.[N] & \longrightarrow   & q^{-1}[N]   \\ \\
\ldots
\end{array}\right.
\ee
The two complexes, associated with the two initial vertices black ($\bullet$)
and white ($\circ$) are:
\be
{\cal K}^\bullet: \ \ \ \ \ 0 \ \longrightarrow \
\bullet \ \stackrel{d^\bullet=\pi}{\longrightarrow}\ \circ \ \longrightarrow 0 \nn \\ \nn \\
{\cal K}^\circ: \ \ \ \ \ 0 \ \longleftarrow \
\bullet \ \stackrel{d^\circ = \sigma}{\longleftarrow}\ \circ \ \longleftarrow 0
\ee
This is a pictorial representation of the table (\ref{redeighttable}).

\subsubsection{Drawing: unreduced case}

A similar representation for (\ref{unredeighttable}) is

\begin{picture}(300,170)(-70,-90)
\put(0,0){\line(1,1){40}}
\put(0,0){\line(1,-1){40}}
\put(80,0){\line(-1,1){40}}
\put(80,0){\line(-1,-1){40}}
\put(10,0){\circle*{3}}
\put(30,0){\circle*{3}}
\put(50,0){\circle*{3}}
\put(70,0){\circle*{3}}
\put(20,10){\circle*{3}}
\put(40,10){\circle*{3}}
\put(60,10){\circle*{3}}
\put(20,-10){\circle*{3}}
\put(40,-10){\circle*{3}}
\put(60,-10){\circle*{3}}
\put(30,20){\circle*{3}}
\put(50,20){\circle*{3}}
\put(30,-20){\circle*{3}}
\put(50,-20){\circle*{3}}
\put(40,30){\circle*{3}}
\put(40,-30){\circle*{3}}
\put(30,-30){\line(1,1){40}}
\put(5,-65){\vector(1,1){25}}
\put(-15,-80){\mbox{Ker($d^\bullet$)}}

\put(100,5){\line(1,1){40}}
\put(100,5){\line(1,-1){40}}
\put(180,5){\line(-1,1){40}}
\put(180,5){\line(-1,-1){40}}
%\put(110,5){\circle*{3}}
\put(130,5){\circle*{3}}
\put(150,5){\circle*{3}}
\put(170,5){\circle*{3}}
%\put(120,15){\circle*{3}}
\put(140,15){\circle*{3}}
\put(160,15){\circle*{3}}
\put(120,-5){\circle*{3}}
\put(140,-5){\circle*{3}}
\put(160,-5){\circle*{3}}
%\put(130,25){\circle*{3}}
\put(150,25){\circle*{3}}
\put(130,-15){\circle*{3}}
\put(150,-15){\circle*{3}}
%\put(140,35){\circle*{3}}
\put(140,-25){\circle*{3}}
\put(110,-5){\line(1,1){40}}

\put(-15,0){\line(1,0){370}}
\put(-5,10){\line(1,0){80}}
\put(5,20){\line(1,0){60}}
\put(15,30){\line(1,0){40}}

\put(-25,-2){\mbox{$0$}}
\put(-15,8){\mbox{$2$}}
\put(-5,18){\mbox{$4$}}
\put(5,28){\mbox{$6$}}

\put(90,5){\line(1,0){105}}
\put(100,15){\line(1,0){85}}
\put(110,25){\line(1,0){65}}

\put(200,3){\mbox{$1$}}
\put(190,13){\mbox{$3$}}
\put(180,23){\mbox{$5$}}

\put(60,45){\vector(1,-1){20}}
\put(73,42){\mbox{$d^\bullet = \pi\otimes {\rm id}$}}
%\put(90,45){\vector(1,-1){5}}
\put(200,45){\vector(1,-1){20}}
\put(213,42){\mbox{$d^\circ = \sigma\otimes {\rm id}$}}

\put(250,0){\line(1,1){40}}
\put(250,0){\line(1,-1){40}}
\put(330,0){\line(-1,1){40}}
\put(330,0){\line(-1,-1){40}}
\put(260,0){\circle*{3}}
\put(280,0){\circle*{3}}
\put(300,0){\circle*{3}}
\put(320,0){\circle*{3}}
\put(270,10){\circle*{3}}
\put(290,10){\circle*{3}}
\put(310,10){\circle*{3}}
\put(270,-10){\circle*{3}}
\put(290,-10){\circle*{3}}
\put(310,-10){\circle*{3}}
\put(280,20){\circle*{3}}
\put(300,20){\circle*{3}}
\put(280,-20){\circle*{3}}
\put(300,-20){\circle*{3}}
\put(290,30){\circle*{3}}
\put(290,-30){\circle*{3}}
\put(260,-10){\line(1,1){40}}
%\put(5,-65){\vector(1,1){25}}
%\put(-15,-80){\mbox{Ker($d$)}}
\put(325,65){\vector(-1,-1){25}}
\put(315,75){\mbox{CoIm($d^\circ$)}}

\put(20,-65){\mbox{$[N]^2=[N]\cdot[N]$}}
\put(80,-50){\mbox{$[N][N-1]= \underbrace{q\Big([N]-q^{N-1}\Big)}\cdot[N]$}}
\put(265,-50){\mbox{$[N]^2=[N]\cdot[N]$}}
\qbezier(60,-72)(80,-100)(160,-65)
\qbezier(200,-65)(230,-90)(300,-57)
\put(156,-67){\vector(2,1){2}}
\put(298,-58){\vector(3,1){2}}
\put(100,-80){\mbox{$\pi$}}
\put(240,-70){\mbox{$\sigma$}}

\end{picture}

The spaces are now "two-dimensional", $[N]\otimes [N]$ and $[N-1]\otimes [N]$.

The complexes this time are:
\be
{\cal K}^\bullet: \ \ \ \ \ 0 \ \longrightarrow \
\bullet \ \stackrel{d^\bullet=\pi\otimes{\rm id}}{\longrightarrow}\ \circ \
\longrightarrow 0 \nn \\ \nn \\
{\cal K}^\circ: \ \ \ \ \ 0 \ \longleftarrow \
\bullet \ \stackrel{d^\circ = \sigma\otimes{\rm id}}{\longleftarrow}\ \circ \
\longleftarrow 0
\ee

Note that using ${\rm id}\otimes\sigma$ for $d^\circ$ instead of
$\sigma\otimes{\rm id}$ would give a wrong answer for ${\rm CoIm}(d^\circ)$.

\subsection{Example of the 2-foil (Hopf link)}

\subsubsection{Hopf: unreduced case}

The picture for the Hopf link in the unreduced case looks like:

\begin{picture}(300,160)(-20,-95)
\put(0,0){\line(1,1){40}}
\put(0,0){\line(1,-1){40}}
\put(80,0){\line(-1,1){40}}
\put(80,0){\line(-1,-1){40}}
\put(10,0){\circle*{3}}
\put(30,0){\circle*{3}}
\put(50,0){\circle*{3}}
\put(70,0){\circle*{3}}
\put(20,10){\circle*{3}}
\put(40,10){\circle*{3}}
\put(60,10){\circle*{3}}
\put(20,-10){\circle*{3}}
\put(40,-10){\circle*{3}}
\put(60,-10){\circle*{3}}
\put(30,20){\circle*{3}}
\put(50,20){\circle*{3}}
\put(30,-20){\circle*{3}}
\put(50,-20){\circle*{3}}
\put(40,30){\circle*{3}}
\put(40,-30){\circle*{3}}
\put(30,-30){\line(1,1){40}}
\put(5,-65){\vector(1,1){25}}
\put(-15,-80){\mbox{Ker($d_1$)}}

\put(100,5){\line(1,1){40}}
\put(100,5){\line(1,-1){40}}
\put(180,5){\line(-1,1){40}}
\put(180,5){\line(-1,-1){40}}
%\put(110,5){\circle*{3}}
\put(130,5){\circle*{3}}
\put(150,5){\circle*{3}}
\put(170,5){\circle*{3}}
%\put(120,15){\circle*{3}}
\put(140,15){\circle*{3}}
\put(160,15){\circle*{3}}
\put(120,-5){\circle*{3}}
\put(140,-5){\circle*{3}}
\put(160,-5){\circle*{3}}
%\put(130,25){\circle*{3}}
\put(150,25){\circle*{3}}
\put(130,-15){\circle*{3}}
\put(150,-15){\circle*{3}}
%\put(140,35){\circle*{3}}
\put(140,-25){\circle*{3}}
\put(110,-5){\line(1,1){40}}

\put(110,-35){\mbox{$2\ \times$}}

\put(250,0){\line(1,1){40}}
\put(250,0){\line(1,-1){40}}
\put(330,0){\line(-1,1){40}}
\put(330,0){\line(-1,-1){40}}
%\put(260,0){\circle*{3}}
\put(280,0){\circle*{3}}
\put(300,0){\circle*{3}}
\put(320,0){\circle*{3}}
%\put(270,15){\circle*{3}}
\put(290,10){\circle*{3}}
\put(310,10){\circle*{3}}
\put(270,-10){\circle*{3}}
\put(290,-10){\circle*{3}}
\put(310,-10){\circle*{3}}
%\put(280,20){\circle*{3}}
\put(300,20){\circle*{3}}
\put(280,-20){\circle*{3}}
\put(300,-20){\circle*{3}}
%\put(290,30){\circle*{3}}
\put(290,-30){\circle*{3}}
\put(260,-10){\line(1,1){40}}
\put(335,10){\mbox{$\oplus$}}
\put(350,10){\line(1,1){40}}
\put(350,10){\line(1,-1){40}}
\put(430,10){\line(-1,1){40}}
\put(430,10){\line(-1,-1){40}}
%\put(360,10){\circle*{3}}
\put(380,10){\circle*{3}}
\put(400,10){\circle*{3}}
\put(420,10){\circle*{3}}
%\put(370,20){\circle*{3}}
\put(390,20){\circle*{3}}
\put(410,20){\circle*{3}}
\put(370,0){\circle*{3}}
\put(390,0){\circle*{3}}
\put(410,0){\circle*{3}}
%\put(380,30){\circle*{3}}
\put(400,30){\circle*{3}}
\put(380,-10){\circle*{3}}
\put(400,-10){\circle*{3}}
%\put(390,40){\circle*{3}}
\put(390,-20){\circle*{3}}
\put(360,0){\line(1,1){40}}
\put(345,-45){\vector(1,1){25}}
\put(315,-60){\mbox{CoIm($d_2$)}}
\put(245,-55){\vector(1,1){25}}
\put(225,-70){\mbox{Im($d_2$)}}
\put(-15,0){\line(1,0){450}}
\put(-5,10){\line(1,0){80}}
\put(5,20){\line(1,0){60}}
\put(15,30){\line(1,0){40}}
\put(-25,-2){\mbox{$0$}}
\put(-15,8){\mbox{$2$}}
\put(-5,18){\mbox{$4$}}
\put(5,28){\mbox{$6$}}
\put(90,5){\line(1,0){105}}
\put(100,15){\line(1,0){85}}
\put(110,25){\line(1,0){65}}
\put(200,3){\mbox{$1$}}
\put(190,13){\mbox{$3$}}
\put(180,23){\mbox{$5$}}
\put(75,45){\vector(1,-1){20}}
\put(90,42){\mbox{$d_1$}}
%\put(90,45){\vector(1,-1){5}}
\put(230,40){\vector(1,-1){20}}
\put(245,37){\mbox{$d_2$}}
\end{picture}

\noindent
In the previous consideration of the eight knot we
showed in the same picture the morphisms $\pi$ and $\sigma$
acting in different directions -- i.e. relevant in the
cases of different initial vertices (channels).
Now only one channel is represented and the picture
shows the complex ${\cal K}(\lD_{c_0})$ for a given
"channel" $c_0=\bu\bu$.
Note that the space $C_1$ at the second place in the picture
consists of two copies of the same rectangular,  only
one is explicitly shown and  $2\times$ is written instead.
$d_1$ consists of tho identical maps which take the first space $[N]^2$ to
the two copies of $[N][N-1]$. It acts just as a shift in the shown direction
and  has a kernel, which is obviously $H_0=q^{1-N}[N]$.
After that $d_2$ acts as another shift -- this time in the same direction.
Because of the conspiracy of gradings the shift itself does not have neither
a kernel nor a co-image, when it acts just between the two rectangles.
However,it acts on one of the two constituents of $C_1$ with a plus sign,
and withminus -- on another: this is the standard way to construct
a complex from Abelian quiver. In result, $d_2$ has a kernel,
which is a diagonal subset in the two-constituent $C_1$ --
and this is exactly the image of $d_1$. This the cohomology $H_1=0$.
As to the target of $d_2$, the space $C_2$ also consists of two
constituents, but this time they are not identical,
but differ by $2$ in grading. $d_2$ maps $C_1$ only into the lower
constituent, while the upper one remain in co-image --
and it forms the cohomology $H_2=q^2[N]\big([N]-q^{N-1}\big) = q[N][N-1]$.

Thus looking at the picture, one straightforwardly concludes that the complex
\be
{\cal K}^{\bu\bu}: \ \ \ \ \  0 \ \longrightarrow \
\l[N]^2\ \stackrel{d_1}{\longrightarrow}\  2[N][N-1]\ \stackrel{d_2}{\longrightarrow}
\ [2][N][N-1] \ \longrightarrow \ 0
\ee
has the Poincare polynomial
\be
{\cal P}^{\bu\bu} = q^{2(N-1)}\Big(q^{1-N}[N] + (qT)\cdot 0
+ (qT)^2\cdot q^2\big([N]-q^{N-1}\big)[N-1]\Big)
= [N]\Big(q^{N-1} + q^{2N+1}T^2[N-1]\Big)
\ee
what reproduces the answer from \cite{CM}.

\subsubsection{Hopf: reduced case}

In reduced case squares and rectangles loose one dimension and are substituted by strips:
in result

\begin{picture}(200,140)(-50,-50)

\put(0,0){\circle*{3}}
\put(0,16){\circle*{3}}
\put(0,32){\circle*{3}}
 \put(0,48){\circle*{3}}
\put(-5,-5){\line(1,0){10}}
\put(-5,-5){\line(0,1){60}}
\put(5,55){\line(-1,0){10}}
\put(5,55){\line(0,-1){60}}
\put(-5,5){\line(1,0){10}}
%
%\put(75,-8){\line(1,0){50}}
\put(-15,0){\line(1,0){300}}
\put(50,8){\line(1,0){145}}
\put(-15,16){\line(1,0){300}}
\put(-25,24){\line(1,0){320}}
\put(-15,32){\line(1,0){300}}
\put(50,40){\line(1,0){145}}
\put(-15,48){\line(1,0){300}}
\put(50,56){\line(1,0){145}}
\put(100,8){\circle*{3}}
\put(100,24){\circle*{3}}
\put(100,40){\circle*{3}}
 %\put(100,56){\circle*{10}}
\put(95,51){\line(1,0){10}}
\put(95,3){\line(1,0){10}}
\put(95,3){\line(0,1){60}}
\put(105,63){\line(-1,0){10}}
\put(105,63){\line(0,-1){60}}
\put(130,8){\circle*{3}}
\put(130,24){\circle*{3}}
\put(130,40){\circle*{3}}
 %\put(100,56){\circle*{10}}
\put(125,51){\line(1,0){10}}
\put(125,3){\line(1,0){10}}
\put(125,3){\line(0,1){60}}
\put(135,63){\line(-1,0){10}}
\put(135,63){\line(0,-1){60}}
\put(230,0){\circle*{3}}
\put(230,16){\circle*{3}}
\put(230,32){\circle*{3}}
%\put(230,48){\circle*{3}}
\put(225,43){\line(1,0){10}}
\put(225,-5){\line(1,0){10}}
\put(225,-5){\line(0,1){60}}
\put(235,55){\line(-1,0){10}}
\put(235,55){\line(0,-1){60}}
\put(270,16){\circle*{3}}
\put(270,32){\circle*{3}}
\put(270,48){\circle*{3}}
 %\put(100,56){\circle*{10}}
\put(265,59){\line(1,0){10}}
\put(265,11){\line(1,0){10}}
\put(265,11){\line(0,1){60}}
\put(275,71){\line(-1,0){10}}
\put(275,71){\line(0,-1){60}}
\put(-6,-35){\mbox{{\footnotesize $[N]$}}}
\put(65,-35){\mbox{{\footnotesize $q\Big([N]-q^{N-1}\Big)=[N-1]$}}}
\put(194,-35){\mbox{{\footnotesize $\Big([N]-q^{N-1}\Big)=\frac{1}{q}[N-1]$}}}
\put(250,-18){\mbox{{\footnotesize $q^2\Big([N]-q^{N-1}\Big)={q}[N-1]$}}}
\qbezier(10,47)(50,44)(90,41)
\put(89,41){\vector(3,-1){2}}
\qbezier(10,31)(50,28)(90,25)
\put(89,25){\vector(3,-1){2}}
\qbezier(10,15)(50,12)(90,9)
\put(89,9){\vector(3,-1){2}}
\qbezier(10,-1)(50,-4)(90,-7)
\put(89,-7){\vector(3,-1){2}}
\put(95,-10){\mbox{$0$}}
\qbezier(140,39)(180,36)(220,33)
\put(219,33){\vector(3,-1){2}}
\qbezier(140,23)(180,20)(220,17)
\put(219,17){\vector(3,-1){2}}
\qbezier(140,7)(180,4)(220,1)
\put(219,1){\vector(3,-1){2}}
\put(-30,-10){\vector(3,1){15}}
\put(-50,-22){\mbox{Ker($d_1$)}}
\put(295,57){\vector(-3,-1){15}}
\put(297,60){\mbox{CoIm($d_2$)}}
\put(48,70){\mbox{$d_1$}}
\put(178,65){\mbox{$d_2$}}
\put(111,60){\mbox{$\oplus$}}
\put(247,60){\mbox{$\oplus$}}
\put(-40,21){\mbox{$0$}}
\put(-27,31){\mbox{$1$}}
\put(-27,47){\mbox{$3$}}
\put(38,36){\mbox{$2$}}
\put(30,2){\mbox{$-2$}}
\put(-35,10){\mbox{$-1$}}
\put(-4,65){\mbox{$C_0$}}
\put(111,73){\mbox{$C_1$}}
\put(245,77){\mbox{$C_2$}}
\end{picture}

\noindent
This time we showed both constituents of $C_1$ explicitly.

From this picture we immediately read:
\be
P^{\bu\bu} = q^{2(N-1)}\Big(q^{1-N} + (qT)^2\cdot q[N-1]\Big) = q^{N-1} + q^{2N+1}T^2[N-1]
\ee
what differs from the answer $q^{1-N} + q^{-1-2N}T^2[N-1]$  of \cite{CM} by a change
$q \longrightarrow 1/q$.

\bigskip

The listing of the spaces this time looks as follows:
\be
\begin{array}{c|cccccc}
&&C_0&&C_1&&C_2\\
&\\
\hline
&\\
q^{N-1} &&1 && && \ \ \ \ \ \boxed{1}   \\
q^{N-2} && &&2\cdot 1 \\
q^{N-3} && 1 && && 1+ \boxed{1} \\
q^{N-4} && &&2\cdot 1 \\
q^{N-5} && 1 && && 1+ \boxed{1} \\
\ldots & \\
q^{5-N} && 1 && && 1+ \boxed{1} \\
q^{4-N}  && &&2\cdot 1 \\
q^{3-N} && 1 && && 1+ \boxed{1} \\
q^{2-N} && &&2\cdot 1 \\
q^{1-N} && \boxed{1} && && 1 \ \ \ \ \ \ \ \, \\
&\\
\hline
& \\
&&[N] && 2\cdot [N-1] && [2][N-1]
\end{array}
\label{redHopftab}
\ee

\noindent
Here we clearly see the advantage of pictures over tables:
if we had just the table, we could alternatively box the entire
first column, getting alternative expression
$q^{N-1}[N]$ for the superpolynomial.
Knowing the morphisms from the pictures, we can easily
reject this option.

\subsubsection{Another channel: two unknots. Reduced case}

The next exercise is to look at the same link diagram ${\lD}$
in another channel,
with initial vertex $\bu\w$.
This requires some morphisms, acting in the other direction.

It will be a little more convenient to begin from the table.
What happens is that now the second column of (\ref{redHopftab})
splits into two, which become the first and the third,
while those instead get combined into the second one:

\be
\begin{array}{c|cccccc}
&&C_0&&C_1&&C_2\\
&\\
\hline
&\\
q^{N-1} && &&\boxed{1}+1 && \ \ \ \ \     \\
q^{N-2} &&1 && && 1  \\
q^{N-3} &&  &&\boxed{1}+2 &&   \\
q^{N-4} &&1 &&  && 1 \\
q^{N-5} &&  &&\boxed{1}+2 &&   \\
\ldots & \\
q^{5-N} &&  && \boxed{1}+2 &&   \\
q^{4-N}  &&1 &&  && 1 \\
q^{3-N} &&  &&\boxed{1}+2 &&   \\
q^{2-N} &&1 && && 1 \\
q^{1-N} &&  &&\boxed{1}+1 &&  \ \ \ \ \ \ \ \, \\
&\\
\hline
& \\
&&[N-1] && [N]+[2][N-1] && [N-1]
\end{array}
\label{redcrossHopftab}
\ee

\noindent
This time there is no ambiguity: non-vanishing cohomology
lies in the middle column and therefore
\be
P^{\bu\w} = \frac{q^{N-1}}{q^NT} \cdot(qT)\cdot [N] = [N] = {\cal P}(O)
\ee
This is the correct answer: for $c_0=\bu\w$ we should get a pair of unknots
and "reduced" means that one of them is eliminated --
thus what we could expect is exactly one {\it unreduced}
unknot -- and this is what we get.

The corresponding pattern of morphisms is

\begin{picture}(200,145)(-80,-50)
\put(0,0){\circle*{3}}
\put(0,16){\circle*{3}}
\put(0,32){\circle*{3}}
 \put(0,48){\circle*{3}}
\put(-5,-5){\line(1,0){10}}
\put(-5,-5){\line(0,1){60}}
\put(5,55){\line(-1,0){10}}
\put(5,55){\line(0,-1){60}}
\put(-5,5){\line(1,0){10}}
%
%\put(75,-8){\line(1,0){50}}
\put(-15,0){\line(1,0){230}}
\put(50,8){\line(1,0){300}}
\put(-15,16){\line(1,0){360}}
\put(-25,24){\line(1,0){380}}
\put(-15,32){\line(1,0){360}}
\put(50,40){\line(1,0){300}}
\put(-15,48){\line(1,0){360}}
%\put(50,56){\line(1,0){75}}
%
\put(100,8){\circle*{3}}
\put(100,24){\circle*{3}}
\put(100,40){\circle*{3}}
 %\put(100,56){\circle*{10}}
\put(95,51){\line(1,0){10}}
\put(95,3){\line(1,0){10}}
\put(95,3){\line(0,1){60}}
\put(105,63){\line(-1,0){10}}
\put(105,63){\line(0,-1){60}}
\put(200,0){\circle*{3}}
\put(200,16){\circle*{3}}
\put(200,32){\circle*{3}}
%\put(200,48){\circle*{3}}
\put(195,43){\line(1,0){10}}
\put(195,-5){\line(1,0){10}}
\put(195,-5){\line(0,1){60}}
\put(205,55){\line(-1,0){10}}
\put(205,55){\line(0,-1){60}}
\put(230,16){\circle*{3}}
\put(230,32){\circle*{3}}
\put(230,48){\circle*{3}}
%\put(200,48){\circle*{3}}
\put(225,59){\line(1,0){10}}
\put(225,11){\line(1,0){10}}
\put(225,11){\line(0,1){60}}
\put(235,71){\line(-1,0){10}}
\put(235,71){\line(0,-1){60}}
\put(330,8){\circle*{3}}
\put(330,24){\circle*{3}}
\put(330,40){\circle*{3}}
 %\put(100,56){\circle*{10}}
\put(325,3){\line(1,0){10}}
\put(325,3){\line(0,1){60}}
\put(335,63){\line(-1,0){10}}
\put(335,63){\line(0,-1){60}}
\put(325,51){\line(1,0){10}}
\put(-4,65){\mbox{$C_1$}}
\put(96,75){\mbox{$C_0$}}
\put(207,65){\mbox{$\oplus$}}
\put(207,80){\mbox{$C_1$}}
\put(326,75){\mbox{$C_2$}}
\put(-6,-25){\mbox{{\footnotesize $[N]$}}}
\put(50,-20){\mbox{{\footnotesize $q\Big([N]-q^{N-1}\Big)=[N-1]$}}}
\put(177,-25){\mbox{{\footnotesize $\frac{1}{q}[N-1]$}}}
\put(240,-5){\mbox{{\footnotesize ${q}[N-1]$}}}
\put(320,-20){\mbox{{\footnotesize $[N-1]$}}}
%
%\qbezier(10,33)(50,37)(90,41)
\qbezier(10,33)(50,36)(90,39)
\put(11,33){\vector(-3,-1){2}}
\qbezier(10,17)(50,20)(90,23)
\put(11,17){\vector(-3,-1){2}}
\qbezier(10,1)(50,4)(90,7)
\put(11,1){\vector(-3,-1){2}}
%\qbezier(10,-1)(50,-4)(90,-7)
%\put(89,-7){\vector(3,-1){2}}
%\put(95,-10){\mbox{$0$}}
\qbezier(110,39)(150,36)(190,33)
\put(189,33){\vector(3,-1){2}}
\qbezier(110,23)(150,20)(190,17)
\put(189,17){\vector(3,-1){2}}
\qbezier(110,7)(150,4)(190,1)
\put(189,1){\vector(3,-1){2}}
\qbezier(240,47)(280,44)(320,41)
\put(319,41){\vector(3,-1){2}}
\qbezier(240,31)(280,28)(320,25)
\put(319,25){\vector(3,-1){2}}
\qbezier(240,15)(280,12)(320,9)
\put(319,9){\vector(3,-1){2}}
%
%
%\put(-30,-10){\vector(3,1){15}}
%\put(-50,-22){\mbox{Ker($\pi$)}}
%\put(225,57){\vector(-3,-1){15}}
%\put(227,60){\mbox{CoIm($\sigma$)}}
\put(48,65){\mbox{$d_1^{\bu\w}$}}
\put(148,62){\mbox{$d_1^{\bu\w}$}}
\put(278,62){\mbox{$d_2^{\bu\w}$}}
\put(-40,21){\mbox{$0$}}
\put(-27,31){\mbox{$1$}}
\put(-27,47){\mbox{$3$}}
\put(38,39){\mbox{$2$}}
\put(30,6){\mbox{$-2$}}
\put(-35,10){\mbox{$-1$}}
\put(-10,-10){\vector(-1,-1){20}}
\put(-35,-17){\mbox{$d_2^{\bu\w}$}}
\put(-40,-40){\mbox{$0$}}
\put(210,-10){\vector(1,-1){20}}
\put(228,-20){\mbox{$d_2^{\bu\w}$}}
\put(237,-40){\mbox{$0$}}
\end{picture}

\noindent
The differential $d_2$ annihilates the two components $N$ and $\frac{1}{q}[N-1]$
of $C_1$ and only the third one, $q[N-1]$, is mapped one-to-one onto $C_2=[N-1]$.
Thus ${\rm Ker}(d_2^{\bu\w}) = [N]+\frac{1}{q}[N-1]$, but part of this space
is also the image of $d_1^{\bu\w}$, so that cohomology $H_1^{\bu\w} = [N]$ --
in full accordance with what we saw in analysis of the table
(\ref{redcrossHopftab}).

\subsubsection{Another channel: two unknots. Unreduced case}

In this case we restrict consideration only to the table:
\be
\begin{array}{c|cccccc}
&&C_0&&C_1&&C_2\\
&\\
\hline
&\\
q^{2N-2} && &&\boxed{1}\ \ \ +\ \ \ 1 && \ \ \ \ \     \\
q^{2N-3} &&1 && && 1  \\
q^{2N-4} &&  &&\boxed{2}+2+2 &&   \\
q^{2N-5} &&2 &&  && 2 \\
q^{2N-6} &&  &&\boxed{3}+2+3 &&   \\
&&&&&& 3\\
\ldots & && \ldots\\
&\\
q^2 && && \boxed{N-1}+(N-2)+(N-1) && \\
q^1  && N-1 && && N-1  \\
q^0  && && \boxed{N} + (N-1) + (N-1) && \\
q^{-1}  &&N-1 && && N-1 \\
q^{-2}  && &&\boxed{N-1}+(N-1)+(N-2) &&\\
q^{-3} &&N-2&&&&N-2 \\
q^{-4} && && \boxed{N-2} + (N-2)+(N-3) && \\
&\\
\ldots & && \ldots \\
&& &&&& N-3\\
q^{6-2N} &&  && \boxed{3}+3+2 &&   \\
q^{5-2N}  &&2 &&  && 2 \\
q^{4-2N} &&  &&\boxed{2}+2+1 &&   \\
q^{3-2N} &&1 && && 1 \\
q^{2-2N} &&  &&\boxed{1}+1 \ \ \ \ \  &&  \ \ \ \ \ \ \ \, \\
&\\
\hline
& \\
&&[N][N-1] && [N]^2+[2][N][N-1] && [N][N-1]
\end{array}
\label{unredcrossHopftab}
\ee
In result
\be
{\cal P}^{\bu\w} = \frac{q^{N-1}}{q^NT}\cdot (qT)\cdot
\Big(q^{2N-2} + 2q^{2N-4} + 3q^{N-6} + \ldots +q^{2-2N}\Big) = [N]^2
= \Big({\cal P}(O)\Big)^2
\ee
The picture is also easy to draw, but it gives nothing new and
we do not present it here.

\subsection{3-foil}

\subsubsection{Two different morphisms on the case of unreduced trefoil}

A new phenomenon in the case of the unreduced trefoil is
the relevance of two different decreasing morphisms and differentials.
The way they arise is clear from the following picture:

\begin{picture}(300,150)(-50,-70)

\put(0,0){\line(1,1){40}}
\put(0,0){\line(1,-1){40}}
\put(80,0){\line(-1,1){40}}
\put(80,0){\line(-1,-1){40}}
\put(10,0){\circle*{3}}
\put(30,0){\circle*{3}}
\put(50,0){\circle*{3}}
\put(70,0){\circle*{3}}
\put(20,10){\circle*{3}}
\put(40,10){\circle*{3}}
\put(60,10){\circle*{3}}
\put(20,-10){\circle*{3}}
\put(40,-10){\circle*{3}}
\put(60,-10){\circle*{3}}
\put(30,20){\circle*{3}}
\put(50,20){\circle*{3}}
\put(30,-20){\circle*{3}}
\put(50,-20){\circle*{3}}
\put(40,30){\circle*{3}}
\put(40,-30){\circle*{3}}
\put(30,-30){\line(1,1){40}}
\put(5,-65){\vector(1,1){25}}
\put(-15,-80){\mbox{Ker($d_1$)}}

\put(100,5){\line(1,1){40}}
\put(100,5){\line(1,-1){40}}
\put(180,5){\line(-1,1){40}}
\put(180,5){\line(-1,-1){40}}
%\put(110,5){\circle*{3}}
\put(130,5){\circle*{3}}
\put(150,5){\circle*{3}}
\put(170,5){\circle*{3}}
%\put(120,15){\circle*{3}}
\put(140,15){\circle*{3}}
\put(160,15){\circle*{3}}
\put(120,-5){\circle*{3}}
\put(140,-5){\circle*{3}}
\put(160,-5){\circle*{3}}
%\put(130,25){\circle*{3}}
\put(150,25){\circle*{3}}
\put(130,-15){\circle*{3}}
\put(150,-15){\circle*{3}}
%\put(140,35){\circle*{3}}
\put(140,-25){\circle*{3}}
\put(110,-5){\line(1,1){40}}

\put(110,-35){\mbox{$3\ \times$}}

\put(230,-15){\mbox{$3\ \times$}}
\put(340,-15){\mbox{$3\ \times$}}

\put(250,0){\line(1,1){40}}
\put(250,0){\line(1,-1){40}}
\put(330,0){\line(-1,1){40}}
\put(330,0){\line(-1,-1){40}}
%\put(260,0){\circle*{3}}
\put(280,0){\circle*{3}}
\put(300,0){\circle*{3}}
\put(320,0){\circle*{3}}
%\put(270,15){\circle*{3}}
\put(290,10){\circle*{3}}
\put(310,10){\circle*{3}}
\put(270,-10){\circle*{3}}
\put(290,-10){\circle*{3}}
\put(310,-10){\circle*{3}}
%\put(280,20){\circle*{3}}
\put(300,20){\circle*{3}}
\put(280,-20){\circle*{3}}
\put(300,-20){\circle*{3}}
%\put(290,30){\circle*{3}}
\put(290,-30){\circle*{3}}
\put(260,-10){\line(1,1){40}}

\put(335,10){\mbox{$\oplus$}}

\put(350,10){\line(1,1){40}}
\put(350,10){\line(1,-1){40}}
\put(430,10){\line(-1,1){40}}
\put(430,10){\line(-1,-1){40}}
%\put(360,10){\circle*{3}}
\put(380,10){\circle*{3}}
\put(400,10){\circle*{3}}
\put(420,10){\circle*{3}}
%\put(370,20){\circle*{3}}
\put(390,20){\circle*{3}}
\put(410,20){\circle*{3}}
\put(370,0){\circle*{3}}
\put(390,0){\circle*{3}}
\put(410,0){\circle*{3}}
%\put(380,30){\circle*{3}}
\put(400,30){\circle*{3}}
\put(380,-10){\circle*{3}}
\put(400,-10){\circle*{3}}
%\put(390,40){\circle*{3}}
\put(390,-20){\circle*{3}}
\put(360,0){\line(1,1){40}}
\put(345,-45){\vector(1,1){25}}
\put(315,-60){\mbox{CoIm($d_2$)}}
\put(245,-55){\vector(1,1){25}}
\put(225,-70){\mbox{Im($d_2$)}}

\put(-15,0){\line(1,0){450}}
\put(-5,10){\line(1,0){80}}
\put(5,20){\line(1,0){60}}
\put(15,30){\line(1,0){40}}

\put(-25,-2){\mbox{$0$}}
\put(-15,8){\mbox{$2$}}
\put(-5,18){\mbox{$4$}}
\put(5,28){\mbox{$6$}}

\put(90,5){\line(1,0){105}}
\put(100,15){\line(1,0){85}}
\put(110,25){\line(1,0){65}}

\put(200,3){\mbox{$1$}}
\put(190,13){\mbox{$3$}}
\put(180,23){\mbox{$5$}}

\put(75,45){\vector(1,-1){20}}
\put(90,42){\mbox{$d_1$}}
%\put(90,45){\vector(1,-1){5}}
\put(230,40){\vector(1,-1){20}}
\put(245,37){\mbox{$d_2$}}

\end{picture}

\begin{picture}(300,190)(287,-120)
\put(230,-15){\mbox{$3\ \times$}}
\put(340,-15){\mbox{$3\ \times$}}

\put(250,0){\line(1,1){40}}
\put(250,0){\line(1,-1){40}}
\put(330,0){\line(-1,1){40}}
\put(330,0){\line(-1,-1){40}}
%\put(260,0){\circle*{3}}
\put(280,0){\circle*{3}}
\put(300,0){\circle*{3}}
\put(320,0){\circle*{3}}
%\put(270,15){\circle*{3}}
\put(290,10){\circle*{3}}
\put(310,10){\circle*{3}}
\put(270,-10){\circle*{3}}
\put(290,-10){\circle*{3}}
\put(310,-10){\circle*{3}}
%\put(280,20){\circle*{3}}
\put(300,20){\circle*{3}}
\put(280,-20){\circle*{3}}
\put(300,-20){\circle*{3}}
%\put(290,30){\circle*{3}}
\put(290,-30){\circle*{3}}
\put(260,-10){\line(1,1){40}}

\put(335,10){\mbox{$\oplus$}}

\put(350,10){\line(1,1){40}}
\put(350,10){\line(1,-1){40}}
\put(430,10){\line(-1,1){40}}
\put(430,10){\line(-1,-1){40}}
%\put(360,10){\circle*{3}}
\put(380,10){\circle*{3}}
\put(400,10){\circle*{3}}
\put(420,10){\circle*{3}}
%\put(370,20){\circle*{3}}
\put(390,20){\circle*{3}}
\put(410,20){\circle*{3}}
\put(370,0){\circle*{3}}
\put(390,0){\circle*{3}}
\put(410,0){\circle*{3}}
%\put(380,30){\circle*{3}}
\put(400,30){\circle*{3}}
\put(380,-10){\circle*{3}}
\put(400,-10){\circle*{3}}
%\put(390,40){\circle*{3}}
\put(390,-20){\circle*{3}}
\put(360,0){\line(1,1){40}}
\put(400,-20){\line(-1,1){30}}
\put(340,-48){\vector(1,1){25}}
\qbezier(360,-4)(367,-23)(384,-28)
\put(315,-60){\mbox{Ker($d_3$)/Im($d_2$)}}
\put(245,-55){\vector(1,1){25}}
\put(225,-70){\mbox{Im($d_2$)}}
\put(435,45){\vector(-1,-1){15}}
\put(420,53){\mbox{CoIm($d_2$)}}
\put(700,45){\vector(-3,-1){30}}
\put(685,52){\mbox{CoIm($d_3$)}}

\put(480,-15){\mbox{$2\ \times$}}

\put(500,5){\line(1,1){40}}
\put(500,5){\line(1,-1){40}}
\put(580,5){\line(-1,1){40}}
\put(580,5){\line(-1,-1){40}}
%\put(110,5){\circle*{3}}
\put(530,5){\circle*{3}}
\put(550,5){\circle*{3}}
\put(570,5){\circle*{3}}
%\put(120,15){\circle*{3}}
\put(540,15){\circle*{3}}
\put(560,15){\circle*{3}}
\put(520,-5){\circle*{3}}
\put(540,-5){\circle*{3}}
\put(560,-5){\circle*{3}}
%\put(130,25){\circle*{3}}
\put(550,25){\circle*{3}}
\put(530,-15){\circle*{3}}
\put(550,-15){\circle*{3}}
%\put(140,35){\circle*{3}}
\put(540,-25){\circle*{3}}
\put(510,-5){\line(1,1){40}}

\put(585,10){\mbox{$\oplus$}}

\put(600,15){\line(1,1){40}}
\put(600,15){\line(1,-1){40}}
\put(680,15){\line(-1,1){40}}
\put(680,15){\line(-1,-1){40}}
%\put(110,5){\circle*{3}}
\put(630,15){\circle*{3}}
\put(650,15){\circle*{3}}
\put(670,15){\circle*{3}}
%\put(120,15){\circle*{3}}
\put(640,25){\circle*{3}}
\put(660,25){\circle*{3}}
\put(620,5){\circle*{3}}
\put(640,5){\circle*{3}}
\put(660,5){\circle*{3}}
%\put(130,25){\circle*{3}}
\put(650,35){\circle*{3}}
\put(630,-5){\circle*{3}}
\put(650,-5){\circle*{3}}
%\put(140,35){\circle*{3}}
\put(640,-15){\circle*{3}}
\put(610,5){\line(1,1){40}}
\put(670,5){\line(-1,1){30}}

\put(680,-10){\mbox{$\oplus$}}

\put(700,-5){\line(1,1){40}}
\put(700,-5){\line(1,-1){40}}
\put(780,-5){\line(-1,1){40}}
\put(780,-5){\line(-1,-1){40}}
%\put(110,5){\circle*{3}}
\put(730,-5){\circle*{3}}
\put(750,-5){\circle*{3}}
\put(770,-5){\circle*{3}}
%\put(120,15){\circle*{3}}
\put(740,5){\circle*{3}}
\put(760,5){\circle*{3}}
\put(720,-15){\circle*{3}}
\put(740,-15){\circle*{3}}
\put(760,-15){\circle*{3}}
%\put(130,25){\circle*{3}}
\put(750,15){\circle*{3}}
\put(730,-25){\circle*{3}}
\put(750,-25){\circle*{3}}
%\put(140,35){\circle*{3}}
\put(740,-35){\circle*{3}}
\put(710,-15){\line(1,1){40}}

\qbezier(293,-50)(350,-150)(710,-30)
\put(708,-31){\vector(3,1){2}}
\qbezier(400,-30)(450,-50)(510,-23)
\put(508,-24){\vector(2,1){2}}
\qbezier(393,-40)(450,-90)(610,-25)
\put(608,-26){\vector(2,1){2}}

\put(380,-100){\vector(1,-1){10}}
\put(450,-22){\vector(1,-1){10}}
\put(480,-45){\vector(-1,-1){10}}
\put(360,-108){\mbox{$d_3:$}}
\put(430,-30){\mbox{$d_3:$}}
\put(450,-53){\mbox{$d_3:$}}

\put(240,0){\line(1,0){555}}
\put(485,5){\line(1,0){300}}
\put(800,-5){\mbox{$0$}}
\put(790,5){\mbox{$1$}}
%\put(-5,10){\line(1,0){80}}
%\put(5,20){\line(1,0){60}}
%\put(15,30){\line(1,0){40}}

%\put(-25,-2){\mbox{$0$}}
%\put(-15,8){\mbox{$2$}}
%\put(-5,18){\mbox{$4$}}
%\put(5,28){\mbox{$6$}}

%\put(90,5){\line(1,0){105}}
%\put(100,15){\line(1,0){85}}
%\put(110,25){\line(1,0){65}}

%\put(200,3){\mbox{$1$}}
%\put(190,13){\mbox{$3$}}
%\put(180,23){\mbox{$5$}}

%\put(75,45){\vector(1,-1){20}}
%\put(90,42){\mbox{$d$}}
%%\put(90,45){\vector(1,-1){5}}
%\put(230,40){\vector(1,-1){20}}
%\put(245,37){\mbox{$d_2$}}

\end{picture}

The space ${\cal C}_2$ is shown two times: at the r.h.s. of the first line and
at the l.h.s. of the second line.
The three arrows at the the second line are constituents of the map $d_3$.

Cohomology
Ker($d_3$)/Im($d_2$)
lies in only one (diagonal) of the three components of Coim($s_2$) --
the rest of it is mapped into the "upper" constituent of the space ${\cal C}_3$,
which also contains Coim($d_3$).
Note that {\bf $d_3$ on this component acts in another direction}.
This allows to make the cohomology {\it smaller}, i.e.subtract as much as
only possible from coimage of $d_3$. This is what we call the {\it maximal subtraction}
rule.

\subsubsection{Morphisms}

Morphisms are shown in the picture.
One can also write them more formally.
For this we introduce the basis $\{e_I\}$ with
${\rm grad}(e_I) = q^{N+1-2I}$, $I=1,\ldots,N$,
$i = 2,\ldots, N$. Then
%(instead one could define $e_i$ with another grading the grading $q^{N+2-2i}$)
\be
d_1:&\ \ \ \ \ \
e_{IJ} = e_I\otimes e_J \longrightarrow  q\cdot e_{I+1}\otimes e_J = q\cdot e_{iJ}
\nn \\ \nn \\
d_2:& \ \ \ \ \ \
qe_{iJ}^{(a)}  \ \longrightarrow    \sum_{b,c}
\epsilon^{abc} e_{iJ}^{(b)},0)
\nn \\
d_3:& \ \ \ \ \ \
\Big(e_{iJ}^{(a)},\ \ q^2  \tilde e_{iJ}^{(b)}\Big) \ \longrightarrow \
 \left(\sum_b  q\omega_b \tilde e_{i|J+1}^{(b)}, \ \ \
 \sum_b     q\bar\omega_b \tilde e_{i|J+1}^{(b)}, \ \ \
 \sum_b q^3 \tilde e_{i|J+1}^{(b)},\ \ \ \ \
 \sum_a q^{-1} \tilde e_{iJ}^{(a)}\right)
\ee
Here $\omega_1 = 1, \ \omega_2 = e^{2\pi i/3} = \bar\omega_3$.
We made a sort of symmetric choice for the mapping in the subspace within $C_2$,
orthogonal to diagonal in the last line -- but this is not
canonical:  what matters is just the mapping of this entire
two-dimensional space onto its two-dimensional counterpart in $C_3$.

In a similar way one can define morphisms in all other channels.
In the rest of this subsection we present the tables of multiplicities,
cohomologies and the superpolynomials in all the four channels
for the trefoil knot diagram, in unreduced and reduced cases.

\subsubsection{Unreduced trefoil in  different channels}

%\bigskip

{\footnotesize
\be
\begin{array}{c|cccccccccccccc}
&& && bbw&&bww&&& \\
{\rm grads}&&bbb&&bwb&&wbw&&www& \\
&&&& wbb &&wwb&&& \\
&&&&&&&&& \\
\hline
&&&&&&&&& \\
2N-1 && && && &&\boxed{1} & \\
2N-2 && \boxed{\boxed{1}} && && 3\cdot 1 && & \\
2N-3 && && 3\cdot 1 && && 2+2\cdot 1=4 &\boxed{(1)} \\
2N-4 && 2\ \ \boxed{\boxed{(1)}} && && 3\cdot(1+2)=9 \ \ \boxed{\boxed{(1)}} && & \\
2N-5 && && 3\cdot 2 && && 3+2\cdot 2 + 1 = 8 &\boxed{(1)} \\
2N-6 && 3\ \ \boxed{\boxed{(1)}} && && 3\cdot(2+3)=15 \ \ \boxed{\boxed{(1)}} && & \\
2N-7 && && 3\cdot 3 && && 4+2\cdot 3+2=12 & \boxed{(1)} \\
2N-8 && 4\ \ \boxed{\boxed{(1)}} && && 3\cdot(3+4)=21\ \ \boxed{\boxed{(1)}} && & \\
2N-9 && && 3\cdot 4 && && 5+2\cdot 4+3=16 & \boxed{(1)} \\ \\
\ldots &&&& \ldots &&&&& \\    \\
3 &&  && 3\cdot(N-2) &&  && N-1+2(N-2)+N-3 & \boxed{(1)} \\
  &&  &&   &&  &&  =4N-8 &  \\
2 && N-1 \ \ \boxed{\boxed{(1)}} && && 3\cdot(N-2+N-1) && & \\
  &&  &&   && =6N-9\ \ \boxed{\boxed{(1)}} &&  &  \\
1 &&  && 3\cdot(N-1) &&  &&N-1+2(N-1)+N-2  &   \\
  &&  &&   &&  &&  =4N-5 &  \\
0 && N \ \ \boxed{(1)}\ \ \boxed{\boxed{(1)}} && && 3\cdot(N-1+N-1) && & \\
  &&  &&   && =6N-6 \ \ \boxed{(1)}\ \ \boxed{\boxed{(1)}} &&  &  \\
-1 &&  && 3\cdot(N-1) &&  &&N-2+2(N-1)+N-1 &  \\
  &&  &&   &&  &&  =4N-5 &  \\
-2 && N-1  \ \ \boxed{(1)} && && 3\cdot(N-2+N-1)  && & \\
  &&  &&   && =6N-9\ \ \boxed{(1)} &&  &  \\
-3 &&  && 3\cdot(N-2) &&  &&N-3+2(N-2)+N-1   &  \\
  &&  &&   &&  &&  =4N-8 &\boxed{\boxed{(1)}}  \\
\ldots &&&& \ldots &&&&& \\    \\
10-2N && 5 \ \ \boxed{(1)} && && 3\cdot(4+5)=27 \ \ \boxed{(1)} && & \\
9-2N && && 3\cdot 4 && && 3+2\cdot 4 +5=16 & \boxed{\boxed{(1)}}\\
8-2N && 4\ \ \boxed{(1)} && && 3\cdot(3+4)=21\ \ \boxed{(1)} && & \\
7-2N && && 3\cdot 3 && && 2+2\cdot 3+4=12 &\boxed{\boxed{(1)}} \\
6-2N && 3\ \ \boxed{(1)} && && 3\cdot(2+3)=15 \ \ \boxed{(1)}&& & \\
5-2N && && 3\cdot 2 && && 1+2\cdot 2+3=8 &\boxed{\boxed{(1)}} \\
4-2N && 2\ \ \boxed{(1)} && && 3\dot(1+2)=9\ \ \boxed{(1)} && & \\
3-2N && && 3\cdot 1  && && 2\cdot 1+2 =4 & \boxed{\boxed{(1)}}\\
2-2N && \boxed{1} && && 3\cdot 1 && & \\
1-2N && &&  && && \boxed{\boxed{(1)}} &  \\ \\ \hline \\
 && [N]^2 && 3\cdot[N][N-1] && 3\cdot[2][N][N-1] &&(1+[3])\cdot[N][N-1]
\end{array}
\nn
\ee
}

\bigskip

\bigskip

For initial vertex $bbb$ we need to look at diagonal lines,
decreasing from left to right.
Contributing to cohomologies will be the lines with non-vanishing
alternating sums.
However, this time in the lower half of the table we have diagonals
with the sums equal to $2$.
This defect of two can be distributed among the two possible columns
in three different ways: $2+0$, $1+1$ and $0+2$.
The relevant choice is $1+1$ and these are the cohomologies contributions a
shown in boxes.
This choice gives rise to the superpolynomial
\be
{\cal P}^{\bullet\bullet\bullet}_{_\Box} =
q^{3(N-1)}\Big( q^{1-N}[N] + 0\cdot(qT) + q^{2-N}[N-1]\cdot (qT)^2 + q^{N+1}[N-1]\cdot(qT)^3\Big)
= \nn \\
= q^{2N-2}\Big( [N]+q^3T^2(1 + q^{2N}T)[N-1]\Big)
\ee

Similarly for initial vertex $www$ we need to pick up
diagonals, decreasing from right to left.
Again there are different possible distributions of the defects,
this time in the upper half of the table.
The relevant choice is again $1+1$,
the corresponding cohomologies are double-boxed and  the superpolynomial is
\be
{\cal P}^{\circ\circ\circ}_{_\Box}(q,T) =
q^{-3N}T^{-3}\Big(
q^{-N-1}[N-1] + q^{N-2}[N-1]\cdot(qT) + 0\cdot (qT)^2 + q^{N-1}[N]\cdot(qT)^3\Big)=\nn\\
= q^{2-2N}\Big( q^{-2N-3}T^{-3}[N-1]+q^{-3}T^{-2}[N-1] + [N]\Big)
= P^{\bullet\bullet\bullet}_{_\Box}\Big(q^{-1},T^{-1}\Big)
\ee

\bigskip

For initial vertex $bbw$ we need to rearrange the columns:
\be
\begin{array}{c|cccrlcrlcccccc}
&& && bww&&&bwb&&& \\
{\rm grads}&&bbw&&wbw&bbb&&wbb&www&&wwb& \\
&&&&&&&&& \\
2N-1 &&     && &     && &1  && \\
2N-2 &&  && 2\cdot 1&\!\!\!+1    &&  &  &&1  \\
2N-3 &&1 && &     &&2\cdot 1 &\!\!\!+ 4 && \\
2N-4 &&  && 2\cdot(1+2)&\!\!\!+2     && &  &&3 \\
2N-5 &&2 && &     && 2\cdot 2&\!\!\!+8  && \\
2N-6 &&  && 2\cdot(2+3)&\!\!\!+3     && &  &&5 \\
2N-7 &&3 && &     && 2\cdot 3&\!\!\!+12  && \\  \\
\ldots &&&&&&&&& \\    \\
8-2N &&  && 2\cdot(4+3)&\!\!\!+4  \ \ \boxed{(1)}   && & &&7 \\
7-2N &&3 && &     && 2\cdot 3&\!\!\!+12 && \\
6-2N &&  && 2\cdot(3+2)&\!\!\!+3 \ \ \boxed{(1)}    && & &&5 \\
5-2N &&2 && &     && 2\cdot 2&\!\!\!+8 && \\
4-2N &&  && 2\cdot(2+1)&\!\!\!+2 \ \ \boxed{(1)}    && & &&3 \\
3-2N &&1 && &     && 2\cdot 1&\!\!\!+4 && \\
2-2N &&  && 2\cdot 1 &\!\!\!+1 \ \ \boxed{(1)}    && & &&1 \\
1-2N  &&  && &     && &1 &&
\end{array}
\ee

\bigskip
Diagonals are decreasing from left to right,
{\it location} of non-vanishing cohomology is boxed, so that
\be
{\cal P}^{\bullet\bullet\circ}_{_\Box}(q,T) =
\frac{q^{2(N-1)}}{q^NT}\Big( 0 + q^{1-N}[N]\cdot(qT) + 0\cdot(qT)^2 + 0\cdot(qT)^3\Big)
= [N] ={\cal P}^{\rm unknot}_{_\Box}(q,T)
\ee

\bigskip

Similarly
for initial vertex $bww$:
\be
\begin{array}{c|cccrlcrlcccccc}
&& && bww&&&bwb&&& \\
{\rm grads}&&bbw&&wbw&bbb&&wbb&www&&wwb& \\
&&&&&&&&& \\
2N-1 &&     && &1     && &  && \\
2N-2 && 1 && &    &&  2\cdot 1&\!\!\!+1  \ \ \boxed{(1)} &&  \\
2N-3 && &&2\cdot 1 &\!\!\!+4     && & &&1 \\
2N-4 &&3  && &     && 2\cdot 3&\!\!\!+2 \ \ \boxed{(1)}  && \\
2N-5 && && 2\cdot 2&\!\!\! +8    && &  &&2 \\
2N-6 &&5  && &     && 2\cdot 5&\!\!\!+3 \ \ \boxed{(1)}  && \\
2N-7 && &&2\cdot 3 &\!\!\!+12     && &  &&3 \\
2N-8 &&7  && &     && 2\cdot 7&\!\!\!+4 \ \ \boxed{(1)}  && \\ \\
\ldots &&&&&&&&& \\    \\
8-2N &&7  && &      && 2\cdot 7&\!\!\!+4 && \\
7-2N && &&2\cdot 3 &\!\!\! +12    &&  &  &&3 \\
6-2N &&5  && &   && 2\cdot 5&\!\!\!+3 && \\
5-2N && && 2\cdot2&\!\!\!+8     && & &&2 \\
4-2N &&3  && &    && 2\cdot 3&\!\!\!+2 && \\
3-2N && &&2\cdot 1 &\!\!\!+4     && & &&1 \\
2-2N &&1  &&  &      && 2\cdot 1&\!\!\!+1 && \\
1-2N  &&  && &1     && & &&
\end{array}
\ee
Diagonals are again decreasing from left to right, and
\be
{\cal P}^{\bullet\circ\circ}_{_\Box}(q,T) =
\frac{q^{N-1}}{(q^NT)^2}\Big( 0 + 0\cdot (qT) + q^{N-1}[N]\cdot(qT)^2 + 0\cdot(qT)^3\Big)
= [N] ={\cal P}^{\rm unknot}_{_\Box}(q,T)
\ee

\subsubsection{Reduced trefoil in three different channels}

\be
\begin{array}{c|cccccccccccccc}
&& && bbw&&bww&&& \\
{\rm grads}&&bbb&&bwb&&wbw&&www& \\
&&&& wbb &&wwb&&& \\
&&&&&&&&& \\
\hline
&&&&&&&&& \\
N && && && &&\boxed{1} & \\
N-1 && \boxed{\boxed{1}} && && 3 && & \\
N-2 && && 3 && && 3 & \\
N-3 && 1 && && \ \ \ \ \ \ \ \ \ \,6\ \ \boxed{\boxed{ (1)}} && & \\
N-4 && && 3 && && 4 & \\
N-5 && 1 && && 6 && & \\
N-6 && && 3 && && 4 & \\
N-7 && 1 && && 6 && & \\ \\
\ldots &&&& \ldots &&&&& \\    \\
7-N && 1 && && 6 && & \\
6-N && && 3 && && 4 & \\
5-N && 1 && && 6 && & \\
4-N && && 3 && && 4 & \\
3-N && 1 && && \ \ \ \ \ \ \ \,6\ \ \boxed{(1)} && & \\
2-N && && 3 && && 3 & \\
1-N && \boxed{1} && && 3 && & \\
-N && &&  && && \boxed{\boxed{ 1}} &  \\ \\ \hline \\
 && [N] && 3\cdot[N-1] && 3\cdot[2][N-1] &&(1+[3])\cdot[N-1]
\end{array}
\label{tabredbbb}
\ee

\bigskip

\bigskip

For initial vertex $bbb$ we need to look at diagonal lines,
decreasing from left to right.
Contributing to cohomologies will be the lines with non-vanishing
alternating sums -- remaining contributions are in boxes.
Collecting all the three we get
\be
P^{\bullet\bullet\bullet}_{_\Box} =
q^{3(N-1)}\Big( q^{1-N} + 0\cdot(qT) + q^{3-N}\cdot (qT)^2 + q^N\cdot(qT)^3\Big)
= q^{2N}\Big( q^{-2}+q^2T^2 + q^{2N}T^3\Big)
\ee
Similarly for initial vertex $www$ we need to do the same with
diagonals, decreasing from right to left.
The corresponding cohomologies are double-boxed, and
\be
P^{\circ\circ\circ}_{_\Box}(q,T) =
q^{-3N}T^{-3}\Big( q^{-N} + q^{N-3}\cdot(qT) + 0\cdot (qT)^2 + q^{N-1}\cdot(qT)^3\Big)=\nn\\
= q^{-2N}\Big( q^{-2N}T^{-3}+q^{-2}T^{-2} + q^{2}\Big)
= P^{\bullet\bullet\bullet}_{_\Box}\Big(q^{-1},T^{-1}\Big)
\ee

For initial vertex $bbw$ we need to rearrange the columns:
\be
\begin{array}{c|cccrlcrlcccccc}
&& && bww&&&bwb&&& \\
{\rm grads}&&bbw&&wbw&bbb&&wbb&www&&wwb& \\
&&&&&&&&& \\
N &&     && &     && &1  && \\
N-1 &&  && 2&1    &&  &  &&1  \\
N-2 &&1 && &     &&2 & 3 && \\
N-3 &&  && 4&1     && &  &&2 \\
N-4 &&1 && &     && 2&4  && \\
N-5 &&  && 4&1     && &  &&2 \\
N-6 &&1 && &     && 2&4  && \\  \\
\ldots &&&&&&&&& \\    \\
7-N &&  && 4&1     && & &&2 \\
6-N &&1 && &     && 2&4 && \\
5-N &&  && 4&1     && & &&2 \\
4-N &&1 && &     && 2&4 && \\
3-N &&  && 4&1     && & &&2 \\
2-N &&1 && &     && 2&3 && \\
1-N &&  && \boxed{2}&\boxed{1}     && & &&1 \\
-N  &&  && &     && &1 &&
\end{array}
\ee

\bigskip
Diagonals are decreasing from left to right,
{\it location} of non-vanishing cohomology is boxed, so that
\be
P^{\bullet\bullet\circ}_{_\Box}(q,T) =
\frac{q^{2(N-1)}}{q^NT}\Big( 0 + q^{1-N}\cdot(qT) + 0\cdot(qT)^2 + 0\cdot(qT)^3\Big) = 1
=P^{\rm unknot}_{_\Box}(q,T)
\ee

\bigskip

Similarly
for initial vertex $bww$:
\be
\begin{array}{c|ccrrllrlcccccc}
&& &&             bbw&&&   wwb&&& \\
{\rm grads}&&bww&&bwb&wbb&&wbw&bbb&&wbb& \\
&&&&&&&&& \\
N &&     && &1     && &  && \\
N-1 &&1  && &    &&  \boxed{2}&\boxed{1}  &&  \\
N-2 && && 2&3     && & &&1 \\
N-3 &&2  && &     && 4&1  && \\
N-4 && && 2&4     && &  &&1 \\
N-5 &&2  && &     && 4&1  && \\
N-6 && && 2&4     && &  &&1 \\  \\
\ldots &&&&&&&&& \\    \\
7-N &&2  && &     && 4&1 && \\
6-N && && 2&4     && & &&1 \\
5-N &&2  && &     && 4&1 && \\
4-N && && 2&4     && & &&1 \\
3-N &&2  && &     && 4&1 && \\
2-N && && 2&3     && & &&1 \\
1-N &&1  && &     && 2&1 && \\
-N  &&  && &1     && & &&
\end{array}
\ee
Diagonals are again decreasing from left to right, and
\be
P^{\bullet\circ\circ}_{_\Box}(q,T) =
\frac{q^{N-1}}{(q^NT)^2}\Big( 0 + 0\cdot (qT) + q^{N-1}\cdot(qT)^2 + 0\cdot(qT)^3\Big) = 1
=P^{\rm unknot}_{_\Box}(q,T)
\ee

\subsection{$k$-folds}

With above experience we are now ready to describe the entire series
of 2-strand knots and links and reproduce the well known result of
\cite{Gor,AgSh,DMMSS,HaLi} from our version of Khovanov's construction.

\subsubsection{Betti numbers for arbitrary 2-strand torus knots}

According to \cite{DMMSS}, {\it reduced} polynomials for 2-strand knots   are:
\be
P^{[2,2k+1]}_{_\Box} \sim \Big\{\Big(1+q^4T^2+q^8T^4+ \ldots + (q^4T^2)^{k}\Big) +
q^{2N+2}T^3\Big(1+q^4T^2+\ldots + (q^4T^2)^{k-1}\Big)\Big\}
\ee
what means that the quantum Betti numbers are
\be
{\rm reduced\ case:}\ \ \ \ \ \ \
1,\ 0,\ q^4,\ q^{2N+2},\ q^8,\ q^{2N+6},\ q^{12},\ q^{2N+10},\ \ldots,\
q^{4k},\ q^{2N+4k-2}
\ee

{\it Unreduced} superpolynomials are less available in the literature,
but from our above considerations it is clear that for the 2-strand knots
one should just introduce factors $[N]$
for the zeroth Betti number and $[N-1]$ for all the rest
and slightly modify the gradings:
\be
{\rm unreduced\ case:}\ \ \ \ \ \ \
\l[N],\ 0,\ [N-1]\times\Big(q^{3},\ q^{2N+3},\ q^7,\ q^{2N+7},\ q^{11},\
q^{2N+11},\
\ldots,  \
q^{4k-1},\ q^{2N+4k-1}\Big)
\ee
For $q=1$ we get an extremely simple pattern:
\be
N, \ 0,\ \underbrace{N-1,\ N-1,\ \ldots,\ N-1}_{2k}
\ee
so that the sum rule (\ref{normH}) is nicely satisfied.

In secs.\ref{how1} and \ref{howl} we demonstrate that
these answers can be easily {\it deduced} from our construction.

\subsubsection{Betti numbers for arbitrary 2-strand torus  links}

For links the structure of generic answers is more subtle.
From \cite{DMMSS} we know that links are associated with super{\it series}
(with all coefficients positive),
rather than polynomials (we denote this quantity by underlined $P$),
and in {\it reduced} case
\be
\underline{P}^{[2,2k]}_{_{\Box\times\Box}}(a,q,T) \sim \frac{1}{1-q^2}
\Big\{\left((1-q^2)\Big(1+q^4T^2+q^8T^4+ \ldots + (q^4T^2)^{k-1}\Big) +
(q^4T^2)^{k}\right) +\nn \\
+ (qT)^{2}\cdot  \left(a^2T (1-q^2)
\Big(1+q^4T^2+\ldots +(q^4T^2)^{k-2}\Big)
+ \Big(\underline{(a^2T+q^{2N})} -q^{2N} \Big)(q^4T^2)^{k-1}\right)\Big\}
\ee
In passing to Khovanov-Rozansky superpolynomial
the underlined term should be eliminated (just erased) to convert the series
into a finite polynomial
(see \cite{DGR} and remark after eq.(34) in \cite{AnoMMM21}).
Thus Khovanov-Rozansky polynomial, implied by \cite{DMMSS} is
\be
P^{[2,2k]}_{_{\Box\times\Box}}(N,q,T) \sim
(1+a^2q^2T^3)\Big(1+q^4T^2 + \ldots + (q^4T^2)^{k-2}\Big)
+ (q^4T^2)^{k-1} + \frac{1-q^{2N-2}}{1-q^2}(q^4T^2)^k
\ee
where one finally substitutes $a=q^N$.

For $k=1$ this gives
$q^{N-1}\Big(\underbrace{1 + \frac{a^2T+q^2}{1-q^2}\cdot(qT)^2}_{\cite{DMMSS}}
- \frac{a^2T+q^{2N}}{1-q^2}(qT)^2\Big)
= \underbrace{q^{N-1}\Big(1+q^{N+2}T^2[N-1]\Big)}_{\cite{CM}}$,

for $k=2$ \  --  \ $q^{3(N-1)}\Big(1+q^4T^2+q^{2N+2}T^3+q^{N+6}T^4[N-1]\Big)$
-- this coincides with $4^2_1$(v2) of \cite{CM} after the substitution $t=1/T$.

Thus quantum Betti numbers are:
\be
{\rm reduced\ case:} \ \ \ \ \ \
1, \ 0,\ q^4 , \ q^{2N+6}, \ q^8,\  q^{2N+10},\ \ldots, \
q^{4k-4}, \ q^{2N+4k-2},\ q^{N+4k-2}[N-1]
\ee
Based on above experience,
in unreduced case we insert $[N]$ into the first term, $[N]$ (in addition to already
existing $[N-1]$) into the last, and $[N-1]$ everywhere else.
Finally, modifying appropriately the gradings, we get:
\be
{\rm unreduced\ case:} \ \ \ \ \ \
[N], \ 0,\ q^{4}[N-1] , \ q^{2N+6}[N-1], \ q^{8}[N-1],\
q^{2N+10}[N-1],\ \ldots, \nn \\
q^{4k-4}[N-1], \ q^{2N+4k-2}[N-1],\ q^{N+4k-2}[N][N-1]
\ee
For $q=1$ we get:
\be
N, 0, \underbrace{N-1,  N-1, \ldots, N-1}_{2k-2}, N(N-1)
\ee
in accordance with (\ref{normH}), because the 2-strand link has
exactly two components.

Now we proceed to the derivation of these results from our
approach.

\subsubsection{How this works. Reduced case
\label{how1}}

In fact it is sufficient just to redraw our pictures in
appropriate way.
Namely, put the lowest (in grading) constituents of all spaces $C_i$
in the first line, then the next -- into the second, and so on.
This makes the structure of morphisms  absolutely transparent
and cohomologies trivial to evaluate.

{\bf Hopf:}
\be
\begin{array}{ccccc}
C_0&& C_1 && C_2\\ \\
\l[N] & \stackrel{d_1}{\longrightarrow} & 2[N-1] &
\stackrel{d_2}{\longrightarrow} & \frac{1}{q}[N-1] \\ \\
&&&& q[N-1]       \\
\end{array}
\ee
The first line has a single non-vanishing cohomology in the first term,
the kernel contains one element of dimension $H_0=q^{1-N}$.
The fact that there is nothing else follows from $1-2+1=0$.
Nothing is mapped to the second line, it is pure cohomology $H_2$.
Thus
\be
P^{\bu\bu} = q^{2(N-1)}\Big(q^{1-N} + (qT)^2 q[N-1]\Big)
\ee

{\bf Trefoil:}
\be
\begin{array}{ccccccc}
C_0&& C_1 && C_2 && C_3\\ \\
\l[N] & \longrightarrow & 3[N-1] & \longrightarrow & \frac{3}{q}[N-1]
& \longrightarrow &  \frac{1}{q^2}[N-1] \\ \\
&&&& (2+1)q[N-1]  & \stackrel{2}{\longrightarrow} &  2[N-1]  \\
 &&&&&\stackrel{1}{\searrow} \\
&&&&&&q^2[N-1]      \\
\end{array}
\ee
The first line has a single non-vanishing cohomology in the first term,
the kernel contains one element of dimension $q^{1-N}$.
The fact that there is nothing else follows from $1-3+3-1=0$.
In the second line the situation is different: $3-2=1\neq 0$.
This what we do, we split the first item in this line $3=2+1$.
Then $2-2=0$ and there is no cohomologies in this {\it reduced} line,
while the remnant gets mapped into the third line,
providing a new cohomology -- because $q[N-1]\longrightarrow q^2[N-1]$
where the map of the weight $-1$ has non-vanishing kernel and coimage:
$q\cdot q^{2-N}$ and $q^2\cdot q^{N-2}$ respectively.

In result
\be
P^{\bu\bu\bu}= q^{3(N-1)}\Big(q^{1-N} + (qT)^2\cdot q\cdot q^{2-N} +
(qT)^3\cdot q^2\cdot q^{N-2}\Big)
\ee

{\bf 4-foil:}

\be
\begin{array}{ccccccccc}
\l[N] & \longrightarrow & 4[N-1] & \longrightarrow & \frac{6}{q}[N-1]
& \longrightarrow &  \frac{4}{q^2}[N-1] & \longrightarrow & \frac{1}{q^3}[N-1] \\ \\
&&&& (5+1)q[N-1]  & \stackrel{5}{\longrightarrow} &  8[N-1]
& \longrightarrow &  \frac{3}{q}[N-1]\\
&&&& &\stackrel{1}{\searrow} \\
&&&& &&4q^2[N-1] & \longrightarrow &  3q[N-1] \\  \\
&&&&&&&& q^3[N-1]
\end{array}
\ee
The balance in lines is now:
\be
\underline{1}-4+8-4+1= 0, \nn \\
5-8+3=0, \nn \\
\underline{1}-(\overline{1}+3)+3 = 0, \nn \\
\underline{1} \neq 0
\ee
and unbalanced cohomologies provide
\be
P^{\bu\bu\bu\bu} = q^{4(N-1)}\Big(q^{1-N} +(qT)^2\cdot q\cdot q^{2-N}
+(qT)^3\cdot q^2\cdot q^{N-2} + (qT)^4\cdot q^3[N-1]\Big)
\ee

{\bf 5-foil:}

\be
\begin{array}{ccccccccccc}
\l[N] & \longrightarrow & 5[N-1] & \longrightarrow & \frac{10}{q}[N-1]
& \longrightarrow &  \frac{10}{q^2}[N-1] & \longrightarrow & \frac{5}{q^3}[N-1]
& \longrightarrow & \frac{1}{q^4}[N-1]       \\ \\
&&&& (9+1)q[N-1]  & \stackrel{9}{\longrightarrow} &  20[N-1]
& \longrightarrow &  \frac{15}{q}[N-1]  & \longrightarrow &
\frac{4}{q^2}[N-1] \\
&&&& &\stackrel{1}{\searrow} \\
&&&& &&10q^2[N-1] & \longrightarrow &  15q[N-1] & \longrightarrow & 6[N-1]\\  \\
&&&&&&&& (4+1)q^3[N-1]    & \stackrel{4}{\longrightarrow} &  4q^2[N-1] \\
&&&& &&&&&\stackrel{1}{\searrow} \\
&&&&&&&&&&q^4[N-1]
\end{array}
\nn
\ee
The balance in lines is now:
\be
\underline{1}-5+10-10+5-1= 0, \nn \\
9-20+15-4=0, \nn \\
\underline{1}-(\overline{1}+9)+15-6 = 0, \nn \\
4-4 =0, \nn \\
\underline{1}-\underline{1} = 0
\ee
and unbalanced cohomologies provide
\be
P^{\bu\bu\bu\bu\bu} = q^{5(N-1)}\Big(q^{1-N} +(qT)^2\cdot q\cdot q^{2-N}
+(qT)^3\cdot q^2\cdot q^{N-2} + (qT)^4\cdot q^3\cdot q^{2-N} +
(qT)^5\cdot q^4\cdot q^{N-2} \Big)
\ee

{\bf Generic case:}

Note that combinatorial factors in above tables are the products of
$C^k_n$ and $C_{k-1}^{j-1}$ for the item at the crossing of
the $k$-th column and $j$-th line.

The balances are:
\be
\begin{array}{c|ccc}
\\
1 & \sum_{k=0}^n(-)^k C^k_n = (1-1)^n = \delta_{n,0} \\
\\
2 &  -1+\sum_{k=2}^n (-)^k (k-1) C^k_n   &
=-1+ \left.\frac{\p}{\p x}\frac{(1-x)^n-1}{x}\right|_{x=1} & = \delta_{n,1} \\ \\
3 & 1+\sum_{k=3}^n(-)^k \frac{)k-1)(k-2)}{2}C^k_n  &
= 1+\left.\frac{1}{2}\frac{\p^2}{\p x^2}\frac{(1-x)^n-1}{x}\right|_{x=1}& = \delta_{n,2} \\ \\
4 & -1+\sum_{k=3}^n(-)^k \frac{)k-1)(k-2)(k-3)}{6}C^k_n  &
= -1+\left.\frac{1}{3!}\frac{\p^3}{\p x^3}\frac{(1-x)^n-1}{x}\right|_{x=1}& = \delta_{n,3} \\ \\
& \ldots   \\ \\
j & (-)^{j+1}+\sum_{k=0}^n (-)^kC^{j-1}_{k-1}C^k_n = \delta_{n,j-1}\\  &
\end{array}
\ee

\subsubsection{How this works. Unreduced case
\label{howl}}

Nothing changes in above reasoning, with the only addition:
all maps {\it within} the lines are provided by the $\searrow$ shifts,
while those {\it between} the lines -- by $\swarrow$ shifts.
This makes all the non-vanishing cohomologies, arising from
the {\it between-the-lines} maps proportional to $[N-1]$,
while the single one in the very first line -- the only one,
non-vanishing {\it within-the-lines}, is proportional to $[N]$.
The unmatched last cohomology in the case of links is
proportional to $[N][N-1]$.

\subsection{Double eight}

As we know from sec.\ref{dei}, this time the resolution at Seifert vertex
involves three cycles, thus in unreduced case the basises of our vector spaces
st the hypercube vertices will be represented by cubes, and in reduced
case they will turn into squares and rectangulars.
Cubes are more difficult to draw, therefore we begin from reduced case.

\subsubsection{Reduced case. The Seifert channel (initial point $\bu\bu$)
\label{debb}}

If initial vertex is $\bu\bu$, i.e. is a Seifert vertex, then

\begin{picture}(300,190)(-30,-100)
\put(0,0){\line(1,1){40}}
\put(0,0){\line(1,-1){40}}
\put(80,0){\line(-1,1){40}}
\put(80,0){\line(-1,-1){40}}
\put(10,0){\circle*{3}}
\put(30,0){\circle*{3}}
\put(50,0){\circle*{3}}
\put(70,0){\circle*{3}}
\put(20,10){\circle*{3}}
\put(40,10){\circle*{3}}
\put(60,10){\circle*{3}}
\put(20,-10){\circle*{3}}
\put(40,-10){\circle*{3}}
\put(60,-10){\circle*{3}}
\put(30,20){\circle*{3}}
\put(50,20){\circle*{3}}
\put(30,-20){\circle*{3}}
\put(50,-20){\circle*{3}}
\put(40,30){\circle*{3}}
\put(40,-30){\circle*{3}}
\put(30,-30){\line(1,1){10}}
\put(50,-30){\line(-1,1){10}}
\put(5,-65){\vector(1,1){25}}
\put(-15,-80){\mbox{Ker($d_1$)}}
\put(100,5){\line(1,1){40}}
\put(100,5){\line(1,-1){40}}
\put(180,5){\line(-1,1){40}}
\put(180,5){\line(-1,-1){40}}
%\put(110,5){\circle*{3}}
\put(130,5){\circle*{3}}
\put(150,5){\circle*{3}}
\put(170,5){\circle*{3}}
%\put(120,15){\circle*{3}}
\put(140,15){\circle*{3}}
\put(160,15){\circle*{3}}
\put(120,-5){\circle*{3}}
\put(140,-5){\circle*{3}}
\put(160,-5){\circle*{3}}
%\put(130,25){\circle*{3}}
\put(150,25){\circle*{3}}
\put(130,-15){\circle*{3}}
\put(150,-15){\circle*{3}}
%\put(140,35){\circle*{3}}
\put(140,-25){\circle*{3}}
\put(110,-5){\line(1,1){40}}
\put(120,-5){\circle{7}}
\put(130,-15){\circle{7}}
\put(140,-25){\circle{7}}
%
%\put(110,-35){\mbox{$2\ \times$}}
\put(186,-15){\mbox{$\oplus$}}
\put(200,5){\line(1,1){40}}
\put(200,5){\line(1,-1){40}}
\put(280,5){\line(-1,1){40}}
\put(280,5){\line(-1,-1){40}}
\put(210,5){\circle*{3}}
\put(230,5){\circle*{3}}
\put(250,5){\circle*{3}}
%\put(270,5){\circle*{3}}
\put(220,15){\circle*{3}}
\put(240,15){\circle*{3}}
%\put(260,15){\circle*{3}}
\put(220,-5){\circle*{3}}
\put(240,-5){\circle*{3}}
\put(260,-5){\circle*{3}}
\put(230,25){\circle*{3}}
%\put(250,25){\circle*{3}}
\put(230,-15){\circle*{3}}
\put(250,-15){\circle*{3}}
%\put(240,35){\circle*{3}}
\put(240,-25){\circle*{3}}
\put(230,35){\line(1,-1){40}}
\put(260,-5){\circle{7}}
\put(250,-15){\circle{7}}
\put(240,-25){\circle{7}}
\put(350,10){\line(1,1){40}}
\put(350,10){\line(1,-1){40}}
\put(430,10){\line(-1,1){40}}
\put(430,10){\line(-1,-1){40}}
%\put(360,10){\circle*{3}}
\put(380,10){\circle*{3}}
\put(400,10){\circle*{3}}
%\put(420,10){\circle*{3}}
%\put(370,20){\circle*{3}}
\put(390,20){\circle*{3}}
%\put(410,20){\circle*{3}}
\put(370,0){\circle*{3}}
\put(390,0){\circle*{3}}
\put(410,0){\circle*{3}}
%\put(380,30){\circle*{3}}
%\put(400,30){\circle*{3}}
\put(380,-10){\circle*{3}}
\put(400,-10){\circle*{3}}
%\put(390,40){\circle*{3}}
\put(390,-20){\circle*{3}}
\put(360,0){\line(1,1){30}}
\put(420,0){\line(-1,1){30}}
%\put(345,-45){\vector(1,1){25}}
%\put(315,-60){\mbox{CoIm($d_2$)}}
%\put(245,-55){\vector(1,1){25}}
%\put(225,-70){\mbox{Im($d_2$)}}
%
\put(-15,0){\line(1,0){450}}
\put(-5,10){\line(1,0){80}}
\put(5,20){\line(1,0){60}}
\put(15,30){\line(1,0){40}}
\put(-25,-2){\mbox{$0$}}
\put(-15,8){\mbox{$2$}}
\put(-5,18){\mbox{$4$}}
\put(5,28){\mbox{$6$}}
\put(90,5){\line(1,0){205}}
\put(100,15){\line(1,0){185}}
\put(110,25){\line(1,0){165}}
\put(300,3){\mbox{$1$}}
\put(290,13){\mbox{$3$}}
\put(280,23){\mbox{$5$}}
\qbezier(45,40)(85,75)(126,40)
\put(124,42){\vector(1,-1){2}}
\qbezier(42,45)(120,100)(200,30)
\put(198,32){\vector(1,-1){2}}
\put(75,45){\vector(1,-1){15}}
\put(90,42){\mbox{$d_1$}}
%\put(90,45){\vector(1,-1){5}}
\put(205,58){\vector(-1,-1){15}}
\put(210,50){\mbox{$d_1$}}
\qbezier(155,-25)(270,-140)(375,-25)
\put(373,-27){\vector(1,1){2}}
\qbezier(255,-30)(300,-70)(365,-15)
\put(363,-16.5){\vector(1,1){2}}
\put(200,-30){\vector(-1,-1){15}}
\put(205,-40){\mbox{$d_2$}}
\put(300,-20){\vector(1,-1){15}}
\put(318,-30){\mbox{$d_2$}}
\put(293,-30){\mbox{$-$}}
\put(38,70){\mbox{$C_0$}}
\put(185,70){\mbox{$C_1$}}
\put(388,70){\mbox{$C_2$}}
\put(35,-60){\mbox{$[N]^2$}}
\put(110,-60){\mbox{$[N][N-1]$}}
\put(225,-60){\mbox{$[N][N-1]$}}
\put(375,-60){\mbox{$[N-1]^2$}}
\end{picture}

\noindent
There is a single non-vanishing cohomology,
consisting of a single point: $H_0=q^{2-2N}$.
We encycled the elements of $C_1$, which
belong to the kernel of $d_2$, but are
independent in two constituents of $C_1$.
The other elements of $C_1$ are the same
in both constituents, when they are obtained
by the action of $d_1$, i.e. belong to
${\rm Im}(d_1)$ -- and they are
annihilated by $d_2$ because of the two
different signs of this map on two
constituents.
In result we have:
\be
P^{\bu\bu} = q^{2(N-1)}\cdot H_0 = 1 = P(O)
\ee

\subsubsection{Reduced case. The anti-Seifert  channel (initial point $\w\w$)
\label{deww}}

If initial vertex is $\w\w$,  then

\begin{picture}(300,190)(-30,-100)
\put(0,10){\line(1,1){40}}
\put(0,10){\line(1,-1){40}}
\put(80,10){\line(-1,1){40}}
\put(80,10){\line(-1,-1){40}}
%\put(10,10){\circle*{3}}
\put(30,10){\circle*{3}}
\put(50,10){\circle*{3}}
%\put(70,10){\circle*{3}}
%\put(20,20){\circle*{3}}
\put(40,20){\circle*{3}}
%\put(60,20){\circle*{3}}
\put(20,-0){\circle*{3}}
\put(40,-0){\circle*{3}}
\put(60,-0){\circle*{3}}
%\put(30,30){\circle*{3}}
%\put(50,30){\circle*{3}}
\put(30,-10){\circle*{3}}
\put(50,-10){\circle*{3}}
%\put(40,40){\circle*{3}}
\put(40,-20){\circle*{3}}
\put(10,-0){\line(1,1){30}}
\put(70,-0){\line(-1,1){30}}
%\put(5,-65){\vector(1,1){25}}
%\put(-15,-80){\mbox{Ker($d_1$)}}
%
%
\put(100,5){\line(1,1){40}}
\put(100,5){\line(1,-1){40}}
\put(180,5){\line(-1,1){40}}
\put(180,5){\line(-1,-1){40}}
%\put(110,5){\circle*{3}}
\put(130,5){\circle*{3}}
\put(150,5){\circle*{3}}
\put(170,5){\circle*{3}}
%\put(120,15){\circle*{3}}
\put(140,15){\circle*{3}}
\put(160,15){\circle*{3}}
\put(120,-5){\circle*{3}}
\put(140,-5){\circle*{3}}
\put(160,-5){\circle*{3}}
%\put(130,25){\circle*{3}}
\put(150,25){\circle*{3}}
\put(130,-15){\circle*{3}}
\put(150,-15){\circle*{3}}
%\put(140,35){\circle*{3}}
\put(140,-25){\circle*{3}}
\put(110,-5){\line(1,1){40}}
\put(170,5){\circle{7}}
\put(160,15){\circle{7}}
\put(150,25){\circle{7}}
%
%\put(110,-35){\mbox{$2\ \times$}}
\put(186,-15){\mbox{$\oplus$}}
\put(200,5){\line(1,1){40}}
\put(200,5){\line(1,-1){40}}
\put(280,5){\line(-1,1){40}}
\put(280,5){\line(-1,-1){40}}
\put(210,5){\circle*{3}}
\put(230,5){\circle*{3}}
\put(250,5){\circle*{3}}
%\put(270,5){\circle*{3}}
\put(220,15){\circle*{3}}
\put(240,15){\circle*{3}}
%\put(260,15){\circle*{3}}
\put(220,-5){\circle*{3}}
\put(240,-5){\circle*{3}}
\put(260,-5){\circle*{3}}
\put(230,25){\circle*{3}}
%\put(250,25){\circle*{3}}
\put(230,-15){\circle*{3}}
\put(250,-15){\circle*{3}}
%\put(240,35){\circle*{3}}
\put(240,-25){\circle*{3}}
\put(230,35){\line(1,-1){40}}
\put(210,5){\circle{7}}
\put(220,15){\circle{7}}
\put(230,25){\circle{7}}
\put(350,0){\line(1,1){40}}
\put(350,0){\line(1,-1){40}}
\put(430,0){\line(-1,1){40}}
\put(430,0){\line(-1,-1){40}}
\put(360,0){\circle*{3}}
\put(380,0){\circle*{3}}
\put(400,0){\circle*{3}}
\put(420,0){\circle*{3}}
\put(370,10){\circle*{3}}
\put(390,10){\circle*{3}}
\put(410,10){\circle*{3}}
\put(370,-10){\circle*{3}}
\put(390,-10){\circle*{3}}
\put(410,-10){\circle*{3}}
\put(380,20){\circle*{3}}
\put(400,20){\circle*{3}}
\put(380,-20){\circle*{3}}
\put(400,-20){\circle*{3}}
\put(390,30){\circle*{3}}
\put(390,-30){\circle*{3}}
\put(390,20){\line(1,1){10}}
\put(390,20){\line(-1,1){10}}
%\put(345,-55){\vector(1,-1){25}}
\put(315,60){\mbox{CoIm($d_2$)}}
\put(365,55){\vector(1,-1){15}}
%\put(225,-70){\mbox{Im($d_2$)}}
%
\put(-15,0){\line(1,0){450}}
\put(-5,10){\line(1,0){80}}
\put(5,20){\line(1,0){60}}
\put(15,30){\line(1,0){40}}
\put(-25,-2){\mbox{$0$}}
\put(-15,8){\mbox{$2$}}
\put(-5,18){\mbox{$4$}}
\put(5,28){\mbox{$6$}}
\put(90,5){\line(1,0){205}}
\put(100,15){\line(1,0){185}}
\put(110,25){\line(1,0){165}}
\put(300,25){\line(1,0){120}}
\put(300,3){\mbox{$1$}}
\put(290,13){\mbox{$3$}}
\put(280,23){\mbox{$5$}}
\qbezier(65,12)(90,30)(115,12)
\put(115,12){\vector(1,-1){2}}
\qbezier(52,25)(120,100)(200,30)
\put(198,32){\vector(1,-1){2}}
\put(80,35){\vector(1,-1){10}}
\put(90,32){\mbox{$d_1$}}
%\put(90,45){\vector(1,-1){5}}
\put(205,58){\vector(-1,-1){15}}
\put(210,50){\mbox{$d_1$}}
\qbezier(155,-27)(270,-140)(370,-30)
\put(368,-32){\vector(1,1){2}}
\qbezier(255,-30)(300,-70)(360,-20)
\put(358,-18.5){\vector(1,1){2}}
\put(200,-30){\vector(-1,-1){15}}
\put(205,-40){\mbox{$d_2$}}
\put(300,-20){\vector(1,-1){15}}
\put(318,-30){\mbox{$d_2$}}
\put(293,-30){\mbox{$-$}}
\put(38,70){\mbox{$C_0$}}
\put(185,70){\mbox{$C_1$}}
\put(388,70){\mbox{$C_2$}}
\put(25,-60){\mbox{$[N-1]^2$}}
\put(110,-60){\mbox{$[N][N-1]$}}
\put(225,-60){\mbox{$[N][N-1]$}}
\put(385,-60){\mbox{$[N]^2$}}
\end{picture}

\noindent
Note that morphisms and differentials are again decreasing
the grading by $-1$.
This time encycled are elements of $C_1$, which do not
belong to the image of $d_1$ -- but they are also not annihilated
by $d_2$, so that the cohomology $H_1=0$.
The only non-vanishing contribution to cohomologies is
${\rm CoIm}(d_2) = q^{2N-2} = H_2$, and
\be
P^{\w\w} = \frac{1}{(q^NT)^2}\cdot (qT)^2\cdot q^{2N-2} = 1 = P(O)
\ee

\subsubsection{Reduced case. Orthogonal channel (initial point $\bu\w$)
\label{debw}}

\begin{picture}(300,170)(-60,-90)
\put(-50,5){\line(1,1){40}}
\put(-50,5){\line(1,-1){40}}
\put(30,5){\line(-1,1){40}}
\put(30,5){\line(-1,-1){40}}
%\put(-40,5){\circle*{3}}
\put(-20,5){\circle*{3}}
\put(0,5){\circle*{3}}
\put(20,5){\circle*{3}}
%\put(-30,15){\circle*{3}}
\put(-10,15){\circle*{3}}
\put(10,15){\circle*{3}}
\put(-30,-5){\circle*{3}}
\put(-10,-5){\circle*{3}}
\put(10,-5){\circle*{3}}
%\put(-20,25){\circle*{3}}
\put(0,25){\circle*{3}}
\put(-20,-15){\circle*{3}}
\put(0,-15){\circle*{3}}
%\put(-10,35){\circle*{3}}
\put(-10,-25){\circle*{3}}
\put(-40,-5){\line(1,1){40}}
\put(100,0){\line(1,1){40}}
\put(100,0){\line(1,-1){40}}
\put(180,0){\line(-1,1){40}}
\put(180,0){\line(-1,-1){40}}
\put(110,0){\circle*{3}}
\put(130,0){\circle*{3}}
\put(150,0){\circle*{3}}
\put(170,0){\circle*{3}}
\put(120,10){\circle*{3}}
\put(140,10){\circle*{3}}
\put(160,10){\circle*{3}}
\put(120,-10){\circle*{3}}
\put(140,-10){\circle*{3}}
\put(160,-10){\circle*{3}}
\put(130,20){\circle*{3}}
\put(150,20){\circle*{3}}
\put(130,-20){\circle*{3}}
\put(150,-20){\circle*{3}}
\put(140,30){\circle*{3}}
\put(140,-30){\circle*{3}}
%\put(130,30){\line(1,-1){40}}
\put(110,-10){\line(1,1){40}}
\put(110,10){\line(1,-1){40}}
\qbezier(100,-8)(120,-40)(133,-45)
\put(95,-65){\vector(1,1){25}}
\put(85,-80){\mbox{Ker($d_2$)}}
\put(75,-35){\vector(1,1){20}}
\put(55,-45){\mbox{CoIm($d_1$)}}
%
%

%
%\put(120,-5){\circle{7}}
%\put(130,-15){\circle{7}}
%\put(140,-25){\circle{7}}
%
%\put(110,-35){\mbox{$2\ \times$}}
\put(186,-15){\mbox{$\oplus$}}
\put(200,10){\line(1,1){40}}
\put(200,10){\line(1,-1){40}}
\put(280,10){\line(-1,1){40}}
\put(280,10){\line(-1,-1){40}}
%\put(210,10){\circle*{3}}
\put(230,10){\circle*{3}}
\put(250,10){\circle*{3}}
%\put(270,10){\circle*{3}}
%\put(220,20){\circle*{3}}
\put(240,20){\circle*{3}}
%\put(260,20){\circle*{3}}
\put(220,0){\circle*{3}}
\put(240,0){\circle*{3}}
\put(260,0){\circle*{3}}
%\put(230,30){\circle*{3}}
%\put(250,30){\circle*{3}}
\put(230,-10){\circle*{3}}
\put(250,-10){\circle*{3}}
%\put(240,40){\circle*{3}}
\put(240,-20){\circle*{3}}
\put(210,0){\line(1,1){30}}
\put(270,0){\line(-1,1){30}}
%\put(345,-45){\vector(1,1){25}}
%\put(315,-60){\mbox{CoIm($d_2$)}}
%\put(245,-55){\vector(1,1){25}}
%\put(225,-70){\mbox{Im($d_2$)}}
%

\put(350,5){\line(1,1){40}}
\put(350,5){\line(1,-1){40}}
\put(430,5){\line(-1,1){40}}
\put(430,5){\line(-1,-1){40}}
\put(360,5){\circle*{3}}
\put(380,5){\circle*{3}}
\put(400,5){\circle*{3}}
%\put(420,5){\circle*{3}}
\put(370,15){\circle*{3}}
\put(390,15){\circle*{3}}
%\put(410,15){\circle*{3}}
\put(370,-5){\circle*{3}}
\put(390,-5){\circle*{3}}
\put(410,-5){\circle*{3}}
\put(380,25){\circle*{3}}
%\put(400,25){\circle*{3}}
\put(380,-15){\circle*{3}}
\put(400,-15){\circle*{3}}
%\put(390,35){\circle*{3}}
\put(390,-25){\circle*{3}}
\put(380,35){\line(1,-1){40}}
%
%\put(260,-5){\circle{7}}
%\put(250,-15){\circle{7}}
%\put(240,-25){\circle{7}}
%

\put(85,0){\line(1,0){300}}
\put(95,10){\line(1,0){200}}
\put(105,20){\line(1,0){180}}
\put(115,30){\line(1,0){160}}
\put(75,-4){\mbox{$0$}}
\put(85,8){\mbox{$2$}}
\put(95,18){\mbox{$4$}}
\put(105,28){\mbox{$6$}}
\put(-55,5){\line(1,0){500}}
\put(-45,15){\line(1,0){85}}
\put(-35,25){\line(1,0){65}}
\put(-35,-25){\line(1,0){300}}
\put(335,-25){\line(1,0){100}}
\put(-65,3){\mbox{$1$}}
\put(-55,13){\mbox{$3$}}
\put(-45,23){\mbox{$5$}}
\put(-55,-27){\mbox{$-5$}}
\qbezier(45,40)(85,75)(126,40)
\put(124,42){\vector(1,-1){2}}
\qbezier(42,45)(120,100)(200,30)
\put(198,32){\vector(1,-1){2}}
\put(100,45){\vector(-1,-1){15}}
\put(80,40){\mbox{$d_1$}}
%\put(90,45){\vector(1,-1){5}}
\put(205,60){\vector(-1,-1){12}}
\put(205,48){\mbox{$d_1$}}
\qbezier(165,-35)(270,-140)(375,-25)
\put(373,-27){\vector(1,1){2}}
\qbezier(255,-30)(300,-70)(365,-15)
\put(363,-16.5){\vector(1,1){2}}
\put(215,-45){\vector(-1,-1){12}}
\put(197,-47){\mbox{$d_2$}}
\put(317,-20){\vector(-1,-1){15}}
\put(315,-30){\mbox{$d_2$}}
\put(293,-30){\mbox{$-$}}
\put(-12,70){\mbox{$C_0$}}
\put(185,70){\mbox{$C_1$}}
\put(388,70){\mbox{$C_2$}}
\put(130,-60){\mbox{$[N]^2$}}
\put(-35,-60){\mbox{$[N][N-1]$}}
\put(375,-60){\mbox{$[N][N-1]$}}
\put(225,-60){\mbox{$[N-1]^2$}}
\end{picture}

\noindent
Note that all the morphisms are the {\it same} as they were
between the corresponding spaces in secs.\ref{debb} and \ref{deww}.
The only non-trivial cohomology is the intersection of
$H_1 = {\rm CoIm}(d_1)\,\cap \,{\rm Ker}(d_2)\,\in\, [N]^2\, =\, q^0$.
Thus
\be
P^{\bu\w} = \frac{q^{N-1}}{q^NT} \cdot (qT)\cdot 1 = 1 = P(O)
\ee

\subsubsection{Reduced case: Another Orthogonal channel $\w\bu$}

This case is literally the same as the previous one,
the differential are again made from morphisms, familiar
secs.\ref{debb} and \ref{deww} -- this time they are all $\ \searrow\ $
instead of $\ \swarrow\ $ in sec.\ref{debw}.
Cohomologies and reduced superpolynomial are also the same
\be
P^{\w\bu} = \frac{q^{N-1}}{q^NT} \cdot (qT)\cdot 1 = 1 = P(O)
\ee

\subsubsection{Unreduced case.}

Unreduced situation is described exactly in the same way.
It is just necessary to add one more dimension of the size $[N]$,
orthogonal to all constituents of $d_1$ and $d_2$.
This simply multiplies everything by $[N]$ and gives:
\be
{\cal P}^{\bu\bu} = {\cal P}^{\bu\w} = {\cal P}^{\w\bu} = {\cal P}^{\w\w}
=[N] = {\cal P}(O)
\ee
The two pictures for the Seifert ($\bu\bu$) and orthogonal ($\bu\w$) channels
are shown in Fig.\ref{doubl8}.
%Figs.1 and 2 respectively.

\begin{figure}[!htb]
\includegraphics[scale=.7]{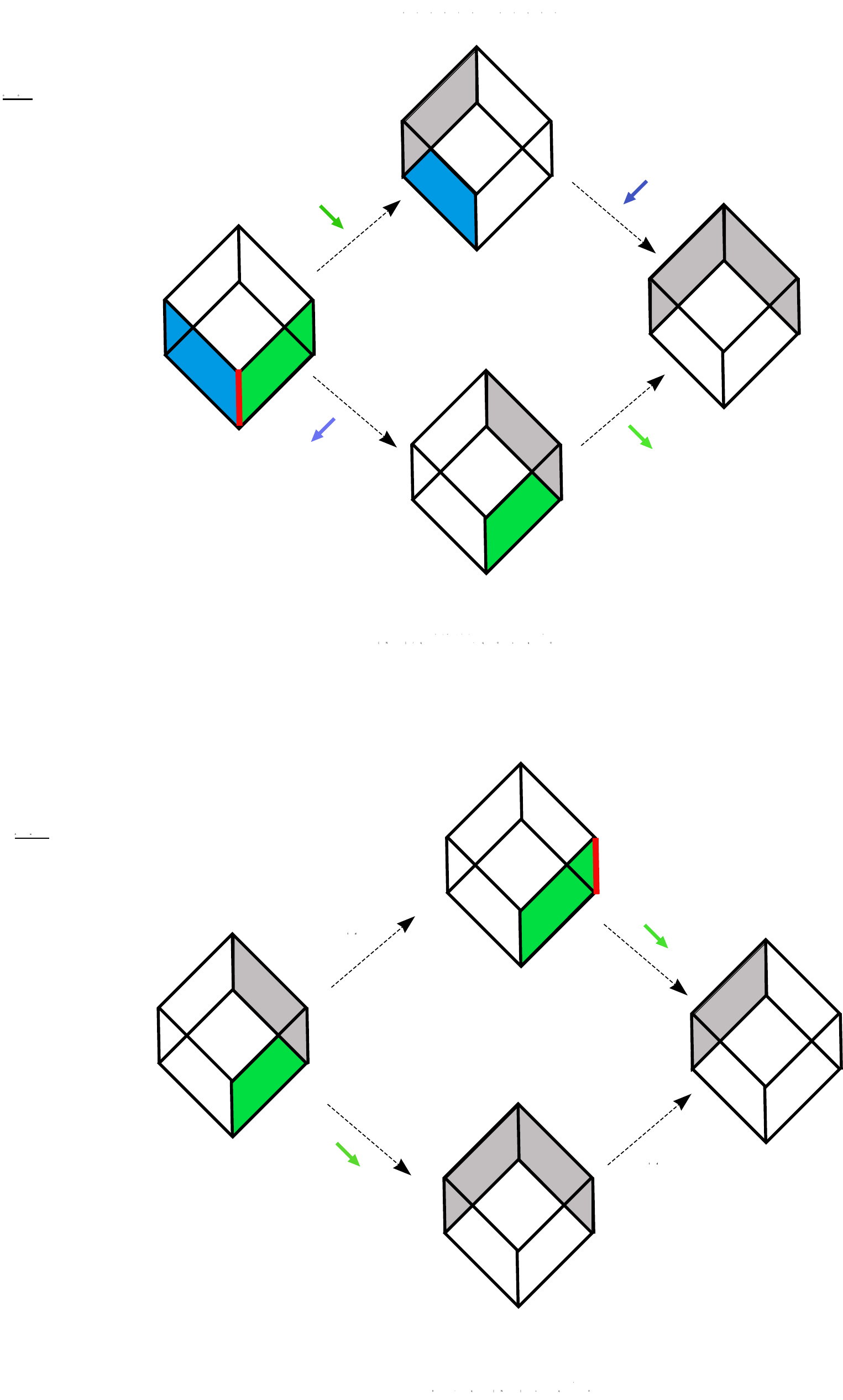}
\caption{Double eight, unreduced case}
\label{doubl8}
\end{figure}

\newpage

\subsection{Twist knots}

This is the series that we already analyzed in sec.\ref{twist},
now we promote our description of HOMFLY to superpolynomials.
According to \cite{inds} and \cite{evo} {\it reduced} superpolynomial
in the fundamental representation is
\be
P_k = 1 +
%F_k\left(A^2q/t)\{Aq\}\{A/t\} =
1 + F_k\left(q^{2N}T^2\right)\left(q^{2N}T^2+ q^2T+\frac{1}{q^2T}+\frac{1}{q^{2N}T^2}\right)
\label{twisu}
\ee
with $F_k(A) = -A^{k+1}\{A^{k}\}/\{A\}$.
We remind that for $k=0$ and $F_0=0$ we get unknot,
for $k=1$ and $F_{1}=-A^2$ -- the trefoil $3_1$
and for $k=-1$ and $F_{-1}=1$ -- the figure eight knot $4_1$.
In fact, for $k>0$ one should multiply the whole expression by $-1$
to make all the terms positive.
With the exception of unknot and trefoil the superpolynomial
contains negative powers of $T$ -- this is because the twisted knots
are {\it not} represented by Seifert vertex in the hypercube:
the corresponding vertices the knot diagram have different colors,
$n_\w\neq 0$ and normalization factor $\alpha_\w^{n_\w} \sim T^{-n_\w}$
provides negative powers of $T$.

In the case of the twisted and $3$-strand torus families we restrict
ourselves just to the simplest example, which lies at the
intersection of two families and can illustrate the both:
that of the knot diagram from s.\ref{fe3str},
which (for different colorings) is either the trefoil $3_1$ or
the figure eight know $4_1$ or the unknot.
The grading tables in the first two cases are:

\bigskip

{\bf Trefoil} $3_1$\\

\centerline{{\footnotesize
$
\begin{array}{c|ccccccrccccc}
2N-2 & & 1 && && \ \ 2 \\
2N-3 && && 4 && && 4 \ \ \  \\
2N-4 && 2 && && 4+4+2 && && 2 \ \ \  \\
2N-5 && && 8 && && 8+4  \\
2N-6 && 3 && && 8+6+4 && && 4+1\\
\ldots & \\
4 && N-2 && \ldots \\
3&& && 4(N-2) && \ldots \\
2&& N-1 && && 4(N-2)+2(N-1)+2(N-2) && && 2(N-2)+(N-3) \\
1 && && 4(N-1) && &\boxed{1}& 4(N-1)+4(N-2) \\
0 && N &&       && 4(N-1) +2(N-1)+2(N-1)&& &&  2(N-1)+(N-3)\\
1 && && 4(N-1) &&&& 4(N-2)+4(N-1) \\
-2 && N-1 && && 4(N-2)+2(N-2)+2(N-1) && && 2(N-2) + (N-3)\\
-3 && && 4(N-2) &&&& 4(N-3)+4(N-2) \\
-4 && N-2 && && 4(N-3) + 2(N-3) + 2(N-2) &&&& 2(N-3) + (N-4) \\
&& &&\ldots && &&  4(N-4) + 4(N-3) && && \\
&& && && \ldots &&  && 2(N-4) + (N-5) \\
\ldots & \\
10-2N && 5 && \ldots \\
9-2N && && 16  && \ldots \\
8-2N && 4 && && 12+ 6 + 8 && \ldots \\
7-2N && && 12 && &&  8+12 \\
6-2N && 3 && && 8+4+6 && && 4+1 \\
5-2n && && 8 && && 4+8  \\
4-2N && 2 && &&\!\!\!\!\!\!\!\!\!\!\!\!\!\!\!\!\!\!\!
\boxed{1} \ \ \ 4+2+4 && && 2 \\
3-2N && && 4 && && 4 \\
2-2N && \boxed{1} && && 2 \\
& \\
\hline
& \\
& & [N]^2 && 4\times [N][N-1] && 4\times [N-1]^2+2\times [2][N][N-1]
&& 4\times [2][N-1]^2 && 2[N-1]^2 +  \\
& & && &&   && && + [N-1][N-3]  \\
\end{array}
$
}}

\bigskip

\be
\Longrightarrow \ \
{\cal P}(3_1) = q^{4N-4}\Big( q^{2-2N} + q^{4-2N}\cdot(qT)^2 +  q\cdot(qT)^3\Big)
= q^{2N-2} + q^{2N+2}T^2 + q^{4N}T^3
\ee

\bigskip

\newpage

{\bf Figure eight} $4_1$ \\

\centerline{ {\footnotesize
$
\begin{array}{c|ccccccrccc}
2N-2 && 1 && && 1 && && \boxed{1} \\
2N-3 && && 2+2 && && 4 \\
2N-4 && 2+1 && && 2+6 && && 3 \\
2N-5 && && 4+ 4+2 && && 10 \\
2N-6 && 3+2 && && 3 +12 +1 && && 5 \\
2N-7 && && 6+6+4 && && 16 \\
2N-8 && 4+3 && && 4+18+2 && && 7 \\
2N-9 && && 8+8+6 && && 22 \\
\ldots &\\
4 && (N-2)+(N-3) && && (N-2) + 6(N-3) + (N-4) && && 2N-5\\
3 && && 2(N-2)+2(N-2) + 2(N-3) && && 6N - 14 \\
2 && (N-1) + (N-2) && && (N-1) + 6(N-2) + (N-3) && && 2N-3\\
1 && && 2(N-1) + 2(N-1)+2(N-1) && &\boxed{1}\!\!\!\!\!\!\!\!\!\!\!
\!\!\!& 6N-8 \\
0 && (N-1)+(N-1) && &\boxed{1}\!\!\!\!\!\!\!\!\!
& N + 6(N-1) + (N-3) && && 2N-2 \\
-1 && &\boxed{1}& 2(N-1) + 2(N-2) + 2(N-1) && && 6N-8 \\
-2 && (N-2)+(N-1) && && (N-1) + 6(N-2) + (N-3) && && 2N-3 \\
-3 && && 2(N-2) + 2(N-3) + 2(N-2) && && 6N-14 \\
-4 && (N-3)+(N-2) && && (N-2) + 6(N-3)+(N-4) && && 2N-5 \\
-5 && && 2(N-3) + 2(N-4) + 2(N-3) && && 6N-20 \\
-6 && \ldots && && (N-3)+6(N-4) + (N-5) && && 2N-7\\
-7 && && && \ldots && 6N-26 \\
-8 && && && && \ldots  && 2N-9  \\
\ldots & \\
9-2N && && 8+6+8 && \ldots &&  \\
8-2N && 3+4 && && 4+18+2 && && 7\\
7-2N && && 6+4+6 && && 16 \\
6-2N && 2+3 && &&  3+12+1 && && 5 \\
5-2N && && 4+2+4 && && 10 \\
4-2N && 1+2 && && 2+6 && && 3 \\
3-2N && && 2 \ + \ 2 && && 4\\
2-2N && \boxed{1} && && 1 && && 1\\ &\\
\hline
& \\
&&[2][N][N-1] && 2[N][N-1]+2\times [2][N-1]^2 &&
\!\!\!\!\!\!\!\! [N]^2+4[N-1]^2+&&
2\times [N][N-1] + && [2][N][N-1]\\
&& &&  && + 2[N-1]^2+[N-1][N-3] && + 2\times [2][N-1]^2  \\
\end{array}
$}}

\bigskip

\be
\Longrightarrow \ \
{\cal P}(4_1) = \frac{q^{2N-2}}{q^{2N}T^2}
\Big( q^{2-2N} + q^{-1}\cdot(qT) +  1\cdot(qT)^2 + q\cdot(qT)^3
+ q^{2N-2}(qT)^4\Big)
= \nn \\
= \frac{1}{q^{2N}T^{2}} + \frac{1}{q^2T} + 1 + q^2T +q^{2N}T^2
\ee

\bigskip

As usual, analysis of these tables does not predict the answers for
Betti numbers unambiguously. The remaining discrete freedom  is
fixed by explicit construction of morphisms.
These, in turn, are severely restricted by the requirement,
that whenever possible (when they are mapping the same spaces
in the same order) morphisms coincide for different colorings,
i.e. for $3_1$, $4_1$ and unknot represented by the same diagram
${\cal D}$.
Since this time even in the reduced case pictorial representations
of vector spaces are multidimensional (squares rather than strips,
i.e. the power of $N$ in dimension of the vector space is
greater than one),  in order to minimize the cohomology,
already the first differential $d_1$ consists of morphisms,
acting in different directions -- like it already happened
above for the double-eight representation of the unknot.

We hope that this kind of ideas, underlying the art of
morphism-construction, is to some extent clarified by the
previous examples and do not go into further details here.
A unified analysis of the whole series of $3$-strand torus
knots and twisted knots, as well as more complicated examples,
is clearly within reach and will be presented elsewhere.
This is important also to demonstrate how things work
in the case of knots, which are not "thin" -- the first
such example is the torus knot $[3,4]$.

In the rest of {\it this} paper we briefly outline a conceptual
approach to definition of morphisms in a systematic way,
from the first principles.
Again, we just formulate the ideas, leaving important details
to further clarification.

\section{
%Towards a rigorous construction:
Appendix: towards the theory of cut-and-join maps
\label{specu}}

In \cite{DM1} we explained that behind the morphisms of the Khovanov construction
for $N=2$ actually stand the cut-and-join operators \cite{MMN},
which nowadays play an increasing role in different branches of quantum field theory.
Now we are going to explain -- without going into too many details -- that this
is also true in our generalization from $N=2$ to arbitrary $N$.
It looks plausible, that a systematic presentation of our approach,
together with all potentially interesting deviations, is best discussed
from this perspective. However,in this paper we give just a brief survey,
leaving the details for another presentation.

This section can be considered as alternative continuation of sec.\ref{unkn}.
Since it is not targeted at concrete results, we allow more deviations
from the main line in the simple examples -- to demonstrate the additional
possibilities, provided by the tensor-algebra approach.
They all open potential new windows to various generalizations.
In particular we attract attention to the freedom in the choice
of morphisms, including their gradings -- which can lead at least
to significant technical simplifications.

\subsection{
An example of eight.
Cut and join operations
\label{eighthom}}

Now we can insert one vertex in the knot diagram $D$,
this means  that there will be two in hypercube $H(D)$.
Topologically this is still an unknot. but now we have two different
representations for it, differing by the choice of the color
of the vertex in $D$ and by the choice of initial vertex in $H(D)$.

As we already know from s.\ref{eightgraphdim},
in this case there are two spaces of dimensions
$[N]$ and $[N][N-1]$, and our purpose is to understand what the spaces and
morphism are.
While the first space at the black-colored vertex/resolution
is clearly just $V^{\otimes 2}$
-- a product of two copies of $V$, associated with the two Seifert cycles,--
the white resolution $\ |\,|\ -\ X\ $ contains subtraction,
what at the level of vector spaces we suggest to interpret as a factor-space.

\subsubsection{Primary hypercube \label{eightprihy}}

\begin{picture}(200,150)(-100,-135)
\put(-60,0){\circle{20}}\put(-40,0){\circle{20}}\put(-50,0){\circle*{5}}
\put(40,0){\circle{20}}\put(60,0){\circle{20}}\put(50,0){\circle{5}}
\put(-62,-40){\circle{20}}\put(-38,-40){\circle{20}}
\put(8,-40){\circle{20}}\put(32,-40){\circle{20}}
\put(48,-42){\mbox{$-$}}
\put(70,-40){\circle{20}}\put(90,-40){\circle{20}}
\put(-50,-65){\line(1,0){100}}\put(-50,-65){\circle*{3}}\put(51,-65){\circle{3}}
\put(-53,-22){\mbox{$||$}}\put(47,-22){\mbox{$||$}}
\put(-60,-85){\mbox{ $[N]^2$}}
\put(27,-85){\mbox{$[N][N-1]$}}
\put(-60,-126){\mbox{{\bf the main hypercube  $H(O\!\!\!\bullet\!\!\!O)$}}}
\put(-70,-105){\mbox{${\cal V}_\bullet=V^{\otimes 2}$} }
\put(15,-105){\mbox{ ${\cal V}_\circ=V^{\otimes 2}/V$}}
\put(200,-65){\line(1,0){100}}\put(200,-65){\circle*{3}}\put(301,-65){\circle{3}}
%\put(250,-65){\vector(-1,0){2}}
\put(188,-40){\circle{20}}\put(212,-40){\circle{20}}
\put(292,-40){\circle{20}}\put(312,-40){\circle{20}}
%\put(230,-120){\mbox{ $\tilde H(O\!\!\!\bullet\!\!\!O)$}}
%
\put(190,-85){\mbox{ $[N]^2$}}
\put(292,-85){\mbox{$[N]$}}
\put(185,-105){\mbox{${\cal U}_\bullet=V^{\otimes 2}$} }
\put(280,-105){\mbox{ ${\cal U}_\circ=V$}}
\put(200,-65){\circle{8}}
\put(128,-80){\mbox{$\Longleftarrow$}}
\put(248,-57){\mbox{$\nabla$}}
\put(244,-63){\mbox{$\longleftarrow$}}
\put(244,-72){\mbox{$\longrightarrow$}}
\put(248,-80){\mbox{$\Delta$}}
\put(-8,-75){\mbox{$\xi$}}
\put(-13,-82){\mbox{$\longrightarrow$}}
\put(-13,-87){\mbox{$\longleftarrow$}}
\put(-8,-93){\mbox{$\eta$}}
\put(170,-15){\line(1,0){160}}
\put(170,-15){\line(0,-1){100}}
\put(330,-115){\line(0,1){100}}
\put(330,-115){\line(-1,0){160}}
\put(185,0){\mbox{{\bf primary hypercube $\tilde H(O\!\!\!\bullet\!\!\!O)$}}}
\end{picture}

\noindent
For taming the emerging world of factor-spaces we suggest the following procedure.

\bigskip

\fbox{\ \ \ \parbox{15cm}{
\bigskip

Introduce an auxiliary  hypercube $\tilde H(D)$
(we call it {\bf primary}, because of its central
and {\it dictating} role in our construction),
where all white vertices of $D$ are resolved
simply as $\ X\ $ (instead of the formal difference $\ ||\ - \ X\ $
in the {\it main} $H(D)$, i.e. no linear combinations and no minus signs).

\bigskip

It has two additional structures.

\bigskip

First, there as a distinguished vertex --
the one with pure black colorings,
i.e. where all vertices of $D$ are resolved as $\ |\,|\ $.
This is the only vertex where the vector spaces are the same in
primary and original hypercubes $\tilde H(D)$ and $H(D)$,
sometime we mark it by additional surrounding circle.

\bigskip

Second, there are arrows at the edges
-- in general quite different from those
on original hypercube (appearing, when initial coloring
is chosen and pointing towards the corresponding initial vertex
of $H(D)$).
The arrows on the edges $\tilde H(D)$   describe embeddings
and they do not obligatory go all in one direction.

\bigskip
}\ \ \ }

\bigskip

\noindent
Additional delicate point is that these are embeddings of {\it graded} vector spaces.
To understand what they are, we need to recall that we want quantum
dimensions to match -- and the two possibilities are:
\be
\underline{[N]^2 =\ q^{1-N}[N] + q[N][N-1]} \  = \ q^{N-1}[N] +\frac{1}{q}[N][N-1]
\label{decodim}
\ee
This means that at the level of basises the embeddings are either
$e_I \ \longrightarrow \ e_I\otimes e_N\pm e_N\otimes e_I$ of degree $q^{1-N}$ or
$e_I \ \longrightarrow \ e_I\otimes e_1\pm e_1\otimes e_N$ of degree $q^{N-1}$ --
choice of any other $e_K$ instead of $e_1$ or $e_N$ would give unappropriate
difference $[N]^2-q^{N+1-2K}[N] \ /\!\!\!\!\!\!\sim [N][N-1]$.
Following our general intention to preserve all the symmetries of the
tensor algebra we avoid considering asymmetric embeddings like
$e_I\rightarrow e_I\otimes e_N$ (in practice they do not give anything new).
Antisymmetric embeddings  have non-vanishing kernels ($e_N$ or $e_1$ respectively)
and are also non-suitable for our purposes.
Thus it remains to choose arbitrarily between the remaining two options.
In what follows we {\it postulate} that embeddings in the primary hypercube
are of degree $q^{1-N}$ (then for $N=2$ we get $q^{-1}$, familiar from
\cite{BN} and \cite{DM1}) and are explicitly given by
\be
\nabla = V\longrightarrow V^{\otimes 2}: \ \ \ \ \ \
\boxed{e_I \ \longrightarrow \ e_I\otimes e_N + e_N\otimes e_I}
\ \ \ \ \ \ \ \ \ \
g(\nabla) = 1-N
\label{cut}
\ee
%(of course one can also permute: $e_N\otimes e_I$ or $e_1\otimes e_I$).
It corresponds to the first (underlined) decomposition in (\ref{decodim}).
For an obvious reason we call $\nabla$ the {\it cut} operation.
Note that the grading $1-N$ for $N=2$ is exactly the standard $-1$.

Now we need a complementary {\it join} operation $\Delta$.
There are different natural choices,
to fix the freedom we ask\footnote{
Another distinguished choice would be an "inverse" to $\nabla$,
with $\ {\rm Ker}(\Delta) =  {\rm CoIm}(\nabla)\ $ and
with grading $N-1$, but it does not seem to
lead to Khovanov-Rozansky homologies.
}
it to have the same (negative) grading as $\nabla$:
\be
\Delta = V^{\otimes 2}\longrightarrow V: \ \ \ \ \ \
\left\{\begin{array}{rcl}
e_1\otimes e_1 & \longrightarrow & e_1 \\
e_1\otimes e_{i+1} & \longrightarrow & e_{i+1} \\
e_{i+1} \otimes e_1 & \longrightarrow & e_{i+1} \\
e_{i+1}\otimes e_{j+1} & \longrightarrow & 0
%\\ \\
%{\footnotesize 2\leq\!\!\!\! & i,j & \!\!\!\! \leq N}
\end{array}\right.
\ \ \ \ \ \ \ \ \ \ \ \ \ \ \ \ \
g(\Delta) = 1-N
\label{join}
\ee
%Here and below we use small Latin letters $i,j,\ldots = 1,\ldots,N-1$
%to label the complement of $e_N$ in the basis of $V^N$,
%while large letters $I,J,\ldots = 1,\ldots, N-1,N$ label entire basis.
Cut operation has no kernel, but
join operation has a huge one:,
\be
{\rm Ker}(\Delta) = \span\Big\{e_{i+1}\otimes e_{j+1},\ \
e_1\otimes e_{i+1}-e_{i+1}\otimes e_1\Big\}
%= \span(ij,\widehat{1i})
\ee
Similarly, coimage\footnote{
Throughout this paper we understand "cokernel" and "coimage"
as complements of the kernel and image in the initial
and target spaces respectively.
In general these are factor-spaces, but
when basises are explicitly specified, thee can actually
be considered as well defined orthogonal complements.
}
 of $\Delta$ is empty, while
\be
{\rm CoIm}(\nabla) = \span\Big\{e_i\otimes e_j,\ e_i\otimes e_N-e_N\otimes e_i\Big\}
= \span(ij,\widehat{iN})
\ee
Here and below we assume that large Latin indices $I,J$ run from $1$ to $N$,
while small ones, $i,j$ -- from $1$ to $N-1$.
We also introduced a shortened notation for the basis elements in $V^{\otimes 2}$.
%Note that $i,j$ always take $N-1$ values, but from two different
%sets: $\{2,\ldots,N\}$ and $\{1,\ldots,N-1\}$ -- we assume that this does not
%cause confusion, but simplifies the formulas.
%For the same reason
To simplify the formulas
we omit brackets between \ dim\  and\  Ker\
in (\ref{dimnade} and below.

The corresponding dimensions are  actually given by (\ref{decodim}):\footnote{
In {\it reduced} case everything looks even simpler:
\be
\nabla =  \ E\longrightarrow V: &\ \ \  E\longrightarrow e_N&
{\rm dim}_q{\rm CoIm}(\nabla) = [N]-q^{1-N} = q[N-1] \nn \\
\Delta =  \ V\longrightarrow E: &\ \ \
\left\{ \begin{array}{ccc} e_1 & \longrightarrow &  E \\
e_i & \longrightarrow & 0 \end{array}\right. \ \ \ \ \ \ \ \
& {\rm dim}_q{\rm Ker}(\Delta) = [N] -q^{N-1} = \frac{1}{q}[N-1]
\nn
\ee
Here $E$  with $g(E)=0$
is the single basis element in $E=C$, and we denote
it by the same letter as the space.
}

\be
{\rm dim}_q{\rm CoIm}(\nabla) =
\l[N]^2 - q^{1-N}[N] = q[N][N-1], \nn \\
{\rm dim}_q{\rm Ker}(\Delta) =
\l[N]^2 - q^{N-1}[N] = \frac{1}{q}[N][N-1]
\label{dimnade}
\ee

\subsubsection{The main hypercube}

Now we can return to the main hypercube.
Resolution at its black vertex is just a pair of Seifert  cycles,
and we associate with this vertex the vector space
${\cal V}_b={\cal V}_\bullet = V^{\otimes 2}$.
Now comes the main point: {\bf with a "difference" of cycles at the white vertex
we associate a factor-space}
%\be
${\cal V}_\circ =V^{\otimes 2}/V$.
%\ee

Now, the two decreasing morphisms $\xi$ and $\eta$ look as follows:
vertical lines are various spaces at the vertices of
$\tilde H(O\!\!\!\bullet\!\!\!O)$ and $H(O\!\!\!\bullet\!\!\!O)$
and arrows are identical (or nullifying) maps on their subspaces.
Clearly, ${\rm Ker}(\xi) = {\rm Im}(\nabla)$ is a result of embedding of
${\cal U}_\circ = V$ into ${\cal U}_\bullet = V^{\otimes 2}$,
but it differs from $V$ itself in a shift of grading.
The same is true for ${\rm CoIm}(\eta)$ -- but the grading shift
in this case is different.

\begin{picture}(200,200)(-200,-45)
\put(-120,40){\line(0,1){20}}
\put(-117,39){\line(3,-1){114}}
\put(-117,59){\line(3,-1){114}}
\put(-117,41){\line(3,1){114}}
\put(-117,61){\line(3,1){114}}
\put(0,0){\line(0,1){100}}
\put(120,40){\line(0,1){80}}
\qbezier(3,1)(60,20)(119,39)
\qbezier(3,21)(60,31)(119,41)
\qbezier(3,81)(60,100)(119,119)
\qbezier(3,101)(60,111)(119,121)
%
%\put(3,19){\line(3,-1){54}}
%\put(3,99){\line(3,-1){54}}
%\put(60,0){\line(0,1){80}}
%\put(63,-1){\line(3,-1){54}}
%\put(63,79){\line(3,-1){54}}
%\put(120,-20){\line(0,1){100}}
%
\put(-70,27){\vector(3,-1){20}}
\put(-65,28){\mbox{$\nabla$}}
\put(-50,70){\vector(-3,-1){20}}
\put(-65,70){\mbox{$\Delta$}}
\put(-135,20){\mbox{${\cal U}_\circ = V$}}
\put(-37,-23){\mbox{${\cal V}_\bullet = {\cal U}_\bullet = V^{\otimes 2}$}}
\put(105,15){\mbox{${\cal V}_\circ = V^{\otimes 2}/V$}}
%\put(105,-39){\mbox{${\cal V}_\bullet= V^{\otimes 2}$}}
%
%
\put(74,118){\vector(3,1){2}}
\qbezier(50,114)(62,116)(74,118)
\put(57,120){\mbox{$\xi$}}
\put(52,90){\vector(-3,-1){2}}
\qbezier(74,98)(62,94)(50,90)
\put(61,85){\mbox{$\eta$}}
\put(-120,50){\circle*{3}}
\put(0,50){\circle*{3}}
%\put(60,40){\circle*{3}}
\put(120,80){\circle*{3}}
\qbezier(3,1)(10,10)(3,19)
\put(9,6){\vector(3,-1){40}}
\put(55,-14){\mbox{$0$}}
\put(35,0){\mbox{$\xi$}}
\qbezier(-3,1)(-15,40)(-3,79)
\put(-15,50){\vector(-3,-1){15}}
\put(-40,42){\mbox{$0$}}
\put(-28,52){\mbox{$\Delta$}}
\put(10,50){\line(1,0){140}}
\put(130,80){\line(1,0){20}}
\put(145,63){\mbox{$m$}}
\put(-60,108){\mbox{${\rm CoKer}(\Delta)$}}
\put(-30,103){\vector(3,-1){25}}
\put(-80,-8){\mbox{${\rm Im}(\nabla)$}}
\put(-45,-5){\vector(3,1){42}}
\linethickness{1mm}
\put(0,0){\line(0,1){20}}
\put(0,80){\line(0,1){20}}
\end{picture}

\noindent
From this picture it is clear that
\be
{\rm Ker}(\xi) = {\rm Im}(\nabla) & \ \ \ \ \ \ \  \ \ \ \ \ \ \
&{\rm CoIm}(\xi) = \emptyset \nn \\
{\rm Im}(\eta) = {\rm Ker}(\Delta) & \ \ \ \ \ \ \
& {\rm Ker}(\eta) = \emptyset
\ee
The two non-vanishing spaces,shown by thick line in the picture, are
\be
 {\rm Im}(\nabla) = \span(e_I\otimes e_N+e_N\otimes e_I)
 \ \ \ \ \ &{\rm dim}_q\Big({\rm Im}(\nabla)\Big) = q^{1-N}[N] \nn \\
{\rm CoKer}(\Delta) = \span(e_1\otimes e_I+e_I\otimes e_1)
\ \ \ \ \ &{\rm dim}_q\Big({\rm CoKer}(\Delta)\Big) = q^{N-1}[N]
\label{eightco}
\ee
As to ${\rm CoKer}(\Delta)$, it is "similar" to ${\cal U}_\circ$,
but is am absolutely different subspace
in ${\cal V}_\bu = {\cal U}_\bu$,
in particular with a very different grading.

\subsubsection{The choice of morphisms. "Gauge" invariance.}

In this picture the vertical "coordinate" is actually the grading degree.
Dots symbolize the elements (subspaces) with grading zero.
The average grading of the space ${\cal V}_\circ$ is  $m$,
which is an arbitrary integer: one can move
${\cal V}_\circ$ arbitrarily along the vertical line --
this reflects the absence of canonical  representative in the
equivalence class of vector subspaces, which the factor-space always is.
In graded case the freedom is substantially restricted, still remains.
Moving ${\cal V}_\circ$ along the vertical line,
one changes the gradings of morphisms $\xi$ and $\eta$,
but two things remain intact:
\be
g(\xi) + g(\eta) = -2
\ee
and cohomologies of $\xi$ and $\eta$ in (\ref{eightco}).
This is somewhat similar to gauge invariance.
% (and for a good reason)
As usual, Euler characteristic can be calculated in two ways --
via dimensions of vector spaces ${\cal V}$ and via cohomologies:
\be
{\rm dim}_q {\cal V}_\bullet - q^{-g(\xi)}{\rm dim}_q {\cal V}_\circ
= {\rm dim}_q{\rm Ker}(\xi) - q^{-g(\xi)}{\rm dim}_q{\rm CoIm}(\xi), \nn \\
{\rm dim}_q {\cal V}_\circ - q^{-g(\eta)}{\rm dim}_q {\cal V}_\bullet
= {\rm dim}_q{\rm Ker}(\eta) - q^{-g(\eta)}{\rm dim}_q{\rm CoIm}(\eta)
\ee
Note that the quantum dimension
${\rm dim}_q {\cal V}_\circ = q^m[N][N-1]$ and the gradings $g(\xi)=m-1$,
$g(\eta)=1-m$ depend on the shift $m$.

The same remains true for arbitrary knot/link diagrams $D$.
However, often it is more convenient to rely upon concrete
(in no way canonical) choice of the representative for
the factor-space.
There are actually three technically distinguished choices:
$g(\xi)=g(\eta)=-1$ or $g(\xi)=0$ or $g(\eta)=0$.
The first case is more symmetric and it makes smoothconnection
with the standard construction at $N=2$ (where no factor-spaces
occur in explicit way).
The other two choices can be natural, if we look at particular
coloring $D_c$ (fix the initial vertex of the hypercube):
then only one of the two morphisms matter, and it is technically
reasonable to simplify it as much as possible.
Choosing grading degree zero for this morphism allows to make
it simply an identity map between its cokernel and coimage --
what makes calculations as simple as only possible.
In other channels (other initial vertices) one can make another
choice, making identical another relevant morphism.
If we take this road, this actually means that we make different
choices of the space $V^{\otimes 2}/V$ (different representatives
of the class) in different channels, i.e. modify slightly the
original definition of the main hypercube.
As we just explained, this has technical advantages.
At the same time, cohomologies of the complex $K(D)$,
their dimensions and thus the Khovanov-Rozansky polynomials
do not feel the difference, if appropriately defined --
as in (\ref{eightco}).
In more technical definitions appropriate adjustements will be needed:
of the parameters in the generating function (it will be
$q^{g(\xi)}T$) and overall coefficient.

\bigskip

We can illustrate the difference between "symmetric" and "identity"
choices already now.

\bigskip

{\bf Symmetric choice.}

\begin{picture}(200,150)(-200,-45)
\put(-120,40){\line(0,1){20}}
\put(-117,39){\line(3,-1){114}}
\put(-117,59){\line(3,-1){114}}
\put(-117,41){\line(3,1){114}}
\put(-117,61){\line(3,1){114}}
\put(0,0){\line(0,1){100}}
\put(120,10){\line(0,1){80}}
\qbezier(3,1)(60,5)(119,9)
\qbezier(3,19)(60,15)(119,11)
\qbezier(3,81)(60,85)(119,89)
\qbezier(3,99)(60,95)(119,91)
%
%\put(3,19){\line(3,-1){54}}
%\put(3,99){\line(3,-1){54}}
%\put(60,0){\line(0,1){80}}
%\put(63,-1){\line(3,-1){54}}
%\put(63,79){\line(3,-1){54}}
%\put(120,-20){\line(0,1){100}}
%
\put(-70,27){\vector(3,-1){20}}
\put(-65,28){\mbox{$\nabla$}}
\put(-50,70){\vector(-3,-1){20}}
\put(-65,70){\mbox{$\Delta$}}
\put(-135,20){\mbox{${\cal U}_\circ = V$}}
\put(-37,-23){\mbox{${\cal V}_\bullet = {\cal U}_\bullet = V^{\otimes 2}$}}
\put(105,-15){\mbox{${\cal V}_\circ = V^{\otimes 2}/V$}}
%\put(105,-39){\mbox{${\cal V}_\bullet= V^{\otimes 2}$}}
%
%
\put(73,98){\vector(3,-1){2}}
\qbezier(50,100)(62,99)(74,98)
\put(57,107){\mbox{$\xi$}}
\put(52,81){\vector(-3,-1){2}}
\qbezier(74,82)(62,81)(50,80)
\put(57,60){\mbox{$\eta$}}
\put(-120,50){\circle*{3}}
\put(0,50){\circle*{3}}
%\put(60,40){\circle*{3}}
\put(120,50){\circle*{3}}
%
%\put(10,5){\vector(3,-1){20}}
%\put(35,-8){\mbox{$0$}}
%
\put(-140,50){\line(1,0){300}}
\linethickness{1mm}
\put(1,0){\line(0,1){20}}
\put(1,80){\line(0,1){20}}
\end{picture}

\noindent
In this case the space $V^{\otimes 2}/V$ has dimension $[N][N-1]$
and distinguished basis $e_{Ii}$, labeled by two indices,
with $g(e_{Ii}) = 2N+1-2I-2i$.
The two morphism, both "decreasing" -- of degree $g(\xi)=g(\eta)=-1$ are:
\be
\xi = V^{\otimes 2} \longrightarrow V^{\otimes 2}/V: &
\left\{\begin{array}{ccc}
e_i\otimes e_j & \longrightarrow & e_{ij} \\
e_i\otimes e_N - e_N\otimes e_i & \longrightarrow & e_{Ni} \\ \\
e_i\otimes e_N + e_N\otimes e_i & \longrightarrow & 0 \\
e_N\otimes e_N & \longrightarrow & 0
\end{array}\right. \nn \\ \nn \\
\eta = V^{\otimes 2}/V \longrightarrow V^{\otimes 2}: &
\left\{\begin{array}{ccc}
e_{ij} &\longrightarrow & e_{i+1}\otimes e_{j+1}  \\
e_{Nj} & \longrightarrow & e_N\otimes e_j - e_j\otimes e_N
\end{array}\right.
\ee
Clearly,
\be
{\rm Ker}(\xi) = \span(e_I\otimes e_N+e_N\otimes e_I) & &
{\rm dim}_q {\rm Ker}(\xi) = q^{1-N}[N] \nn \\
{\rm CoIm}(\eta) = \span(e_1\otimes e_I+e_I\otimes e_1) & &
{\rm dim}_q {\rm CoIm}(\eta) = q^{N-1}[N]
\ee
The typical feature of symmetric construction
is that, say,  $g(e_{11})=2N-3 = g(e_1\otimes e_1) -1$.

\bigskip

{\bf Identity choice for $\xi$}

\begin{picture}(200,150)(-200,-45)
\put(-120,40){\line(0,1){20}}
\put(-117,39){\line(3,-1){114}}
\put(-117,59){\line(3,-1){114}}
\put(-117,41){\line(3,1){114}}
\put(-117,61){\line(3,1){114}}
\put(0,0){\line(0,1){100}}
\put(120,20){\line(0,1){80}}
\qbezier(3,1)(60,10)(119,19)
\qbezier(3,20)(60,20)(119,20)
\qbezier(3,81)(60,90)(119,99)
\qbezier(3,100)(60,100)(119,100)
%
%\put(3,19){\line(3,-1){54}}
%\put(3,99){\line(3,-1){54}}
%\put(60,0){\line(0,1){80}}
%\put(63,-1){\line(3,-1){54}}
%\put(63,79){\line(3,-1){54}}
%\put(120,-20){\line(0,1){100}}
%
\put(-70,27){\vector(3,-1){20}}
\put(-65,28){\mbox{$\nabla$}}
\put(-50,70){\vector(-3,-1){20}}
\put(-65,70){\mbox{$\Delta$}}
\put(-135,20){\mbox{${\cal U}_\circ = V$}}
\put(-37,-23){\mbox{${\cal V}_\bullet = {\cal U}_\bullet = V^{\otimes 2}$}}
\put(105,-5){\mbox{${\cal V}_\circ = V^{\otimes 2}/V$}}
%\put(105,-39){\mbox{${\cal V}_\bullet= V^{\otimes 2}$}}
%
%
\put(50,103){\vector(1,0){20}}
%\qbezier(50,100)(62,99)(74,98)
\put(57,108){\mbox{$\xi$}}
\put(52,81){\vector(-3,-1){2}}
\qbezier(74,84)(62,82)(50,80)
\put(60,70){\mbox{$\eta$}}
\put(-120,50){\circle*{3}}
\put(0,50){\circle*{3}}
%\put(60,40){\circle*{3}}
\put(120,60){\circle*{3}}
%
%\put(10,5){\vector(3,-1){20}}
%\put(35,-8){\mbox{$0$}}
%
\qbezier(3,21)(15,60)(3,99)
\put(15,57){\mbox{${\rm CoIm}(\nabla)$}}
\put(-140,50){\line(1,0){300}}
\put(130,57){\mbox{$1$}}
\linethickness{1mm}
\put(1,0){\line(0,1){20}}
\put(1,80){\line(0,1){20}}
\end{picture}

\noindent
In this case we just take ${\rm CoIm}(\nabla)$
to represent $V^{\otimes 2}/V = \span(ij,\widehat{iN})$
of dimension $q[N][N-1]$, i.e. by definition $\xi$
is just an identity map on ${\rm CoIm}(\nabla)$, so that $g(\xi)=0$.
At the same time $\eta$ maps it  one-to-one onto ${\rm Ker}(\Delta)$
and has $g(\eta)=-2$:
\be
\xi = V^{\otimes 2} \longrightarrow V^{\otimes 2}/V: &
\left\{\begin{array}{ccc}
e_i\otimes e_j & \longrightarrow & e_i\otimes e_j \\
e_i\otimes e_{N} - e_{N}\otimes e_i & \longrightarrow &
e_i\otimes e_{N} - e_{N}\otimes e_i \\ \\
e_i\otimes e_N + e_N\otimes e_i & \longrightarrow & 0 \\
e_N\otimes e_N & \longrightarrow & 0
\end{array}\right. \nn \\ \nn \\
\eta = V^{\otimes 2}/V \longrightarrow V^{\otimes 2}: &
\left\{\begin{array}{ccc}
e_i\otimes e_j &\longrightarrow & e_{i+1}\otimes e_{j} - e_i\otimes e_{j+1}  \\
e_i\otimes e_{N} - e_{N}\otimes e_i & \longrightarrow &
e_{i+1}\otimes e_N + e_N\otimes e_{i+1}
\end{array}\right.
\label{eightxiid}
\ee
Clearly,
\be
{\rm Ker}(\xi) = \span(e_I\otimes e_N+e_N\otimes e_I) & &
{\rm dim}_q {\rm Ker}(\xi) = q^{1-N}[N] \nn \\
{\rm CoIm}(\eta) = \span(e_1\otimes e_I+e_I\otimes e_1) & &
{\rm dim}_q {\rm CoIm}(\eta) = q^{N-1}[N]
\ee
i.e. just the same as in symmetric case above -- in full accordance
with (\ref{eightco}).

\bigskip

{\bf Identity choice for $\eta$}

\begin{picture}(200,150)(-200,-45)
\put(-120,40){\line(0,1){20}}
\put(-117,39){\line(3,-1){114}}
\put(-117,59){\line(3,-1){114}}
\put(-117,41){\line(3,1){114}}
\put(-117,61){\line(3,1){114}}
\put(0,0){\line(0,1){100}}
\put(120,0){\line(0,1){80}}
\qbezier(3,0)(60,0)(119,0)
\qbezier(3,19)(60,10)(119,1)
\qbezier(3,80)(60,80)(119,80)
\qbezier(3,99)(60,90)(119,81)
%
%\put(3,19){\line(3,-1){54}}
%\put(3,99){\line(3,-1){54}}
%\put(60,0){\line(0,1){80}}
%\put(63,-1){\line(3,-1){54}}
%\put(63,79){\line(3,-1){54}}
%\put(120,-20){\line(0,1){100}}
%
\put(-70,27){\vector(3,-1){20}}
\put(-65,28){\mbox{$\nabla$}}
\put(-50,70){\vector(-3,-1){20}}
\put(-65,70){\mbox{$\Delta$}}
\put(-135,20){\mbox{${\cal U}_\circ = V$}}
\put(-37,-23){\mbox{${\cal V}_\bullet = {\cal U}_\bullet = V^{\otimes 2}$}}
\put(105,-23){\mbox{${\cal V}_\circ = V^{\otimes 2}/V$}}
%\put(105,-39){\mbox{${\cal V}_\bullet= V^{\otimes 2}$}}
%
%
\qbezier(50,96)(62,94)(74,92)
\put(73,92){\vector(3,=1){2}}
\put(57,99){\mbox{$\xi$}}
\put(70,75){\vector(-1,0){20}}
%\qbezier(74,84)(62,82)(50,80)
\put(60,65){\mbox{$\eta$}}
\put(-120,50){\circle*{3}}
\put(0,50){\circle*{3}}
%\put(60,40){\circle*{3}}
\put(120,40){\circle*{3}}
%
%\put(10,5){\vector(3,-1){30}}
%\put(43,-12){\mbox{$0$}}
%
\qbezier(3,1)(15,40)(3,79)
\put(15,38){\mbox{${\rm Ker}(\Delta)$}}
\put(-140,50){\line(1,0){300}}
\put(130,38){\mbox{$-1$}}
\linethickness{1mm}
\put(1,0){\line(0,1){20}}
\put(1,80){\line(0,1){20}}
\end{picture}

\noindent
In this case $V^{\otimes 2}/V$
is represented by ${\rm Ker}(\Delta) =
\span\Big\{e_{i+1}\otimes e_{j+1},\ e_1\otimes e_{j+1} - e_{j+1}\otimes e_1\Big\}$.
It has dimension $q^{-1}[N][N-1]$.
This time identical on ${\rm Ker}(\Delta)$ is the map $\eta$,
so that $g(\eta)=0$.
As to  $\xi$, it is now  one-to-one on ${\rm CoIm}(\nabla)$
and has $g(\xi)=-2$:
\be
\xi = V^{\otimes 2} \longrightarrow V^{\otimes 2}/V: &
\left\{\begin{array}{ccc}
e_i\otimes e_j & \longrightarrow & e_{i+1}\otimes e_j - e_i\otimes e_{j+1} \\
e_i\otimes e_{N} - e_{N}\otimes e_i & \longrightarrow &
e_{i+1}\otimes e_{N} + e_{N}\otimes e_{i+1} \\ \\
e_i\otimes e_N + e_N\otimes e_i & \longrightarrow & 0 \\
e_N\otimes e_N & \longrightarrow & 0
\end{array}\right. \nn \\ \nn \\
\eta = V^{\otimes 2}/V \longrightarrow V^{\otimes 2}: &
\left\{\begin{array}{ccc}
e_{i+1}\otimes e_{j+1} & \longrightarrow & e_{i+1}\otimes e_{j+1} \\
e_1\otimes e_{j+1} - e_{j+1}\otimes e_1 & \longrightarrow &
e_1\otimes e_{j+1} - e_{j+1}\otimes e_1
\end{array}\right.
\label{eightetaid}
\ee
Again,
\be
{\rm Ker}(\xi) = \span(e_I\otimes e_N+e_N\otimes e_I) & &
{\rm dim}_q {\rm Ker}(\xi) = q^{1-N}[N] \nn \\
{\rm CoIm}(\eta) = \span(e_1\otimes e_I+e_I\otimes e_1) & &
{\rm dim}_q {\rm CoIm}(\eta) = q^{N-1}[N]
\ee
in accordance with (\ref{eightco}).

\subsubsection{A $c\,$-dependent gauge choice and general procedure}

Clearly, identity maps of grading zero are much simpler to deal with.
Moreover, using them we make a conceptual simplification:
instead of arbitrary factor-spaces we consider canonically
defined ${\rm Ker}(\Delta)$ and ${\rm CoIm}(\nabla)$.
The price to pay for this is to allow the vector spaces at the hypercube
vertices to depend on initial vertex, i.e. on the coloring
$c$ of vertices in the knot/link diagram $D_c$.
For each particular $c$ we deal either with $\xi$ or with $\eta$,
but not with the both together.
Therefore, for each particular $c$ we can choose the spaces ${\cal V}$
so that the relevant maps are of the grading degree zero.
For another coloring we shift the spaces so that the new maps
have vanishing degree.

In our current example, we have just two choices of $c$:
with initial vertex black (left picture) and white (right picture):

\begin{picture}(200,230)(0,-130)
\put(20,20){\line(0,1){60}}
\put(100,40){\line(0,1){40}}
\put(30,40){\line(1,0){60}}
%\put(110,65){\mbox{${\cal V}_\circ} = $}}
\put(110,75){\mbox{{\footnotesize ${\cal V}_\circ
={\rm CoIm}(\nabla) = {\cal V}_\bullet/\nabla$}}}
\put(0,75){\mbox{{\footnotesize ${\cal V}_\bullet$}}}
\put(45,-20){\mbox{ $H_\bullet (O\!\!\!\bullet\!\!\!O)$}}
\qbezier(23,21)(30,30)(23,39)
\put(35,28){\vector(3,-1){30}}
%\put(25,23){\mbox{$\longrightarrow$}}
\put(70,14){\mbox{$0$}}
\put(50,28){\mbox{$\xi$}}
\put(52,55){\mbox{$\longrightarrow$}}
\put(58,62){\mbox{$\xi$}}
\qbezier(17,21)(10,30)(17,39)
\put(-18,28){\mbox{{\footnotesize $\im(\nabla)$}}}
\put(20,0){\line(1,0){78}}
\put(70,0){\vector(1,0){2}}
\put(20,0){\circle*{4}}
\put(100,0){\circle{4}}
\put(300,20){\line(0,1){60}}
\put(380,20){\line(0,1){40}}
\put(310,60){\line(1,0){60}}
%\put(310,40){\mbox{${\cal V}_\circ} = $}}
\put(390,54){\mbox{{\footnotesize
${\cal V}_\circ = {\rm Ker}(\Delta) = {\cal V}_\bullet\backslash\Delta$}}}
\put(280,75){\mbox{{\footnotesize ${\cal V}_\bullet$}}}
\put(325,-20){\mbox{ $H_\circ (O\!\!\!\bullet\!\!\!O)$}}
\qbezier(297,21)(285,40)(297,59)
\put(285,35){\vector(-1,-1){20}}
\put(250,10){\mbox{$0$}}
\put(265,33){\mbox{$\Delta$}}
\put(332,25){\mbox{$\longleftarrow$}}
\put(338,32){\mbox{$\eta$}}
\put(300,0){\line(1,0){78}}
\put(330,0){\vector(-1,0){2}}
\put(300,0){\circle*{4}}
\put(380,0){\circle{4}}
\put(150,-80){\line(1,0){78}}
\put(180,-80){\vector(-1,0){2}}
\put(150,-80){\circle*{4}}
\put(230,-80){\circle{4}}
\put(150,-80){\circle{8}}
\put(138,-60){\circle{20}}
\put(162,-60){\circle{20}}
\put(220,-60){\circle{20}}
\put(240,-60){\circle{20}}
\put(175,-100){\mbox{ $\tilde H_\bullet (O\!\!\!\bullet\!\!\!O)$}}
\end{picture}

\noindent
The difference between the two pictures is that the direction
of the arrow in the hypercube $H(O\!\!\!\bullet\!\!\!O)$:
it coincides with direction in the {\it primary} hypercube
$\tilde H(O\!\!\!\bullet\!\!\!O)$ (it is shown in the bottom)
in the right picture and it is opposite in the left one.
At the same time, it is clear that the left picture
(morphism $\xi$) is clearly {\it associated} with the cut operation
$\nabla$, while the right picture -- with the join operation $\Delta$.
Therefore in what follows
%{\bf
\bigskip

\fbox{\parbox{16cm}{
\noindent
we mark the edge of the hypercube $H_c(D)$ with a given
initial vertex by $\Delta$ or $\nabla$
if its direction is the same or the opposite to direction
of the same edge in the primary hypercube $\tilde H(D)$.
}}

\bigskip

\noindent
The new notation for the two colored hypercubes in
our current case of the knot eight is:

%\fbox{\parbox{16cm}{
\begin{picture}(200,150)(-50,-35)
\put(0,0){\vector(1,0){78}}
\put(0,0){\circle*{4}}
\put(80,0){\circle{4}}
\put(35,10){\mbox{$\nabla$}}
\put(-3,10){\mbox{${\cal V}_\bullet$}}
\put(70,10){\mbox{{ $
%{\cal V}_\circ} ={\rm CoIm}(\nabla) =
{\cal V}_\bullet/\nabla={\rm CoIm}(\nabla)$}}}
\put(25,-20){\mbox{$H_\bullet (O\!\!\!\bullet\!\!\!O)$}}
\put(298,0){\vector(-1,0){77}}
\put(220,0){\circle*{4}}
\put(300,0){\circle{4}}
\put(255,10){\mbox{$\Delta$}}
\put(217,10){\mbox{${\cal V}_\bullet$}}
\put(290,10){\mbox{{ $
%{\cal V}_\circ} ={\rm CoIm}(\nabla) =
{\cal V}_\bullet\backslash\Delta ={\rm Ker}(\Delta)$}}}
\put(245,-20){\mbox{$H_\circ (O\!\!\!\bullet\!\!\!O)$}}
\put(188,80){\vector(-1,0){77}}
\put(110,80){\circle*{4}}
\put(190,80){\circle{4}}
\put(145,90){\mbox{$\nabla$}}
\put(107,90){\mbox{${\cal U}_\bullet$}}
\put(180,90){\mbox{${\cal U}_\circ$ }}
\put(135,60){\mbox{$\tilde H(O\!\!\!\bullet\!\!\!O)$}}
\put(100,60){\vector(-1,-1){30}}
\put(200,60){\vector(1,-1){30}}
\put(-30,-30){\line(1,0){400}}
\put(-30,-30){\line(0,1){140}}
\put(370,110){\line(-1,0){400}}
\put(370,110){\line(0,-1){140}}
\put(335,95){\circle{10}}
\put(345,95){\circle{10}}
\put(335,78){\mbox{$D$}}
\put(310,88){\vector(-1,0){85}}
\end{picture}
%}}

\noindent

Note that the meaning of $\nabla$ in the primary hypercube
at the top of the picture and in the colored hypercube on the left
is different: at the top it is indeed a map, at the left it is
just a label, implying the action of the morphism $\xi$
in the way, which was shown in the previous picture.

Note also, that our new choice substitutes ${\cal U}_\bullet/{\cal U}_\circ$
over the white vertex by a vector space, depending on the choice
of coloring. In $\nabla$-case this is indeed a factor-space
${\cal U}_\bullet/\Delta = {\rm CoIm}(\nabla)$,
in the $\Delta$-case it is simply a subspace
${\cal U}_\bullet\backslash\Delta = {\rm Ker}(\Delta)$.
We remind that ${\cal V}_\bullet={\cal U}_\bullet$.

Finally. the morphisms can be read from (\ref{eightxiid}) and (\ref{eightetaid}):
\be
\xi = V^{\otimes 2} \longrightarrow V^{\otimes 2}/\nabla: &
\left\{\begin{array}{ccc}
e_i\otimes e_j & \longrightarrow & e_i\otimes e_j \\
e_i\otimes e_{N} - e_{N}\otimes e_i & \longrightarrow &
e_i\otimes e_{N} - e_{N}\otimes e_i \\ \\
e_i\otimes e_N + e_N\otimes e_i & \longrightarrow & 0 \\
e_N\otimes e_N & \longrightarrow & 0
\end{array}\right. \nn \\ \nn \\
\eta = V^{\otimes 2}\backslash \Delta \longrightarrow V^{\otimes 2}: &
\left\{\begin{array}{ccc}
e_{i+1}\otimes e_{j+1} & \longrightarrow & e_{i+1}\otimes e_{j+1} \\
e_1\otimes e_{j+1} - e_{j+1}\otimes e_1 & \longrightarrow &
e_1\otimes e_{j+1} - e_{j+1}\otimes e_1
\end{array}\right.
\ee
Sometime we will simply write $\nabla$ and $\Delta$
instead of  $\xi_\nabla$ and $\eta_\Delta$.

\subsubsection{Associated complex and unreduced superpolynomials}

In this particular case of a one-dimensional hypercube,
the last step -- building a complex from a commutative quiver
-- is just trivial ($H(O\!\!\!\bullet\!\!\!O)$ is
one-dimensional and does not contain any squares
to get rid of for promoting commutativity to nilpotency).
The complexes $K_\bullet(O\!\!\!\bullet\!\!\!O)$ and
$K_\circ(O\!\!\!\bullet\!\!\!O)$, associated with two possible
colorings of a single vertex in $D = O\!\!\!\bullet\!\!\!O$,
are just the hypercube $H(O\!\!\!\bullet\!\!\!O)$ itself with
the differentials $d_b = \xi$ and $d_w = \eta$ respectively:
\be
K_\bullet(O\!\!\!\bullet\!\!\!O) = \Big\{H_\bullet(O\!\!\!\bullet\!\!\!O), \xi\Big\}\ \ \
= \ \ \ \ 0\ \longrightarrow\ V^{\otimes 2}\ \stackrel{\xi}{\longrightarrow}\
V^{\otimes 2}/V \
\longrightarrow\ 0 \nn \\
K_\circ(O\!\!\!\bullet\!\!\!O) = \Big\{H_\circ(O\!\!\!\bullet\!\!\!O), \eta\Big\}\ \ \
= \ \ \ \ 0\ \longrightarrow\ V^{\otimes 2}\backslash V\ \stackrel{\eta}{\longrightarrow}\
V^{\otimes 2} \
\longrightarrow\ 0
\ee
Therefore the cohomologies of these complexes and their Poincare polynomials
can be just read from (\ref{eightco}).

Khovanov-Rozansky superpolynomial is obtained from Poincare polynomial
by adding a simple overall factor:
\be
q^{(N-1)\#({\rm black\ vertices\ in}\ D)}\cdot
(q^{N-1}T)^{-\#({\rm white\ vertices\ in}\ D)}
\label{normaf}
\ee
Since our morphisms along the edges of $H(D)$ have grading $0$,
the weights in the sum in Poincare polynomial
are powers of $T$, not $qT$.\footnote{
Note that in sec.\ref{modif} we tried to keep close to sec.\ref{Jones},
thus the factors were different and the weight was made out of $qT$.
It is an interesting question, if one can construct some other set
of morphisms in $H(D)$, with non-vanishing grading, to match those formulas.
However, the morphisms of grading zero seem extremely natural in
our construction.
To restore matching with sec.\ref{modif} it is sufficient to change

(i) eq.(\ref{weightsEu}) for $q^{(N-1)\#({\rm black\ vertices\ in}\ D)}\cdot
(-q^{N-1})^{\#({\rm white\ vertices\ in}\ D)} $,

(ii)  $(qT) \longrightarrow T$ in the weights,

(iii) the prescriptions for quantum dimensions.

\noindent
For example, in eq.(\ref{HOeight}) we should now assume that
${\rm dim}_q {\cal V}_\circ = q[N][N-1]$ with an extra factor of $q$, then
by the new rules
$$
H^{\bullet}_{_\Box} = q^{N-1}\Big([N]^2 - q[N][N-1]\Big) = [N] = \frac{\{A\}}{\{q\}},
\ \ \ \ \ \ \ \ \ \ \ \ \ \
H^{\circ}_{_\Box} = -q^{N-1}\Big([qN][N-1]-[N]^2\Big) = [N] = \frac{\{A\}}{\{q\}}
$$
and coincidence between the two polynomials is now just explicit term-by term.
The price to pay is explicit appearance of $q$-factors in the space dimensions --
which would be un-understandable in sec.\ref{modif}, but gets clear now, when
the spaces and morphisms are explicitly defined.
}
Actually, the value of this factor can be {\it obtained}
from the requirement that superpolynomials for the eight are the same as
for the circle  -- and then used in calculation for all other examples.

Finally, the superpolynomials
for the eight for initial vertex black an white are respectively
\be
{\cal P}^\bullet_{_\Box}(O\!\!\!\bullet\!\!\!O) \ =\  q^{N-1}\left\{
{\rm dim}_q{\rm Ker}
\Big( V^{\otimes 2} \ \stackrel{\xi}{\longrightarrow} \ V^{\otimes 2}/V \Big)
+ T \cdot{\rm dim}_q{\rm CoIm}
\Big( V^{\otimes 2} \ \stackrel{\xi}{\longrightarrow} \ V^{\otimes 2}/V \Big)
\right\}
= \nn\\ \nn\\
= q^{N-1}\Big( q^{1-N}[N] + (qT)\cdot 0\Big) = [N] = {\cal P}_{_\Box}(O)
\ \ \ \ \ \ \ \ \ \ \ \ \ \ \ \ \ \ \ \ \ \ \ \ \ \
\ee
and
\be
{\cal P}^\circ_{_\Box}(O\!\!\!\bullet\!\!\!O) \ =\  \frac{q^{1-N}}{T}\left\{
{\rm dim}_q{\rm Ker}
\Big( V^{\otimes 2}\backslash V \ \stackrel{\eta}{\longrightarrow} \ V^{\otimes 2} \Big)
+ T \cdot{\rm dim}_q{\rm CoIm}
\Big( V^{\otimes 2}\backslash V  \ \stackrel{\eta}{\longrightarrow} \  V^{\otimes 2} \Big)
\right\}
= \nn\\ \nn\\
= \frac{q^{1-N}}{T}\Big( 0 + T \cdot q^{N-1}[N]\Big) = [N] = {\cal P}_{_\Box}(O)
\ \ \ \ \ \ \ \ \ \ \ \ \ \ \ \ \ \ \ \ \ \ \ \ \ \
\ee

\subsubsection{Reduced superpolynomials}

As explained in sec.\ref{redunknot},  reduced superpolynomial
is obtained by the same construction, only one vector space
per vertex of the hypercube, associated with a cycle, passing
through a marked edge in $D$,
should be reduced from $N$-dimensional $V$ to $1$-dimensional $E$.
Clearly, in our construction this should be done
at the level of primary hypercube,
which is now (for the eight)
\be
E \stackrel{\nabla}{\longrightarrow} V
\ \ \ \ \ {\rm instead \ of} \ \ \ \ \
V \stackrel{\nabla}{\longrightarrow} V^{\otimes 2}
\ee
Since we want the cut-operation $\nabla$ to always have grading
$q^{1-N}$, the choice is predefined,
see s.\ref{eightprihy}:
\be
\nabla =  \ E\longrightarrow V: &\ \ \  E\longrightarrow e_N&
{\rm dim}_q{\rm CoIm}(\nabla) = [N]-q^{1-N} = q[N-1] \nn \\
\Delta =  \ V\longrightarrow E: &\ \ \
\left\{ \begin{array}{ccc} e_1 & \longrightarrow &  E \\
e_i & \longrightarrow & 0 \end{array}\right. \ \ \ \ \ \ \ \
& {\rm dim}_q{\rm Ker}(\Delta) = [N] -q^{N-1} = \frac{1}{q}[N-1]
\ee
Here $E$  with $g(E)=0$
is the single basis element in $E=C$, and we denote
it by the same letter as the space.

At two vertices of the main hypercube we now have
vector spaces $V\otimes E = V$ and $V\otimes E/E=V/E$ with the maps
(morphisms)
\be
\xi = V \longrightarrow V/\nabla: \ \ \
\left\{\begin{array}{rlc}
e_i & \longrightarrow & e_i \\
e_N & \longrightarrow &0
\end{array}
\right.
\ee
and
\be
\eta = V\backslash \Delta \longrightarrow V: \ \ \
\left\{\begin{array}{rlc}
e_{i+1} & \longrightarrow & e_{i+1}
\end{array}
\right.
\ee
with
\be
{\rm Ker} (\xi) = \{e_N\}  & {\rm dim}_q{\rm Ker}(\xi) = q^{1-N}\nn \\
{\rm CoIm} (\xi) = \emptyset & {\rm dim}_q{\rm CoIm}(\xi) = 0 \nn \\ \\
{\rm Ker} (\eta) = \emptyset & {\rm dim}_q{\rm Ker}(\eta) = 0\nn \\
{\rm CoIm} (\xi) = \{e_1\} & {\rm dim}_q{\rm CoIm}(\xi) = q^{N-1} \nn
\ee
This gives the proper reduced superpolynomials
\be
P^\bullet_{_\Box}(O\!\!\!\bullet\!\!\!O) = q^{N-1}\Big(q^{1-N}+T\cdot 0\Big)
= 1 = P_{_\Box}(O), \nn \\
P^\circ_{_\Box}(O\!\!\!\bullet\!\!\!O) = \frac{q^{1-N}}{T}
\Big(0 + T \cdot q^{N-1}\Big) = 1 = P_{_\Box}(O)
\ee

\subsection{Double eight. Combinations of cut and join}

In the next example the number of vertices in $D$ is $2$,
thus hypercube $H(D)$ is  a $2$-dimensional square (or rhombus).
In fact there are two different $D$ with two vertices:
double eight, consisting of three circles,
which is the unknot for any coloring, and
two circle intersecting at two points --
depending on coloring this is either a Hopf link
ar two disconnected unknots.
We begin in this section from the double eight example.

\subsubsection{Primary hypercube}

The starting point of our construction is the primary hypercube:

\begin{picture}(200,160)(-140,-90)
% rombus
\put(130,0){\line(-3,1){50}}\put(130,0){\vector(-3,1){30}}
\put(130,0){\line(-3,-1){50}}\put(130,0){\vector(-3,-1){30}}
\put(80,-17){\line(-3,1){50}} \put(80,-17){\vector(-3,1){30}}
\put(80,17){\line(-3,-1){50}} \put(80,17){\vector(-3,-1){30}}
\put(30,0){\circle*{3}}\put(130,0){\circle*{3}}
\put(80,17){\circle*{3}}\put(80,-17){\circle*{3}}
\put(30,0){\circle{8}}
\put(70,0){\circle{10}}\put(80,0){\circle{10}}\put(90,0){\circle{10}}
%
% bb
\put(27,10){\circle*{3}}
\put(33,10){\circle*{3}}
\put(-20,0){\circle{10}}\put(-7,0){\circle{10}}\put(6,0){\circle{10}}
\put(-25,-20){\mbox{{\footnotesize ${\cal U}_{\bu\bu}=V^{\otimes 3} $}}}
%
% wb
\put(65,22){\circle{3}}
\put(71,22){\circle*{3}}
\put(70,35){\circle{10}}\put(80,35){\circle{10}}\put(94,35){\circle{10}}
\put(60,52){\mbox{{\footnotesize ${\cal U}_{\w\bu}= V^{\otimes 2} $}}}
%
% bw
\put(65,-22){\circle*{3}}
\put(71,-22){\circle{3}}
\put(70,-35){\circle{10}}\put(84,-35){\circle{10}}\put(94,-35){\circle{10}}
\put(60,-52){\mbox{{\footnotesize ${\cal U}_{\bu\w} = V^{\otimes 2} $}}}
%
% ww
\put(127,10){\circle{3}}
\put(133,10){\circle{3}}
\put(155,0){\circle{10}}\put(165,0){\circle{10}}\put(175,0){\circle{10}}
\put(155,-20){\mbox{{\footnotesize ${\cal U}_{\w\w}=V $}}}
\put(60,-75){\mbox{$\tilde H(O\!\!\!\bullet\!\!\!O\!\!\!\bullet\!\!\!O)$}}
\put(45,14){\mbox{$\nabla_1$}}
\put(45,-19){\mbox{$\nabla_2$}}
\put(105,14){\mbox{$\nabla_3$}}
\put(105,-19){\mbox{$\nabla_4$}}
\end{picture}

\noindent
with embedding maps of degree $1-N$ are:
\be
\w\w\ \longrightarrow\ \w\bu:   &  e_I \ \longrightarrow \ e_I\otimes e_N + e_N\otimes e_I, \nn \\
\w\w\ \longrightarrow\ \bu\w:   &  e_I \ \longrightarrow \ e_I\otimes e_N + e_N\otimes e_I, \nn \\
\nn \\
\w\bu \longrightarrow\ \bu\bu:   &  e_I\otimes e_K \ \longrightarrow \
\Big(e_I\otimes e_N + e_N\otimes e_I\Big)\otimes e_K, \nn \\
\bu\w \longrightarrow\ \bu\bu:   &  e_I\otimes e_K  \ \longrightarrow \
e_I\otimes\Big(e_N\otimes e_K + e_K\otimes e_N\Big)
\label{doubleeightmaps}
\ee
The result of two consequetive embeddings
${\cal U}_{\w\w}\hookrightarrow {\cal U}_{\w\bu}\hookrightarrow {\cal U}_{\bu\bu}$
along the upper side of the rhombus
is the same of
${\cal U}_{\w\w}\hookrightarrow {\cal U}_{\bu\w}\hookrightarrow {\cal U}_{\bu\bu}$
along the lower side and coincides with the intersection
\be
{\rm Im}(\nabla \otimes I)\cap{\rm Im}(I\otimes \nabla) =
\span\Big(e_I\otimes e_N\otimes e_N + e_N\otimes e_I\otimes e_N +
e_N\otimes e_N\otimes e_I\Big)
\label{doubleeightintersect}
\ee
%It is easy to check that the corresponding operations $\xi$ and $\eta$

\subsubsection{The main hypercube
and the superpolynomials with initial vertices $\bullet\bullet$ and $\w\w$ }

According to our new strategy we should consider separately
the four versions of the main hypercube
$H_c(O\!\!\!\bullet\!\!\!O\!\!\!\bullet\!\!\!O)$
with four different initial vertices $v_c$.

The simplest are the two cases with $v_c=v_b=bb=\bullet\bullet$ and
$v_c = v_w=ww=\circ\circ$. In the first case all the arrows in
the hypercube are opposite to those in the primary hypercube,
while in the second case they coincide with those.
Accordingly all the arrows in the first case are of the $\nabla$ (cut) type,
and of the $\Delta$ (join) in the second
(note that the maps in the two pictures are actually $\xi$ and $\eta$,
while $\nabla$ and $\Delta$ play here the role of {\it labels},
marking the {\it type} of the spaces -- factor or sub -- and of the
morphisms):

\begin{picture}(200,180)(-60,-100)
% rombus
\put(30,0){\line(3,1){50}} \put(30,0){\vector(3,1){30}}
\put(30,0){\line(3,-1){50}} \put(30,0){\vector(3,-1){30}}
\put(80,-17){\line(3,1){50}} \put(80,-17){\vector(3,1){30}}
\put(80,17){\line(3,-1){50}}  \put(80,17){\vector(3,-1){30}}
\put(30,0){\circle*{3}}\put(130,0){\circle*{3}}
\put(80,17){\circle*{3}}\put(80,-17){\circle*{3}}
\put(70,0){\circle{10}}\put(80,0){\circle{10}}\put(90,0){\circle{10}}
%
% bb
\put(27,10){\circle*{3}}
\put(33,10){\circle*{3}}
\put(-20,0){\circle{10}}\put(-7,0){\circle{10}}\put(6,0){\circle{10}}
\put(-20,-20){\mbox{{\footnotesize ${\cal V}_{\bullet\bullet}=V^{\otimes 3} $}}}
\put(30,0){\circle{5}}\put(30,0){\circle{7}}

%
% wb
\put(65,22){\circle{3}}
\put(71,22){\circle*{3}}
\put(44,35){\circle{10}}\put(57,35){\circle{10}}\put(70,35){\circle{10}}
\put(80,35){\mbox{$\Big/$}}
\put(95,35){\circle{10}}\put(105,35){\circle{10}}\put(119,35){\circle{10}}
\put(40,52){\mbox{{\footnotesize ${\cal V}_{\circ\bullet} = (V^{\otimes 2}/V)\otimes V  $}}}
%
% bw
\put(65,-22){\circle*{3}}
\put(71,-22){\circle{3}}
\put(44,-35){\circle{10}}\put(57,-35){\circle{10}}\put(70,-35){\circle{10}}
\put(80,-35){\mbox{$\Big/$}}
\put(95,-35){\circle{10}}\put(109,-35){\circle{10}}\put(119,-35){\circle{10}}
\put(40,-52){\mbox{{\footnotesize ${\cal V}_{\bullet\circ}=V\otimes (V^{\otimes 2}/V)  $}}}
%
% ww
\put(127,10){\circle{3}}
\put(133,10){\circle{3}}
\put(152,0){\circle{10}}\put(165,0){\circle{10}}\put(178,0){\circle{10}}
\put(190,0){\mbox{$\Big/\ \Big($}}
\put(215,0){\circle{10}}\put(225,0){\circle{10}}\put(239,0){\circle{10}}
\put(250,-3){\mbox{$+$}}
\put(269,0){\circle{10}}\put(283,0){\circle{10}}\put(293,0){\circle{10}}
\put(302,0){\mbox{$\Big)$}}
\put(160,-20){\mbox{{\footnotesize ${\cal V}_{\circ\circ}=
V^{\otimes 3}/\big(V^{\otimes 2}+V^{\otimes 2}\big) $}}}
\put(60,-80){\mbox{$H_{\bullet\bullet}(O\!\!\!\bullet\!\!\!O\!\!\!\bullet\!\!\!O)$}}
\put(45,14){\mbox{$\nabla_1$}}
\put(45,-19){\mbox{$\nabla_2$}}
\put(105,14){\mbox{$\nabla_3$}}
\put(105,-19){\mbox{$\nabla_4$}}
\end{picture}

\begin{picture}(200,180)(-140,-100)
% rombus
\put(130,0){\line(-3,1){50}} \put(130,0){\vector(-3,1){30}}
\put(130,0){\line(-3,-1){50}} \put(130,0){\vector(-3,-1){30}}
\put(80,-17){\line(-3,1){50}} \put(80,-17){\vector(-3,1){30}}
\put(80,17){\line(-3,-1){50}}  \put(80,17){\vector(-3,-1){30}}
\put(30,0){\circle*{3}}\put(130,0){\circle*{3}}
\put(80,17){\circle*{3}}\put(80,-17){\circle*{3}}
\put(70,0){\circle{10}}\put(80,0){\circle{10}}\put(90,0){\circle{10}}
%
% bb
\put(27,10){\circle*{3}}
\put(33,10){\circle*{3}}
\put(-20,0){\circle{10}}\put(-7,0){\circle{10}}\put(6,0){\circle{10}}
\put(-20,-20){\mbox{{\footnotesize ${\cal V}_{\bullet\bullet}=V^{\otimes 3} $}}}
%
% wb
\put(65,22){\circle{3}}
\put(71,22){\circle*{3}}
\put(44,35){\circle{10}}\put(57,35){\circle{10}}\put(70,35){\circle{10}}
\put(80,35){\mbox{$\!\Big\backslash$}}
\put(95,35){\circle{10}}\put(105,35){\circle{10}}\put(119,35){\circle{10}}
\put(40,52){\mbox{{\footnotesize ${\cal V}_{\circ\bullet} =
(V^{\otimes 2}\backslash V)\otimes V  $}}}
%
% bw
\put(65,-22){\circle*{3}}
\put(71,-22){\circle{3}}
\put(44,-35){\circle{10}}\put(57,-35){\circle{10}}\put(70,-35){\circle{10}}
\put(80,-35){\mbox{$\!\Big\backslash$}}
\put(95,-35){\circle{10}}\put(109,-35){\circle{10}}\put(119,-35){\circle{10}}
\put(40,-52){\mbox{{\footnotesize
${\cal V}_{\bullet\circ}=V\otimes (V^{\otimes 2}\backslash V)  $}}}
%
% ww
\put(127,10){\circle{3}}
\put(133,10){\circle{3}}
\put(152,0){\circle{10}}\put(165,0){\circle{10}}\put(178,0){\circle{10}}
\put(190,0){\mbox{$\!\Big\backslash\ \Big($}}
\put(215,0){\circle{10}}\put(225,0){\circle{10}}\put(239,0){\circle{10}}
\put(250,-3){\mbox{$\cap$}}
\put(269,0){\circle{10}}\put(283,0){\circle{10}}\put(293,0){\circle{10}}
\put(302,0){\mbox{$\Big)$}}
\put(160,-20){\mbox{{\footnotesize ${\cal V}_{\circ\circ}=
V^{\otimes 3}\backslash\big(V^{\otimes 2} \cap V^{\otimes 2}\big) $}}}
\put(130,0){\circle{5}}\put(130,0){\circle{7}}
\put(30,0){\circle{7}}
\put(60,-80){\mbox{$H_{\circ\circ}(O\!\!\!\bullet\!\!\!O\!\!\!\bullet\!\!\!O)$}}
\put(45,14){\mbox{$\Delta_1$}}
\put(45,-19){\mbox{$\Delta_2$}}
\put(105,14){\mbox{$\Delta_3$}}
\put(105,-19){\mbox{$\Delta_4$}}
\end{picture}

\noindent
The choice of spaces ${\cal V}_{\circ\bullet}$ and ${\cal V}_{\bullet\circ}$ follows our
previous example for a single eight in sec.\ref{eighthom}  --
they are defined just at the segments $(\bu\w,\w\bu)$ and $(\bu\bu,\bu\w)$ respectively,
which are equivalent to the corresponding $H_c(O\!\!\!\bullet\!\!\!O)$ --
with no reference to the rest of $H_c(O\!\!\!\bullet\!\!\!O\!\!\!\bullet\!\!\!O)$.
A new thing is the space ${\cal V}_{\w\w}$ -- it knows about the
entire $H_c(O\!\!\!\bullet\!\!\!O\!\!\!\bullet\!\!\!O)$.
and now matters everything, what is embedded into
${\cal U}_{\bu\bu}$  -- i.e. the  both spaces  ${\cal U}_{\bu\w}$  and ${\cal U}_{\w\bu}$
at once.

However, they are combined in different ways:
When two $\nabla$ arrows are entering the vertex $\circ\circ$,
we associate with this vertex a factor space over a {\it union} of these two spaces:
${\cal U}_{\bu\w} + {\cal U}_{\w\bu} = \span\Big({\cal U}_{\bu\w},\ {\cal U}_{\w\bu}\Big)$
which we denote by an ordinary plus sign in what follows.
When two $\Delta$ arrows are quitting the vertex $\circ\circ$,
we associate with this vertex a subspace, complementing
the {\it intersection} of the two:
${\cal U}_{\bu\w}\cap{\cal U}_{\w\bu}$.

From now our consideration of the two cases splits for a while,
only to merge again at the end of this section.

\bigskip

The following picture is for the case of $\boxed{c=\bullet\bullet}$.
If one wants all the spaces in this picture to be represented
by segments (rather than consists of two pieces, cometime)
it is enough to imagine that the space ${\cal U}_{\bu\bu}$
is a circle, i.e. the points $a$ and $e=a$ coincide,
and segment $yx$ is  a complement of $xy$.

\begin{picture}(300,160)(-70,-22)
\put(-80,120){\mbox{$\boxed{c=\bu\bu}$}}
\put(0,0){\line(0,1){120}}
\put(5,60){\line(0,1){60}}
\put(10,20){\line(0,1){60}}
\put(-10,117){\mbox{$a$}}
\put(-10,77){\mbox{$b$}}
\put(-10,57){\mbox{$c$}}
\put(-10,17){\mbox{$d$}}
\put(-30,-3){\mbox{$e=a$}}
\put(-75,65){\mbox{$ae = {\cal V}_{\bu\bu}$}}
\put(-70,53){\mbox{$ = {\cal U}_{\bu\bu}$}}
\put(155,40){\mbox{$ce = {\rm Im}(\nabla_2) \approx {\cal U}_{\bu\w}$}}
\put(35,102){\mbox{$ac = {\cal V}_{\bu\bu}\big/\nabla_2 =$}}
\put(45,90){\mbox{$ =\coim(\nabla_2) = {\cal V}_{\bu\w}$}}
\put(175,110){\mbox{$db = {\rm Ker}(\xi_1)=$}}
\put(177,98){\mbox{${\rm Im}(\nabla_1) \approx {\cal U}_{\w\bu}$}}
\put(35,42){\mbox{$bd = {\cal V}_{\bu\bu}\big/\nabla_1 = $}}
\put(45,30){\mbox{$=\coim(\nabla_1) = {\cal V}_{\w\bu}$}}
\put(270,5){\mbox{$de={\rm Im}(\nabla_1)\,\cap\,{\rm Im}(\nabla_2)\approx {\cal U}_{\w\w}$}}
\put(255,100){\mbox{$cb=\span\Big({\rm Im}(\nabla_1),{\rm Im}(\nabla_2)\Big)$}}
\put(50,65){\mbox{$bc={\cal V}_{\bu\bu}\big/(\nabla_1+\nabla_2)= {\cal V}_{\w\w}$}}
\qbezier(-13,110)(-30,60)(-13,10)
\qbezier(13,63)(25,90)(13,117)
\qbezier(13,23)(25,50)(13,77)
\qbezier(23,61)(30,70)(23,79)
\put(28,60){\line(1,0){240}}
\put(28,80){\line(1,0){240}}
\put(20,20){\line(1,0){260}}
\put(170,120){\vector(0,-1){38}}
\put(170,0){\vector(0,1){18}}
\put(150,2){\vector(0,1){56}}\put(150,4){\vector(0,-1){2}}
\put(250,120){\vector(0,-1){38}}
\put(250,0){\vector(0,1){58}}
\put(265,2){\vector(0,1){16}}\put(265,4){\vector(0,-1){2}}
\end{picture}

\noindent
Since we choose our morphisms $\xi_i$ to be of degree zero,
all the factor-spaces can be {\it de facto} identified
with the subsets of ${\cal V}_{\bu\bu}$, where they act
as identities, so that our quiver is obviously Abelian:
\be
\xi_3\xi_1 = \xi_4\xi_2
\label{doubleeightAb}
\ee

With this Abelian quiver one naturally associates a complex
\be
K^{\bu\bu}(O\!\!\!\bullet\!\!\!O\!\!\!\bullet\!\!\!O)
= \ \ 0 \ \longrightarrow \ {\cal V}_{\bu\bu} \ \stackrel{d_0}{\longrightarrow}
 \ {\cal V}_{\bu\w}\oplus {\cal V}_{\w\bu} \ \stackrel{d_1}{\longrightarrow}
 \ {\cal V}_{\w\w} \ \longrightarrow \ 0
\ee
were the two differentials are
\be
d_0 = \left(\begin{array}{cc} \xi_1 &\xi_2  \end{array}\right), \ \ \ \ \ \ \
d_1 = \left(\begin{array}{c} \xi_3 \\ -\xi_4  \end{array}\right)
\ee
and $d_1d_0=0$ as a corollary of (\ref{doubleeightAb}).
Superpolynomial is just a Poincare polynomial of this complex,
multiplied by additional factor (\ref{normaf}):
\be
{\cal P}^{\bu\bu}_{_\Box}(O\!\!\!\bullet\!\!\!O\!\!\!\bullet\!\!\!O)
= q^{2(N-1)}\Big\{ {\rm dim}_q{\rm Ker}(d_0) \ + \
T\cdot\Big(\dq{\rm Ker}(d_1)-\dq\im(d_0)\Big) \ + \
T^2\cdot\dq\coim(d_1)\Big\}
\ee
It is clear from the picture that
\be
\ker(d_0) = \ker(\xi_1)\cap\ker(\xi_2) \ \stackrel{{\rm def}}{=}\
\im(\nabla_1)\cap\im(\nabla_2) = (de)
\ee
while $\im(d_0) = \ker(d_1)$ and $\coim(d_1)=\emptyset$, so that
\be
{\cal P}^{\bu\bu}_{_\Box}(O\!\!\!\bullet\!\!\!O\!\!\!\bullet\!\!\!O)
= q^{2(N-1)}\Big\{ {\rm dim}_q(de) + T\cdot 0 + T^2\cdot 0\Big\}
\ \stackrel{(\ref{doubleeightintersect})}{=} =
q^{2(N-1)}\cdot q^{2-2N}[N] = [N] = {\cal P}_{_\Box}(O)
\ee
is indeed equal to the superpolynomial for the unknot.
The same is true in the {\it reduced} case, where the only change is that
the intersection (\ref{doubleeightintersect}) reduces to just
a single element $e_N\otimes e_N$, which has quantum dimension $q^{2(1-N)}$,
so that the reduced superpolynomial is
\be
{P}^{\bu\bu}_{_\Box}(O\!\!\!\bullet\!\!\!O\!\!\!\bullet\!\!\!O)
= q^{2(N-1)}\Big\{ {\rm dim}_q(e_N\otimes e_N) + T\cdot 0 + T^2\cdot 0\Big\}
\ \stackrel{(\ref{doubleeightintersect})}{=}
q^{2(N-1)}\cdot q^{2-2N} = 1 = {P}_{_\Box}(O)
\ee

\bigskip

Remarkably in the case of $\ \boxed{c=\w\w}\ $ we can draw just the same picture,
only upside-down.
What is different is just {\it interpretation} of different segments:
one changes $\nabla$ for $\Delta$,
coimages for kernels and factor-spaces for subspaces  -- and of course
the subspaces themselves are quite different from the $\bu\bu$ case:
are spanned by the basises of quite different grading.
%In fact, symbolically more correct picture could be its upside-down
%version.

\begin{picture}(300,160)(-50,-140)
\put(-75,-120){\mbox{$\boxed{c=\w\w}$}}
\put(0,0){\line(0,-1){120}}
\put(5,-60){\line(0,-1){60}}
\put(10,-20){\line(0,-1){60}}
\put(-10,-123){\mbox{$a'$}}
\put(-10,-82){\mbox{$b'$}}
\put(-10,-62){\mbox{$c'$}}
\put(-10,-22){\mbox{$d'$}}
\put(-30,-2){\mbox{$e'=a'$}}
\put(-75,-58){\mbox{$a'e' = {\cal V}_{\bu\bu}$}}
\put(-70,-70){\mbox{$ = {\cal U}_{\bu\bu}$}}
\put(155,-40){\mbox{$c'e' = {\rm CoKer}(\Delta_2) \approx {\cal U}_{\bu\w}$}}
\put(35,-95){\mbox{$a'c' = {\cal V}_{\bu\bu}\big\backslash\Delta_2 = \im(\eta_1)$}}
\put(45,-107){\mbox{$ =\ker(\Delta_2) = {\cal V}_{\bu\w}$}}
\put(175,-103){\mbox{$d'b' = {\rm CoKer}(\Delta_1)$}}
\put(190,-115){\mbox{$\approx {\cal U}_{\w\bu}$}}
\put(35,-35){\mbox{$b'd' = {\cal V}_{\bu\bu}\big\backslash\Delta_1 = \im(\eta_2) $}}
\put(45,-47){\mbox{$=\ker(\Delta_1) = {\cal V}_{\w\bu}$}}
\put(285,-10){\mbox{$d'e'={\rm Co}\Big({\rm Ker}(\Delta_1)\cup{\rm Ker}(\Delta_2)\Big)$}}
\put(352,-24){\mbox{$\approx {\cal U}_{\w\w}$}}
\put(275,-105){\mbox{$c'b'={\rm Co}\Big({\rm Ker}(\Delta_1)\cap{\rm Ker}(\Delta_2)\Big)$}}
\put(50,-73){\mbox{$b'c'={\cal V}_{\bu\bu}\big\backslash(\Delta_1\wedge\Delta_2)
= \ker(\Delta_1)\cap \ker(\Delta_2) = {\cal V}_{\w\w}$}}
\qbezier(-13,-110)(-30,-60)(-13,-10)
\qbezier(13,-63)(25,-90)(13,-117)
\qbezier(13,-23)(25,-50)(13,-77)
\qbezier(23,-61)(30,-70)(23,-79)
\put(28,-60){\line(1,0){250}}
\put(28,-80){\line(1,0){250}}
\put(20,-20){\line(1,0){270}}
\put(170,-120){\vector(0,1){38}}
\put(170,0){\vector(0,-1){18}}
\put(150,-2){\vector(0,-1){56}}\put(150,-4){\vector(0,1){2}}
\put(270,-120){\vector(0,1){38}}
\put(270,0){\vector(0,-1){58}}
\put(280,-2){\vector(0,-1){16}}\put(280,-4){\vector(0,1){2}}
\end{picture}

\noindent
Since our agreement is that this time of degree zero are the
morphisms $\eta_i$,
they act on the subspaces
as identities, and the quiver is again obviously Abelian:
\be
\eta_1\eta_3 = \eta_2\eta_4
\label{doubleeightAb1}
\ee
Associated complex is now
\be
K^{\w\w}(O\!\!\!\bullet\!\!\!O\!\!\!\bullet\!\!\!O)
= \ \ 0 \ \longleftarrow \ {\cal V}_{\bu\bu} \ \stackrel{d_1^{\w\w}}{\longleftarrow}
 \ {\cal V}_{\bu\w}\oplus {\cal V}_{\w\bu} \ \stackrel{d_0^{\w\w}}{\longleftarrow}
 \ {\cal V}_{\w\w} \ \longleftarrow \ 0
\ee
were the two differentials are
\be
d_0^{\w\w} = \left(\begin{array}{cc} \eta_3 &-\eta_4  \end{array}\right), \ \ \ \ \ \ \
d_1^{\w\w} = \left(\begin{array}{cc} \eta_1 &\eta_2  \end{array}\right)
\ee
and $d_1^{\w\w}d_0^{\w\w}=0$ as a corollary of (\ref{doubleeightAb1}).
Multiplying the   Poincare polynomial of this complex
by additional factor (\ref{normaf}), we obtain the Khovanov-Rozansky superpolynomial:
\be
{\cal P}^{\w\w}_{_\Box}(O\!\!\!\bullet\!\!\!O\!\!\!\bullet\!\!\!O)
= \frac{1}{q^{2(N-1)T^2}}\Big\{ {\rm dim}_q{\rm Ker}(d_0^{\w\w}) \ + \
T\cdot\Big(\dq{\rm Ker}(d_1^{\w\w})-\dq\im(d_0^{\w\w})\Big) \ + \
T^2\cdot\dq\coim(d_1^{\w\w})\Big\}
\ee
It is clear from the picture that
$\ \ker(d_1^{\w\w})=\emptyset\ $ and
$\ \im(d_0^{\w\w}) = \ker(d_1^{\w\w})$,\ while
\be
\coim(d_1^{\w\w}) = {\rm Co}\Big(\im(\eta_1)\cap\im(\eta_2)\Big)
\ \stackrel{{\rm def}}{=}\
{\rm Co}\Big(\ker(\Delta_1)\cap\ker(\Delta_2)\Big) =
{\rm CoKer}(\Delta_1) \cap {\rm CoKer}(\Delta_2) = (d'e')
\ee
so that
\be
{\cal P}^{\w\w}_{_\Box}(O\!\!\!\bullet\!\!\!O\!\!\!\bullet\!\!\!O)
= \frac{1}{q^{2(N-1)}T^2}\Big\{ 0 + T\cdot 0 + T^2\cdot {\rm dim}_q(de)\Big\}
= \frac{1}{q^{2(N-1)}T^2}\cdot T^2q^{2N-2}[N] = [N] = {\cal P}_{_\Box}(O)
\ee
Note, that in our pictures $d'e'$ can look similar to $de$,
but in fact they have different gradings -- different by a change $q\longrightarrow q^{-1}$.
In the {\it reduced} case, where the only change is that
the intersection of cokernels  reduces to just
a single element $e_1\otimes e_1$, which has quantum dimension $q^{2(1-N)}$,
so that the reduced superpolynomial is
\be
{P}^{\w\w}_{_\Box}(O\!\!\!\bullet\!\!\!O\!\!\!\bullet\!\!\!O)
= q^{2(N-1)}\Big\{ 0+ T\cdot 0 + T^2\cdot {\rm dim}_q(e_1\otimes e_1) \Big\}
\ \stackrel{(\ref{doubleeightintersect})}{=}
q^{-2(N-1)}T^{-2}\cdot T^2 q^{2N-2} = 1 = {P}_{_\Box}(O)
\ee

\bigskip

For two other initial vertices, $\w\bu$ and $\bu\w$,
the situation is a little more tricky, because
where two merging edges at the vertices of the hypercube
are of different types -- one $\nabla$ and one $\Delta$.
In order to handle such configurations we need still one
more reformulation of our approach.

\begin{picture}(200,130)(-120,-80)
% rombus
\put(30,0){\line(3,1){50}}
\put(30,0){\line(3,-1){50}}
\put(80,-17){\line(3,1){50}}
\put(80,17){\line(3,-1){50}}
\put(30,0){\circle*{3}}\put(130,0){\circle*{3}}
\put(80,17){\circle*{3}}\put(80,-17){\circle*{3}}
\put(30,0){\vector(3,1){30}}
\put(80,-17){\vector(-3,1){30}}
\put(80,-17){\vector(3,1){30}}
\put(130,0){\vector(-3,1){30}}
\put(70,0){\circle{10}}\put(80,0){\circle{10}}\put(90,0){\circle{10}}
%
% bb
\put(27,10){\circle*{3}}
\put(33,10){\circle*{3}}
\put(-20,0){\circle{10}}\put(-7,0){\circle{10}}\put(6,0){\circle{10}}
\put(-20,-20){\mbox{{\footnotesize ${\cal V}_{bb}=V^{\otimes 3} $}}}
%%
% wb
\put(65,22){\circle{3}}
\put(71,22){\circle*{3}}
%\put(44,35){\circle{10}}\put(57,35){\circle{10}}\put(70,35){\circle{10}}
%\put(80,35){\mbox{$\Big/$}}
%\put(95,35){\circle{10}}\put(105,35){\circle{10}}\put(119,35){\circle{10}}
%\put(40,52){\mbox{{\footnotesize ${\cal V}_{wb} = (V^{\otimes 2}/V)\otimes V  $}}}
%
% bw
\put(65,-22){\circle*{3}}
\put(71,-22){\circle{3}}
%\put(44,-35){\circle{10}}\put(57,-35){\circle{10}}\put(70,-35){\circle{10}}
%\put(80,-35){\mbox{$\Big/$}}
%\put(95,-35){\circle{10}}\put(109,-35){\circle{10}}\put(119,-35){\circle{10}}
%\put(40,-52){\mbox{{\footnotesize ${\cal V}_{bw}=V\otimes (V^{\otimes 2}/V)  $}}}
%
% ww
\put(127,10){\circle{3}}
\put(133,10){\circle{3}}
%\put(152,0){\circle{10}}\put(165,0){\circle{10}}\put(178,0){\circle{10}}
%\put(190,0){\mbox{$\Big/\ \Big($}}
%\put(215,0){\circle{10}}\put(225,0){\circle{10}}\put(239,0){\circle{10}}
%\put(250,-3){\mbox{$+$}}
%\put(269,0){\circle{10}}\put(283,0){\circle{10}}\put(293,0){\circle{10}}
%\put(302,0){\mbox{$\Big)$}}
%\put(160,-20){\mbox{{\footnotesize ${\cal V}_{ww}=
%V^{\otimes 3}/\big(V^{\otimes 2}+V^{\otimes 2}\big) $}}}
%
\put(80,-17){\circle{5}}\put(80,-17){\circle{7}}
\put(130,0){\circle{7}}
\put(45,14){\mbox{$\nabla_1$}}
\put(45,-19){\mbox{$\Delta_2$}}
\put(105,14){\mbox{$\Delta_3$}}
\put(105,-19){\mbox{$\nabla_4$}}
\put(60,-60){\mbox{$H_{\bu\w}(O\!\!\!\bullet\!\!\!O\!\!\!\bullet\!\!\!O)$}}
\end{picture}

\subsubsection{A slight reformulation}

We now draw the same pictures in a slightly different form

\begin{picture}(300,180)(-250,-100)
\put(-230,55){\mbox{$\boxed{c=\bu\bu}$}}
%
% left half
\put(-50,-60){\line(0,1){120}}
\put(-130,-40){\line(0,1){70}}
\put(-210,-20){\line(0,1){40}}
\qbezier(-50,-60)(-130,-40)(-210,-20)
\qbezier(-50,-20)(-130,0)(-210,20)
\qbezier(-130,30)(-90,20)(-50,10)
%\qbezier(-130,40)(-90,50)(-50,60)
%\qbezier(-130,0)(-90,10)(-50,20)
\put(-175,-5){\mbox{$\nabla_4$}}
\qbezier(-181,-6)(-173,-8)(-165,-10)
\put(-163,-10.5){\vector(4,-1){2}}
\put(-95,5){\mbox{$\nabla_2$}}
\qbezier(-101,4)(-93,2)(-85,0)
\put(-83,-0.5){\vector(4,-1){2}}
\put(-95,-25){\mbox{$\nabla_2$}}
\qbezier(-101,-26)(-93,-28)(-85,-30)
\put(-83,-30.5){\vector(4,-1){2}}
\put(-215,-40){\mbox{${\cal U}_{\w\w}$}}
\put(-135,-60){\mbox{${\cal U}_{\bu\w}$}}
\put(-75,-75){\mbox{${\cal U}_{\bu\bu}={\cal V}_{\bu\bu}$}}
%
%
%
%right half
\put(50,-60){\line(0,1){120}}
\put(130,-40){\line(0,1){70}}
\put(210,-20){\line(0,1){40}}
\qbezier(50,-60)(130,-40)(210,-20)
\qbezier(50,-20)(130,0)(210,20)
%\qbezier(130,40)(90,30)(50,20)
\qbezier(130,30)(90,45)(50,60)
\qbezier(130,0)(90,15)(50,30)
\put(175,-5){\mbox{$\nabla_3$}}
\qbezier(191,-8)(183,-10)(175,-12)
\put(173,-12.5){\vector(-4,-1){2}}
\put(95,25){\mbox{$\nabla_1$}}
\qbezier(108,16)(100,19)(92,22)
\put(90,22.5){\vector(-3,1){2}}
\put(95,-25){\mbox{$\nabla_1$}}
\qbezier(111,-28)(103,-30)(95,-32)
\put(93,-32.5){\vector(-4,-1){2}}
\put(205,-40){\mbox{${\cal U}_{\w\w}$}}
\put(125,-60){\mbox{${\cal U}_{\w\bu}$}}
\put(25,-75){\mbox{${\cal U}_{\bu\bu}={\cal V}_{\bu\bu}$}}
\put(-50,10){\line(1,0){100}}
\put(-50,30){\line(1,0){100}}
\put(-5,18){\mbox{${\cal V}_{\w\w}$}}
\put(-40,-60){\line(1,0){80}}
\put(-40,-20){\line(1,0){80}}
\put(-15,-43){\mbox{$\ker(d_0^{\bu\bu})$}}
\put(-40,-60){\line(0,1){40}}
\put(40,-60){\line(0,1){40}}
\put(-70,35){\mbox{${\cal V}_{\bu\w}$}}
\put(57,3){\mbox{${\cal V}_{\w\bu}$}}
\linethickness{1mm}
\put(-50,10){\line(0,1){50}}
\put(50,-20){\line(0,1){50}}
\end{picture}

\noindent
From (\ref{doubleeightmaps}) we know explicit expressions for the maps:
\be
\nabla_3 \Big(\w\w\ \longrightarrow\ \w\bu\Big):
&  e_I \ \longrightarrow \ e_I\otimes e_N + e_N\otimes e_I, \nn \\
\nabla_4\Big( \w\w\ \longrightarrow\ \bu\w\Big):   &  e_I \ \longrightarrow \ e_I\otimes e_N + e_N\otimes e_I, \nn \\
\nn \\
\nabla_1\Big( \w\bu \longrightarrow\ \bu\bu\Big):   &  e_I\otimes e_K \ \longrightarrow \
\Big(e_I\otimes e_N + e_N\otimes e_I\Big)\otimes e_K, \nn \\
\nabla_2\Big(\bu\w \longrightarrow\ \bu\bu\Big):   &  e_I\otimes e_K  \ \longrightarrow \
e_I\otimes\Big(e_N\otimes e_K + e_K\otimes e_N\Big)
\label{doubleeightmaps1}
\ee
%The result of two consequetive embeddings
%${\cal U}_{\w\w}\hookrightarrow {\cal U}_{\w\bu}\hookrightarrow {\cal U}_{\bu\bu}$
%along the upper side of the rhombus
%is the same of
%${\cal U}_{\w\w}\hookrightarrow {\cal U}_{\bu\w}\hookrightarrow {\cal U}_{\bu\bu}$
%along the lower side and coincides with the intersection
%\be
%{\rm Im}(\nabla \otimes I)\cap{\rm Im}(I\otimes \nabla) =
%\span\Big(e_I\otimes e_N\otimes e_N + e_N\otimes e_I\otimes e_N +
%e_N\otimes e_N\otimes e_I\Big)
%\label{doubleeightintersect1}
%\ee
In fact, the non-trivial cohomology in this case is given by
\be
\boxed{
\ker(d_0^{\bu\bu}) = \nabla_2\Big(\im(\nabla_4)\Big)\cap\nabla_1\Big(\im(\nabla_3)\Big)
} \ = \nn \\
= \span\Big(e_I\otimes e_N\otimes e_N + e_N\otimes e_I\otimes e_N +
e_N\otimes e_N\otimes e_I\Big)
\ee
In this particular case the two intersecting consequtive images are actually the same.

To avoid possible confusion, we note that in our picture a
part of $\nabla_1$ looks like a gradation-increasing map --
this is an artefact of the drawing, actually all $\nabla$ maps
are of degree $1-N$. The same applies to a part of $\Delta_1$
in the next picture.

\begin{picture}(300,180)(-250,-100)
\put(-230,55){\mbox{$\boxed{c=\w\w}$}}
%
% left half
\put(-50,-60){\line(0,1){120}}
\put(-130,-30){\line(0,1){70}}
\put(-210,-20){\line(0,1){40}}
\qbezier(-50,60)(-130,40)(-210,20)
\qbezier(-50,20)(-130,0)(-210,-20)
\qbezier(-130,-30)(-90,-20)(-50,-10)
%\qbezier(-130,40)(-90,50)(-50,60)
%\qbezier(-130,0)(-90,10)(-50,20)
\put(-180,10){\mbox{$\Delta_4$}}
\qbezier(-181,3)(-173,5)(-165,7)
\put(-183,2.5){\vector(-4,-1){2}}
\put(-100,25){\mbox{$\Delta_2$}}
\qbezier(-101,18)(-93,20)(-85,22)
\put(-103,17.5){\vector(-4,-1){2}}
\put(-100,-8){\mbox{$\nabla_2$}}
\qbezier(-101,-15)(-93,-13)(-85,-11)
\put(-103,-15.5){\vector(-4,-1){2}}
\put(-215,-40){\mbox{${\cal U}_{\w\w}$}}
\put(-135,-60){\mbox{${\cal U}_{\bu\w}$}}
\put(-75,-75){\mbox{${\cal U}_{\bu\bu}={\cal V}_{\bu\bu}$}}
%
%
%
%right half
\put(50,-60){\line(0,1){120}}
\put(130,-30){\line(0,1){70}}
\put(210,-20){\line(0,1){40}}
\qbezier(50,60)(130,40)(210,20)
\qbezier(50,20)(130,0)(210,-20)
%\qbezier(130,40)(90,30)(50,20)
\qbezier(130,-30)(90,-45)(50,-60)
\qbezier(130,0)(90,-15)(50,-30)
\put(175,5){\mbox{$\Delta_3$}}
\qbezier(191,-2)(183,0)(175,2)
\put(193,-2.5){\vector(4,-1){2}}
\put(95,25){\mbox{$\Delta_1$}}
\qbezier(108,18)(100,20)(92,22)
\put(110,17.5){\vector(4,-1){2}}
\put(95,-25){\mbox{$\Delta_1$}}
\qbezier(111,-28)(103,-31)(95,-34)
\put(113,-27.5){\vector(3,1){2}}
\put(205,-40){\mbox{${\cal U}_{\w\w}$}}
\put(125,-60){\mbox{${\cal U}_{\w\bu}$}}
\put(25,-75){\mbox{${\cal U}_{\bu\bu}={\cal V}_{\bu\bu}$}}
\put(-50,-10){\line(1,0){100}}
\put(-50,-30){\line(1,0){100}}
\put(-5,-23){\mbox{${\cal V}_{\w\w}$}}
\put(-40,60){\line(1,0){80}}
\put(-40,20){\line(1,0){80}}
\put(-25,37){\mbox{$\coim(d_1^{\w\w})$}}
\put(-40,20){\line(0,1){40}}
\put(40,20){\line(0,1){40}}
\put(-70,-45){\mbox{${\cal V}_{\bu\w}$}}
\put(57,-13){\mbox{${\cal V}_{\w\bu}$}}
\linethickness{1mm}
\put(-50,-60){\line(0,1){50}}
\put(50,-30){\line(0,1){50}}
\end{picture}

\noindent
The join operations act as follows:
\be
\Delta_3: & e_I\otimes e_K \longrightarrow e_I\otimes e_K & {\rm if}\ I\ {\rm or}\ K =1\nn \\
\Delta_4: & e_I\otimes e_K \longrightarrow e_I\otimes e_K & {\rm if}\ I\ {\rm or}\ K =1\nn \\
\nn \\
\Delta_1: & e_I\otimes e_J\otimes e_K \longrightarrow e_I\otimes e_J\otimes e_K &
{\rm if}\ I\ {\rm or}\ J =1\nn \\
\Delta_2: & e_I\otimes e_J\otimes e_K \longrightarrow e_I\otimes e_J\otimes e_K &
{\rm if}\ J\ {\rm or}\ K =1
\ee

\begin{picture}(300,180)(-250,-100)
\put(-230,55){\mbox{$\boxed{c=\bu\w}$}}
%
% left half
\put(-50,-60){\line(0,1){120}}
\put(-130,-30){\line(0,1){70}}
\put(-210,-20){\line(0,1){40}}
\qbezier(-50,60)(-90,50)(-130,40)
%\qbezier(-58,28)(-90,20)(-122,12)
\qbezier(-210,20)(-170,15)(-130,10)
\qbezier(-50,20)(-90,10)(-130,0)
\qbezier(-210,-20)(-170,-25)(-130,-30)
\qbezier(-130,-30)(-90,-20)(-50,-10)
%\qbezier(-130,40)(-90,50)(-50,60)
%\qbezier(-130,0)(-90,10)(-50,20)
\put(-180,0){\mbox{$\nabla_4$}}
\qbezier(-185,-3)(-177,-4)(-169,-5)
\put(-167,-5.25){\vector(4,-1){2}}
\put(-100,30){\mbox{$\Delta_2$}}
\qbezier(-101,23)(-93,25)(-85,27)
\put(-103,22.5){\vector(-4,-1){2}}
\put(-100,-8){\mbox{$\nabla_2$}}
\qbezier(-101,-15)(-93,-13)(-85,-11)
\put(-103,-15.5){\vector(-4,-1){2}}
\put(-215,-40){\mbox{${\cal U}_{\w\w}$}}
\put(-135,-60){\mbox{${\cal U}_{\bu\w}$}}
\put(-75,-75){\mbox{${\cal U}_{\bu\bu}={\cal V}_{\bu\bu}$}}
%
%
%
%right half
\put(50,-60){\line(0,1){120}}
\put(130,-40){\line(0,1){70}}
\put(210,-20){\line(0,1){40}}
\qbezier(50,-60)(90,-50)(130,-40)
\qbezier(58,-28)(90,-20)(122,-12)
\qbezier(210,-20)(170,-15)(130,-10)
\qbezier(50,-20)(90,-10)(130,0)
\qbezier(210,20)(170,25)(130,30)
%\qbezier(130,40)(90,30)(50,20)
\qbezier(130,30)(90,45)(50,60)
\qbezier(130,0)(90,15)(50,30)
\put(175,5){\mbox{$\Delta_3$}}
\qbezier(186,0)(178,1)(170,2)
\put(188,-0.25){\vector(4,-1){2}}
\put(95,25){\mbox{$\nabla_1$}}
\qbezier(108,16)(100,19)(92,22)
\put(90,22.5){\vector(-3,1){2}}
\put(95,-28){\mbox{$\nabla_1$}}
\qbezier(111,-31)(103,-33)(95,-35)
\put(93,-35.5){\vector(-4,-1){2}}
\put(205,-40){\mbox{${\cal U}_{\w\w}$}}
\put(125,-60){\mbox{${\cal U}_{\w\bu}$}}
\put(25,-75){\mbox{${\cal U}_{\bu\bu}={\cal V}_{\bu\bu}$}}
\put(-40,60){\line(1,0){80}}
\put(-40,30){\line(1,0){80}}
\put(-35,37){\mbox{${\footnotesize \ker(d_1)\Big/\im(d_0)}$}}
\put(-40,30){\line(0,1){30}}
\put(40,30){\line(0,1){30}}
\put(-40,-30){\line(1,0){80}}
\put(-40,-20){\line(1,0){80}}
%\put(-25,37){\mbox{$\coim(d_1^{\w\w})$}}
\put(-40,-30){\line(0,1){10}}
\put(40,-30){\line(0,1){10}}
\put(-70,-35){\mbox{${\cal V}_{\bu\w}$}}
\put(57,3){\mbox{${\cal V}_{\w\bu}$}}
\linethickness{1mm}
\put(-50,-60){\line(0,1){50}}
\put(50,-20){\line(0,1){50}}
\end{picture}

\noindent
The non-trivial cohomology in this case is
\be
\ker d_1^{\bu\w}\Big/ \im d_0^{\bu\w} =
 \cup \nabla_1\Big({\rm CoKer}(\Delta_3)\Big) = \nn \\
= \span\Big((e_1\otimes e_N+e_N\otimes e_1)\otimes e_I +
(e_I\otimes e_N+e_N\otimes e_I)\otimes e_1\Big)
\ee

\subsubsection{The main hypercube
and the superpolynomials with initial vertices $\bullet\w$ and $\w\bu$ }

Now the situation is a little more tricky, because we have vertices,
where two edges of different type -- $\nabla$ and $\Delta$ -- merge together.

\begin{picture}(200,180)(-60,-100)
% rombus
\put(30,0){\line(3,1){50}} \put(30,0){\vector(3,1){30}}
\put(30,0){\line(3,-1){50}} \put(30,0){\vector(3,-1){30}}
\put(80,-17){\line(3,1){50}} \put(80,-17){\vector(3,1){30}}
\put(80,17){\line(3,-1){50}}  \put(80,17){\vector(3,-1){30}}
\put(30,0){\circle*{3}}\put(130,0){\circle*{3}}
\put(80,17){\circle*{3}}\put(80,-17){\circle*{3}}
\put(70,0){\circle{10}}\put(80,0){\circle{10}}\put(90,0){\circle{10}}
%
% bb
\put(27,10){\circle*{3}}
\put(33,10){\circle*{3}}
\put(-20,0){\circle{10}}\put(-7,0){\circle{10}}\put(6,0){\circle{10}}
\put(-20,-20){\mbox{{\footnotesize ${\cal V}_{\bullet\bullet}=V^{\otimes 3} $}}}
%
% wb
\put(65,22){\circle{3}}
\put(71,22){\circle*{3}}
\put(44,35){\circle{10}}\put(57,35){\circle{10}}\put(70,35){\circle{10}}
\put(80,35){\mbox{$\Big/$}}
\put(95,35){\circle{10}}\put(105,35){\circle{10}}\put(119,35){\circle{10}}
\put(40,52){\mbox{{\footnotesize ${\cal V}_{\circ\bullet} = (V^{\otimes 2}/V)\otimes V  $}}}
%
% bw
\put(65,-22){\circle*{3}}
\put(71,-22){\circle{3}}
\put(44,-35){\circle{10}}\put(57,-35){\circle{10}}\put(70,-35){\circle{10}}
\put(80,-35){\mbox{$\Big/$}}
\put(95,-35){\circle{10}}\put(109,-35){\circle{10}}\put(119,-35){\circle{10}}
\put(40,-52){\mbox{{\footnotesize ${\cal V}_{\bullet\circ}=V\otimes (V^{\otimes 2}/V)  $}}}
\put(80,-17){\circle{5}}\put(80,-17){\circle{7}}
%
% ww
\put(127,10){\circle{3}}
\put(133,10){\circle{3}}
\put(152,0){\circle{10}}\put(165,0){\circle{10}}\put(178,0){\circle{10}}
\put(190,0){\mbox{$\Big/\ \Big($}}
\put(215,0){\circle{10}}\put(225,0){\circle{10}}\put(239,0){\circle{10}}
\put(250,-3){\mbox{$+$}}
\put(269,0){\circle{10}}\put(283,0){\circle{10}}\put(293,0){\circle{10}}
\put(302,0){\mbox{$\Big)$}}
\put(160,-20){\mbox{{\footnotesize ${\cal V}_{\circ\circ}=
V^{\otimes 3}/\big(V^{\otimes 2}+V^{\otimes 2}\big) $}}}
\put(60,-80){\mbox{$H_{\bullet\bullet}(O\!\!\!\bullet\!\!\!O\!\!\!\bullet\!\!\!O)$}}
\put(45,14){\mbox{$\nabla_1$}}
\put(45,-19){\mbox{$\nabla_2$}}
\put(105,14){\mbox{$\nabla_3$}}
\put(105,-19){\mbox{$\nabla_4$}}
\end{picture}

\noindent
The choice of spaces ${\cal V}_{wb}$ and ${\cal V}_{bw}$ follows our
previous example for a single eight in sec.\ref{eighthom}  --
they are defined just at the segments $(bb,wb)$ and $(bb,bw)$ respectively,
which are equivalent to $H(O\!\!\!\bullet\!\!\!O)$ --
with no reference to the rest of $H(O\!\!\!\bullet\!\!\!O\!\!\!\bullet\!\!\!O)$.
A new thing is the space ${\cal V}_{ww}$ -- it knows about the
entire $H(O\!\!\!\bullet\!\!\!O\!\!\!\bullet\!\!\!O)$ and factors
${\cal U}_{bb}$ over everything what is embedded into it,
i.e. over $\span({\cal U}_{wb},{\cal U}_{bw})$, which we will often
denote simply by ${\cal U}_{wb}+{\cal U}_{bw}$.
Existence of ${\cal U}_{ww}$ implies that the images of these two subspaces,
within ${\cal U}_{bb}$ can intersect.

In other words, the embedding pattern is as follows:

\begin{picture}(200,180)(-100,-25)
\put(0,0){\line(0,1){130}}
\put(150,0){\line(0,1){130}}
\put(152,0){\line(0,1){50}}
\put(148,30){\line(0,1){70}}
\put(15,30){\line(1,0){120}}
\put(15,50){\line(1,0){120}}
\put(15,100){\line(1,0){120}}
\put(-90,60){\mbox{${\footnotesize {\cal V}_{bb}={\cal U}_{bb}=V^{\otimes 3}}$}}
\put(167,54){\mbox{${\footnotesize {\cal U}_{bw}}$}}
\put(170,12){\mbox{${\footnotesize {\cal U}_{wb}}$}}
%{\cal V}_{bw}={\cal U}_{bb}/{\cal U}_{bw}
%=V\otimes(V^{\otimes 2}/V)}$}}
\put(185,81){\mbox{${\footnotesize {\cal V}_{wb}={\cal U}_{bb}/{\cal U}_{wb}
= (V^{\otimes 2}/V)\otimes V}$}}
\put(165,123){\mbox{${\footnotesize {\cal V}_{ww}={\cal U}_{bb}/({\cal U}_{bw}+{\cal U}_{wb})=}$}}
\put(178,110){\mbox{${\footnotesize
=V^3\Big/\span\Big(V\otimes(V^{\otimes 2}/V),(V^{\otimes 2}/V)\otimes V\Big)}$}}
\qbezier(153,103)(165,115)(153,125)
%\put(170,40){\vector(-1,0){10}}
\qbezier(153,33)(165,65)(153,93)
\qbezier(153,53)(210,88)(153,128)
\qbezier(153,3)(170,25)(153,47)
\put(100,38){\mbox{${\footnotesize {\cal U}_{wb}\cap{\cal U}_{bw}}$}}
\end{picture}

\noindent
Here all spaces ${\cal U}_v$ and ${\cal V}_v$ are shown as embedded
into the largest one,  ${\cal V}_{bb} = {\cal U}_{bb}$
(in this picture the factor-space
${\cal V}_{bw}={\cal U}_{bb}/{\cal U}_{bw} =V\otimes(V^{\otimes 2}/V)$
consists of two segments and is omitted). \\

Now it is clear what the factor-spaces are,
we can identify the basis vectors in them:
\be
{\rm vertex}\ bb:  &  V^{\otimes 3} = \span\{e_I\otimes J\otimes e_K\} = \span(IJK), \nn \\
{\rm vertex}\ wb:  &  (V^{\otimes 2}/V)\otimes V) =
\span\{e_i\otimes e_j\otimes e_K, (e_i\otimes e_N-e_N\otimes e_I)\otimes e_K\} =
\span(ijK, \widehat{iN}K), \nn \\
{\rm vertex}\ bw:  &  V\otimes(V^{\otimes 2}/V) =
\span\{e_I\otimes e_j\otimes e_k, \ e_I\otimes(e_j\otimes e_N-e_N\otimes e_j)\} =
\span(Ijk, I\widehat{Nk}), \nn \\
{\rm vertex}\ ww:  &  V^{\otimes 3}/(V^{\otimes 2}+V^{\otimes 2}) =
\span\{e_i\otimes e_j\otimes e_k, \ 2e_i\otimes e_N\otimes e_k-
e_i\otimes e_k\otimes e_N  - e_N\otimes e_i\otimes e_k\} = \nn\\
& \ \ \ \ \ \ \ \ \ \ \ \ \ \ \ \ \ \ \ \ \ \ \ \ \ \ \ \ \ \ \ \ \ \ \ \ \ \ \ \ \ \
\ \ \ \ \ \ \ \ \ \ \
= \span(ijk, \widehat{iN}k+i\widehat{Nk})
\ee
The only comment can be needed in the case of the last space ${\cal V}_{bb}$:
it is a complement in $V^{\otimes 3}$ to
$V^{\otimes 2} + V^{\otimes 2} \equiv \span(V^{\otimes 2},V^{\otimes 2})
= \span\Big((e_I\otimes e_N + e_N\otimes e_I)\otimes e_K, \
e_I\otimes(e_N\otimes e_K + e_K\otimes e_N)\Big)$,
and these two sets have a non-trivial intersection,
when embedded into ${\cal U}_{bb}$
\be
{\rm Emb}({\cal U}_{bw})\cap{\rm Emb}({\cal U}_{wb}) =
\span(\widetilde{INN}) = \span(e_I\otimes e_N\otimes e_N +
e_N\otimes e_I \otimes e_N + e_N\otimes e_N\otimes e_I)
\ee
of dimension
\be
\boxed{\ \
{\rm dim}_q\Big({\rm Emb}({\cal U}_{bw})\cap{\rm Emb}({\cal U}_{wb})\Big) =
q^{2-2N}[N]
\ \  }
\label{dimUU}
\ee
Therefore the dimension of the factor-space at the $ww$ vertex,
calculated in two ways, is
\be
{\rm dim}_q{\cal V}_{ww} = [N]^3 - 2q^{1-N}[N]^2+q^{2-2N}[N]
= q^2\underline{[N][N-1]^2} = q^3[N-1]^3 + q^2q^{1-N}[N-1]^2
\ee
Underlined product is what we took for dimension of this space in
(\ref{dimsde}), while $q^2$ appeared there as additional weight  in the
definition of the HOMFLY polynomial.
Similarly, dimensions
\be
{\rm dim}_q{\cal V}_{wb} = {\rm dim}_q{\cal V}_{bw}
= q^2[N-1]^2[N] + q\cdot q^{1-N}[N-1][N] = q[N][N-1]
\ee
include the factor $q$.

\subsubsection{Morphisms and superpolynomials}

Pictorially the structure of hypercube and morphisms is:

\begin{picture}(200,180)(-200,-15)
\put(0,0){\line(0,1){130}}
\put(200,0){\line(0,1){130}}
\put(202,0){\line(0,1){50}}
\put(198,30){\line(0,1){70}}
\put(-200,0){\line(0,1){130}}
\put(-202,0){\line(0,1){50}}
\put(-198,30){\line(0,1){70}}
\put(100,0){\line(0,1){130}}
\put(102,0){\line(0,1){50}}
\put(-100,0){\line(0,1){130}}
\put(-98,30){\line(0,1){70}}
%
%\put(15,30){\line(1,0){70}}
\put(15,50){\line(1,0){70}}
\put(115,100){\line(1,0){70}}
\put(-85,30){\line(1,0){70}}
%\put(-85,50){\line(1,0){70}}
\put(-85,100){\line(1,0){70}}
%
%\put(-85,50){\line(1,0){70}}
%\put(-185,30){\line(1,0){70}}
%\put(-185,50){\line(1,0){70}}
\put(-185,100){\line(1,0){70}}
\put(25,90){\vector(1,0){50}} \put(27,90){\vector(-1,0){2}}
\put(27,95){\mbox{{\footnotesize ${\cal V}_{bb} \longleftrightarrow {\cal V}_{wb}$}}}
\qbezier(5,53)(20,90)(5,127) \qbezier(95,53)(80,90)(95,127)
\qbezier(5,3)(15,25)(3,47) \put(12,7){\vector(2,-1){20}}
\put(35,-7){\mbox{$0$}}
\put(27,10){\mbox{{\footnotesize ${\cal V}_{bb} \longrightarrow {\cal V}_{wb}$}}}
\put(125,115){\vector(1,0){50}} \put(127,115){\vector(-1,0){2}}
\put(127,120){\mbox{{\footnotesize ${\cal V}_{wb} \longleftrightarrow {\cal V}_{ww}$}}}
\qbezier(105,103)(115,115)(105,127) \qbezier(195,103)(185,115)(195,127)
\qbezier(105,53)(115,75)(105,97) \put(114,71){\vector(2,-1){20}}
\put(137,57){\mbox{$0$}}
\put(128,70){\mbox{{\footnotesize ${\cal V}_{wb} \longrightarrow {\cal V}_{ww}$}}}
\put(-205,138){\mbox{${\cal V}_{ww}$}}
\put(-105,138){\mbox{${\cal V}_{bw}$}}  \put(-105,-12){\mbox{${\cal V}_{bw}$}}
\put(-5,138){\mbox{${\cal V}_{bb}$}}
\put(95,138){\mbox{${\cal V}_{wb}$}}
\put(195,138){\mbox{${\cal V}_{ww}$}}
\qbezier(-5,3)(-15,15)(-5,27)\qbezier(-5,33)(-20,65)(-5,97)\qbezier(-5,103)(-15,115)(-5,127)
\qbezier(-95,3)(-85,15)(-95,27)\qbezier(-95,103)(-85,115)(-95,127)
\qbezier(-105,103)(-115,115)(-105,127) \qbezier(-195,103)(-185,115)(-195,127)
\qbezier(-105,3)(-115,15)(-105,27)
\put(-175,115){\vector(1,0){50}} \put(-173,115){\vector(-1,0){2}}
\put(-175,120){\mbox{{\footnotesize ${\cal V}_{ww} \longleftrightarrow {\cal V}_{bw}$}}}
\put(-75,115){\vector(1,0){50}} \put(-73,115){\vector(-1,0){2}}
\put(-73,120){\mbox{{\footnotesize ${\cal V}_{bw} \longleftrightarrow {\cal V}_{bb}$}}}
\put(-75,10){\vector(1,0){50}} \put(-73,10){\vector(-1,0){2}}
\put(-73,15){\mbox{{\footnotesize ${\cal V}_{bw} \longleftrightarrow {\cal V}_{bb}$}}}
 \put(-112,7){\vector(-2,-1){20}}
\put(-140,-7){\mbox{$0$}}
\put(-170,10){\mbox{{\footnotesize ${\cal V}_{ww} \longleftarrow {\cal V}_{bw}$}}}
  \put(-15,55){\vector(-2,-1){20}}
\put(-45,35){\mbox{$0$}}
\put(-77,50){\mbox{{\footnotesize ${\cal V}_{bw} \longleftarrow {\cal V}_{bb}$}}}
\put(-175,105){\mbox{$-$}}

\end{picture}

\noindent
In this picture the $bb$ vertex is at the center,
$wb$ and $bw$ are to the right and to the left of it respectively,
and the $ww$ vertex is shown twice -- in the very right and the very left
column.

\bigskip

From this picture we immediately see  what happens when
{\it initial} vertex is, say, $bb$. Then
\be
{\rm Ker}\Big({\cal V}_{bb} \longrightarrow {\cal V}_{bw} \oplus {\cal V}_{wb}\Big)
= {\cal U}_{bw}\cap{\cal U}_{wb} =  \span(\widetilde{INN}), \nn \\ \nn \\
{\rm Im}\Big({\cal V}_{bb} \longrightarrow {\cal V}_{bw} \oplus {\cal V}_{wb}\Big)
= {\cal V}_{bw} \oplus {\cal V}_{wb}, \nn \\
{\rm Ker}\Big( {\cal V}_{bw} \oplus {\cal V}_{wb} \longrightarrow {\cal V}_{ww} \Big)
= {\cal V}_{bw} \oplus {\cal V}_{wb}, \nn \\ \nn \\
{\rm Im}\Big( {\cal V}_{bw} \oplus {\cal V}_{wb} \longrightarrow {\cal V}_{ww} \Big)
= {\cal V}_{ww}
\ee
Only the third line here requires a comment:
from the picture it can seem that there is no kernel at all,
but in fact, because the $ww$ is shown twice, there are {\it two} arrows
directed to it, moreover the left arrow acts with the minus sign --
therefore the kernel is non-vanishing and given by above formula.
Thus unreduced superpolynomial
\be
{\cal P}^{\bullet\bullet}_{_\Box}(O\!\!\!\bullet\!\!\!O\!\!\!\bullet\!\!\!O) \ =\  q^{2(N-1)}
\left(
{\rm dim}_q{\rm Ker}\Big({\cal V}_{bb} \longrightarrow {\cal V}_{bw} \oplus {\cal V}_{wb}\Big)
+ T\Big\{{\rm dim}_q{\rm Ker}\Big( {\cal V}_{bw} \oplus {\cal V}_{wb} \longrightarrow {\cal V}_{ww} \Big)
- \right.\nn \\ \left.
- {\rm dim}_q{\rm Im}\Big({\cal V}_{bb} \longrightarrow {\cal V}_{bw} \oplus {\cal V}_{wb}\Big)\Big\}
+ T^2{\rm dim}_q{\rm CoIm}\Big( {\cal V}_{bw} \oplus {\cal V}_{wb} \longrightarrow {\cal V}_{ww} \Big)
\right) = \nn \\
=q^{2(N-1)}\Big(q^{2-2N}[N]+T\cdot 0 +T^2\cdot 0\Big) = [N] = {\cal P}_{_\Box}(O)
\ee
As to reduced polynomial,
note, that this time there are two {\it a priori} inequivalent choices of the marked edge
in $D$:  on external circle and on internal circles of double eight.
However, in both cases the  spaces at the   vertices of
auxiliary hypercube are reduced the same way:
to
\be
{\cal U}_{bb} = V^{\otimes 2}, \ \ \ \ \ \
{\cal U}_{wb}=E\otimes V = V , \ \ \ \ \ \
{\cal U}_{bw} = V\otimes E=V, \ \ \ \ \ \
{\cal U}_{ww}=E
\ee
Actually such are the spaces, when a middle edge in $D$ is marked.
If it was instead an edge with both ends at $bb$, we would rather get
${\cal U}_{bw} = E\otimes V$ -- however it is again the same as $V$,
and in both cases
${\rm Emb}({\cal U}_{bw})\cap{\rm Emb}({\cal U}_{wb}) = {\rm Emb}(E)
= \span(e_N\otimes e_N)$
with
\be
\boxed{\ \
{\rm dim}_q\Big( {\rm Emb}({\cal U}_{bw})\cap{\rm Emb}({\cal U}_{wb})\Big)
= q^{2-2N}
\ \ } \ \ \ \ \ \ \
{\rm reduced\ case}
\label{dimUUred}
\ee
so that the reduced superpolynomial is
\be
 P^{\bullet\bullet}_{_\Box}(O\!\!\!\bullet\!\!\!O\!\!\!\bullet\!\!\!O) \ =\
q^{2N-2}\left({\rm dim}_q \Big({\rm Emb}({\cal U}_{bw})\cap{\rm Emb}({\cal U}_{wb})\Big)
+ T\cdot 0 + T^2\cdot 0\right) = 1 = P_{_\Box}(O)
\ee

\bigskip

Similarly, for the other choices of initial vertex:
\be
{\cal P}^{\circ\bullet}_{_\Box}(O\!\!\!\bullet\!\!\!O\!\!\!\bullet\!\!\!O) \ =\
\frac{q^{2(N-1)}}{T}
\left(
{\rm dim}_q{\rm Ker}\Big({\cal V}_{bw} \longrightarrow {\cal V}_{bb} \oplus {\cal V}_{ww}\Big)
+ T\Big\{{\rm dim}_q{\rm Ker}\Big( {\cal V}_{bb} \oplus {\cal V}_{ww} \longrightarrow {\cal V}_{wb} \Big)
- \right.\nn \\ \left.
- {\rm dim}_q{\rm Im}\Big({\cal V}_{bw} \longrightarrow {\cal V}_{bb} \oplus {\cal V}_{ww}\Big)\Big\}
+ T^2{\rm dim}_q{\rm CoIm}\Big( {\cal V}_{bb} \oplus {\cal V}_{ww} \longrightarrow {\cal V}_{wb} \Big)
\right) = \nn \\
=\frac{q^{2N-2}}{T}\Big(0 + T\cdot{\rm dim}_q
\Big({\rm Emb}({\cal U}_{bw})\cap{\rm Emb}({\cal U}_{wb})\Big)
+T^2\cdot 0\Big) = [N] = {\cal P}_{_\Box}(O),
\ee
\be
P^{\bullet\circ}_{_\Box}(O\!\!\!\bullet\!\!\!O\!\!\!\bullet\!\!\!O) \ =\
1 = P_{_\Box}(O)
\ee
and
\be
{\cal P}^{\circ\circ}_{_\Box}(O\!\!\!\bullet\!\!\!O\!\!\!\bullet\!\!\!O) \ =\
\frac{q^{2(N-1)}}{T^2}
\left(
{\rm dim}_q{\rm Ker}\Big({\cal V}_{ww} \longrightarrow {\cal V}_{bw} \oplus {\cal V}_{wb}\Big)
+ T\Big\{{\rm dim}_q{\rm Ker}\Big( {\cal V}_{bw} \oplus {\cal V}_{wb} \longrightarrow {\cal V}_{bb} \Big)
- \right.\nn \\ \left.
- {\rm dim}_q{\rm Im}\Big({\cal V}_{ww} \longrightarrow {\cal V}_{bw} \oplus {\cal V}_{wb}\Big)\Big\}
+ T^2{\rm dim}_q{\rm CoIm}\Big( {\cal V}_{bw} \oplus {\cal V}_{wb} \longrightarrow {\cal V}_{bb} \Big)
\right) = \nn \\
=\frac{q^{2N-2}}{T^2}\Big(0 + T\cdot 0
+T^2\cdot {\rm dim}_q
\Big({\rm Emb}({\cal U}_{bw})\cap{\rm Emb}({\cal U}_{wb})\Big)\Big) = [N] = {\cal P}_{_\Box}(O),
\ee
\be
P^{\circ\circ}_{_\Box}(O\!\!\!\bullet\!\!\!O\!\!\!\bullet\!\!\!O) \ =\
1 = P_{_\Box}(O)
\ee
Note that -- as clear from the above pictures --
the only non-vanishing contribution to all these formulas comes
from (\ref{dimUU}) and (\ref{dimUUred}), this is why we put them in boxes.

\subsection{The general procedure for the choice of the vector spaces
\label{gen}}

We end this preliminary presentation of the cut-and-join formalism
behind our version of KR calculus by formulating  the general rule,
for the choice of vector spaces at the vertices  of the hypercube $H(D)$.
This choice is straightforwardly dictated by the structure of
the primary hypercube $\tilde H(D)$.

\bigskip

\fbox{\parbox{16cm}{
%Now we need to construct the hypercube $H(O\!\!\!\bullet\!\!\!O\!\!\!\bullet\!\!\!O)$
%itself: namely, to define the vector spaces at its vertices.
Primary hypercube $\tilde H(D)$ has a preferred vertex  $v_b$ -- with all vertices
in $D$ black.
If we take any other vertex $v$ in $H(D)$, then it is also a vertex in
$\tilde H(D)$ and it cuts out a sub-cube $\tilde h_v \subset \tilde H(D)$ -- the  one,
between $v_b$ and this $v$.
%In our example this will be just a point $\bullet\bullet$, if $v=\bullet\bullet=v_b=bb$,
%a 1-dimensional sub-cube (segment) if $v=\bullet\circ=bw$ or $v=\circ\bullet=wb$
%and the entire 2-dimensional hypercube $\tilde H(O\!\!\!\bullet\!\!\!O\!\!\!\bullet\!\!\!O)$
%if $v=\circ\circ=ww$.

For the totally-black (Seifert) vertex the sub-cube is just this vertex itself:
$h_{v_b} = v_b$.
For the total-white vertex the sub-cube coincides with the entire hypercube:
$h_{v_w} = H(D)$.
If a straight path from $v$ to $v_b$ can pass through a vertex $v'$,
then $h_{v'}\subset h_v$: the sub-cubes $h_v$ form an embedded hierarchy
around $v_b$ (similar to a flag variety).

In the sub-cube $h_v$ there will be {\it drain} vertices -- where all arrows enter
and no one exits. Since arrows describe embeddings, one can factor a vector
space at the drain vertex over a  span  of all embedded spaces at the origins
of the entering edges.
Finally we associate with a hypercube vertex $v\in H(D)$ (in original hypercube)
a sum of these factor-spaces over all drain-vertices $d$, belonging to
the corresponding sub-cube $h_v$:
}}
\be
\boxed{
{\cal V}_v = \oplus_{d\in h_v\subset \tilde H(D)}  \ \ \ {\cal U}_d\Big/\Big(\span_{
\stackrel{w\rightarrow d}{w\in h_v}} {\cal U}_w\Big)
}
\label{defW}
\ee
The space in denominator of the factor is spanned by a combination ${\cal U}_w$
at the vertices $w$ in $h_v$, which are preimages of the given drain point $d$,
i.e. by definition of the drain point all of them are embedded into ${\cal U}_d$,
but can intersect  --
and after that we take a direct sum of such factor-spaces over all drain points
in $h_v$.
What is important, the drain points and their preimages are taken not from
entire hypercube, but from the $v$-dependent sub-cube $h_v$.
Note that all vector spaces ${\cal U}_v$ at the vertices of $\tilde H(D)$ are just
tensor powers of $V=C^N$.

Eq.(\ref{defW}) looks like a terribly complicated formula,
but, hopefully, after working through several examples
from sec.\ref{KRco}, its simple meaning gets perfectly clear.

\bigskip

\fbox{\parbox{16cm}{
The r.h.s. of (\ref{defW}) is somewhat symbolic, because it is important how the
vector spaces in "denominators" are embedded in those in the "numerators",
i.e. how the factor-spaces are actually defined.
This is, however, straightforwardly dictated by the embeddings
of spaces  ${\cal U}_v$  along the edges of the primary hypercube,
i.e. by the cut operation (\ref{cut}).
After that the cut and join operations (\ref{cut}) and (\ref{join}) define
what are all the morphisms in all directions along the edges of the
main hypercube.
They form a commuting set -- an {\it Abelian quiver}. Therefore,
once initial vertex $c_0$ is chosen in the main hypercube,
one can always construct associated complex ${\cal K}(D_{c_0})$.
After appropriate normalization its Poincare polynomial
{\bf coincides with Khovanov-Rozansky polynomial}, obtained by
{\bf a very different and far more complicated}
{\it matrix-factorization} technique.
}}

\bigskip

\section{Conclusion}

In this paper we suggested an {\it alternative} construction
of Khovanov-Rozansky superpolynomials for arbitrary
knots and links and for arbitrary gauge group $GL(N)$.
It is completely different from
%In variance with
the original matrix-factorization
construction of \cite{KhR} and our calculations
have nothing in common with those of \cite{CM} --
except for the answers.
Moreover, in our way we get the answers for all
values of $N$ at once.
Also calculations are extremely simple and easily
computerizeable -- probably even the programs,
used for the calculations of Jones superpolynomials
(i.e. for $N=2$), can be easily modified and used
for arbitrary $N$.
In the paper the simplest examples are done "by hands",
and these include the big part of the list of \cite{CM},
obtained by extremely tedious computer calculus
at particular $N$.
Moreover, we explained how the entire series of 2-strand
$k$-folds can be handled.
Extensions to other series, beginning from twist and
3-strand torus knots would be a natural next step to do.

The paper concentrates on the ideas,
and does not present the story as a systematic algorithmic
approach -- the ways to do this are outlined but nor
developed to the very end.
Accordingly, no general proof is given of the
\Rede\ invariance.
These issues, as well as the relation to the matrix-factorization
formalism, to Hecke-algebra \cite{Che,GoChe}
and to refined-Chern-Simons \cite{RCS1}-\cite{RCSl} approaches
will be discussed elsewhere.

Of more importance, however, would be practical calculations,
making the list of the Khovanov-Rozansky polynomials
as rich as that of HOMFLY. Moreover, this approach can
probably be more than competitive in HOMFLY calculations
themselves, like it already is for $N=2$, see s.4 in \cite{evo}.
The next step should be extension from fundamental to antisymmetric,
symmetric and arbitrary representations -- where already for
HOMFLY the standard ${\cal R}$-matrix
approach \cite{TR,MoSmi,MMMkn12,Anopaths} gets extremely tedious and too few results
are available, what slows down the progress in the field.
It looks like these extensions can also be found, by application
of the same tensor-algebra vision of \cite{nonlin}
which led to the important success, reported in the present paper.

\section*{Acknowledgements}

Our work is partly supported by the
Ministry of Education and Science of the Russian Federation
under the contract 8207,
by  the Brazil National Counsel of Scientific and Technological Development
and by the grants
NSh-3349.2012.2,
RFBR-13-02-00478,
12-02-92108-Yaf-a,
13-02-91371-ST-a,
14-01-93004-Viet.

\end{document}